\documentclass[%
aps,
pre,
preprint,%
longbibliography%
]{revtex4-1}

\usepackage{graphicx}   																												% need for figures & tables
\usepackage[percent]{overpic}																										% need for overlaying items on a figure
\usepackage{transparent}																												% need for creating transparent \colorbox{} item using overpic package
\usepackage[space]{grffile}																											% need for importing figure with spaces in the filename
\usepackage[hang, nooneline]{subfigure}  																				% use for creating subfigures
\usepackage{afterpage}																													% need for placing figure on the next immediate page

\usepackage[usenames, dvipsnames]{xcolor}                                       % need for using colors in document
\usepackage{amsmath}    																												% need for equations
\usepackage{amssymb}																														% need for displaying \therefore
\usepackage{amsbsy}																															% need for providing better bold math support (with \boldsymbol).
\usepackage{amsfonts}																														% need for displaying Mathbb fonts
\usepackage{mathrsfs}																														% need for Euler caligraphic uppercase fonts \mathscr{}
\usepackage{threeparttable}																											% for columns to be aligned on decimal points
\usepackage{multirow}																														% need for merging rows and columns in a table
\usepackage{dcolumn}																														% need for producing decimal-aligned tabular math columns
\usepackage{booktabs}																														% need for producing professional table
\usepackage{array}																															% need for resizing table
\usepackage{MnSymbol}																														% need for displaying \smallstar in math environments
\usepackage{dashrule}																														% need for drawing a variety of dashed or dotted line styles using the command \hdashrule
\usepackage[sort&compress]{natbib}     																					% need for in-text citation, entry in square bracket denotes the in-text citation style
\usepackage[colorlinks=false, allbordercolors=white]{hyperref}									% need for generating hyperlink on cross-references, always load as the last package

\newcommand{\dsty}{\displaystyle}																												% short form for \displaystyle
\DeclareMathAlphabet{\mathbbit}{U}{bbm}{m}{sl}																					% blackboard, italic font in mathmode

\DeclareMathAlphabet\mathsfbi{OT1}{cmss}{m}{sl}								% For tensor math font
\newcommand{\Rey}{\mbox{\textit{Re}}}  												% Reynolds number
\providecommand{\upi}{\pi}																		% Upright Pi symbol
\newcommand{\Um}{U_m}																					% Bulk streamwise velocity
\newcommand{\Utau}{u_{\tau}}																	% Friction velocity
\newcommand{\ReT}{\Rey_{\tau}}																% Friction Reynolds number
\newcommand{\Rem}{\Rey_m}																			% Bulk Reynolds number
\newcommand{\Sp}{s^+}																					% Spacing of riblets
\newcommand{\Hp}{h^+}																					% Height of riblets
\newcommand{\lgp}{l_g^+}																			% Characteristic length scale of riblets
\definecolor{tecplotgray}{gray}{0.275}														

\begin{document}

\title{On the turbulent flow field over riblets of various groove sizes at low Reynolds number} %Title of paper

\author{J. H. Ng}
\email[]{ngjeehann@u.nus.edu}
%\thanks{}
\affiliation{Department of Mechanical Engineering, National University of Singapore}

\author{R. K. Jaiman}
\email[Corresponding author: ]{mperkj@nus.edu.sg}
%\thanks{}
\affiliation{Department of Mechanical Engineering, National University of Singapore}

\author{T. T. Lim}
\email[]{mpelimtt@nus.edu.sg}
%\thanks{}
\affiliation{Department of Mechanical Engineering, National University of Singapore}

\begin{abstract}
	
	In this work, low Reynolds number turbulent flow through a corrugated channel, formed 
by standard V-groove riblets, are investigated via direct numerical simulations (DNS). 
The simulations attempt to assess the variation of flow dynamics caused by a change in the 
characteristic size of the V-groove riblets. Such characterization of size effects provides 
a means to identify the local flow features arising from fluid-riblet interaction, and to 
investigate their relevance to the change in viscous drag and turbulence statistics. The 
present simulations confirmed an improved variation trend when the flow dynamics is 
examined in the light of the groove size. This suggests that the transition across the 
range of local flow regimes perceived by the V-groove riblets may be better characterized 
by the groove size, as compared to the spacing or height. At the lower end of the range of 
groove size considered, the profiles of turbulence statistics more or less resemble the 
plane channel flow, except for a systematic shift with the groove size. When the groove 
size becomes increasingly large, the lodging of near-wall flow structures and the 
generation of mean secondary flows tend to be more apparent. The collective impact of 
the lodging of flow structures and the mean secondary flow correlates with the increase 
in viscous drag, and leads to significant alterations of the turbulence statistics. In 
addition, the invigorated near-wall fluid motions due to their closer proximity with the 
groove surface can cause the formation of humps on the velocity fluctuation profiles. 
Lastly, two correlations as a function of the groove size are explored to illustrate their 
potential to capture the overall effects of riblets. Such correlations involving riblets 
of various sizes may provide insights on modeling fluid-riblet interaction in low Reynolds 
number turbulent channel flows.

\end{abstract}


\maketitle %\maketitle must follow title, authors, abstract and \pacs

\section{Introduction}
\label{sec:Introduction}

	The ability to control the multitude of fluid flow phenomena has been a perpetual quest 
that mankind has delved into since the dawn of fluid study. In particular, flow over 
patterned surface is a fundamental problem that continues to be an active field of 
research owing to profound practical implications. Riblets, which are longitudinal 
striations of different cross-sections, have been a potential candidate for applications 
ranging from aircraft drag reduction at high Reynolds number~\citep{Gad-el-Hak00, 
Bushnell03}, to the mitigation of blood cell damage caused by flow-induced stress at 
moderate Reynolds number~\citep{Ramachandran16}. Grasping the flow field evolution with 
the characteristic size of riblets is pivotal to achieve the desired flow attributes in all 
these applications. Of particular interest in this work is to gain a more systematic 
understanding of fluid-riblet interaction in low Reynolds number turbulent flows, which 
may provide useful insights for improving the blood compatibility of cardiovascular 
devices via surface modification~\cite{Chen11}. Interested readers may recap about the 
research progress from a few comprehensive reviews~\citep{Pollard98, Bechert00Review, 
Choi01, Viswanath02, Karniadakis03, Dean10}.

	The immense amount of research in the past few decades has established that the drag 
reduction performance of riblets has a dependency on their size measured in wall 
units~\citep{Walsh83, Bruse93}, i.e. normalized by the viscous length scale $\nu/\Utau$ 
where $\nu$ is the kinematic viscosity and $\Utau = \sqrt{\tau_w/\rho}$ is the friction 
velocity derived from the wall shear stress $\tau_w$ and fluid density $\rho$. Typically, 
the spacing $s$ or the height $h$ of riblets measured in wall units is taken as the 
reference length scale to denote the size. As the size gets progressively larger, the 
flow field over riblets can exhibit different physics ranging from one dominated by 
viscous effects (\emph{viscous regime}) to another strongly influenced by non-linear 
inertial effects ($k$-\emph{roughness regime}). Such nature gives rise to the diverse 
viscous drag behavior of riblets with their size~\cite{Garcia11Review}, and in turn it 
renders the study and understanding of the physics of flow around riblets an arduous 
endeavor. Apart from the size, the shape of the riblets can lead to a varying degrees of 
performance~\cite{Walsh84, Walsh90A}. To date, the thin blade geometry at the optimum 
size produces up to 9.9\% reduction of turbulent skin friction drag~\citep{Bechert97}. 

	In most of the earlier works, the focus has been on relating the change in drag with 
either $\Sp$ or $\Hp$ to determine the optimum riblet geometry for drag reduction, while 
only a handful of studies are devoted to elucidate the drag reduction mechanisms. The 
\emph{protrusion height} concept~\citep{Bechert89, Bechert90, Luchini91} states that 
riblets produce viscous drag reduction owing to their higher cross-flow resistance. 
Unfortunately, this concept rests on the premise that the riblets are small enough to 
not be affected by inertial effects. As the size gets larger, the possibility that 
additional mechanisms or flow features that arise from inertial effects come into play 
cannot be discounted. \citet{Choi93} postulated that near-wall streamwise vortices tend 
to lodge in the groove when the riblet spacing $s$ becomes larger than their average 
diameter of 30 wall units. On the other hand, \citet{Goldstein98} reported that widely 
spaced riblets generates secondary flow due to the less effective damping of cross-flow 
fluctuations. Although the nature of influence of these proposed mechanisms on the viscous 
drag are mainly said to be contingent on $\Sp$, their variation trend with the 
characteristic size of riblets is still unclear.

	Although the conventional description of the riblet size in terms of $\Sp$ and $\Hp$ 
may be immaterial when considering the scaling of drag reduction curves, the choice 
becomes imperative if the aim is to elucidate the flow dynamics across a range of riblet 
geometries. The rationale is that the fluid-riblet interaction of various geometries could 
exhibit a diverse nature. In this connection, there is a need to assign a suitable 
dimensionless length scale, which also acts as the effective \emph{local} Reynolds number, 
to delineate the overall effects of riblets. \citet{Garcia11} have recently proposed an 
alternative length scale $\lgp$, defined as the square-root of the riblet groove 
cross-sectional area $A_g$ measured in wall units (see the schematic diagram alongside 
Fig.~\ref{Channel-V}), to characterize the riblet size. The improved scaling of drag 
reduction curves from past experiments enabled them to pinpoint the onset of breakdown 
of the \emph{viscous regime} at $\lgp \approx 11$. Unfortunately, their simulations 
are limited to thin blade riblets with a height-to-spacing ratio $h/s$ of 0.5, i.e. 
$l_g/s$ is constant. Hence, it is rather difficult to confirm whether scaling with $\lgp$ 
offers any benefit over $\Sp$ or $\Hp$, in terms of characterizing the flow dynamics 
imposed by V-groove riblets.

	The first objective of the present work is to investigate the evolution of the flow 
dynamics at low Reynolds number when the characteristic size of V-groove riblets is changed 
systematically. Once the variation trend is established, the next objective is to explore 
the relevance of the various flow features stemming from fluid-riblet interaction on the 
viscous drag and turbulence statistics. The objectives are achieved by performing direct 
numerical simulations (DNS) of fully-developed turbulent channel flow over six V-groove 
(triangular) riblet configurations. These configurations have groove sizes denoted by 
$\lgp$ that span both drag-reducing and drag-increasing regimes when simulated at a bulk 
Reynolds number $\Rem$ of either 1842 or 2800. Of particular note is that the riblet 
configurations comprise of two different $h/s$ ratios to ascertain the effectiveness of 
$\lgp$ in capturing the influence of both $s$ and $h$. It is worthwhile to mention that 
the characterization of the flow dynamics imposed by riblets, or any other patterned 
surface, is essential to achieve the desired flow attributes in various applications of 
interest. It can also provide some prospects for the development of turbulence models to 
ease the computational intensity of simulations of high Reynolds number flows over riblets.

\section{Methodology}
\label{sec:Methodology}

\begin{figure*}[t!]
	\centering
	% Insert 'grid, tics=10' in the square bracket to show the grid in 10% intervals.
	% trim option's parameter order: left bottom right top
	\begin{overpic}[trim = 30mm 7.5mm 30mm 10mm, clip, width = 0.7\textwidth]{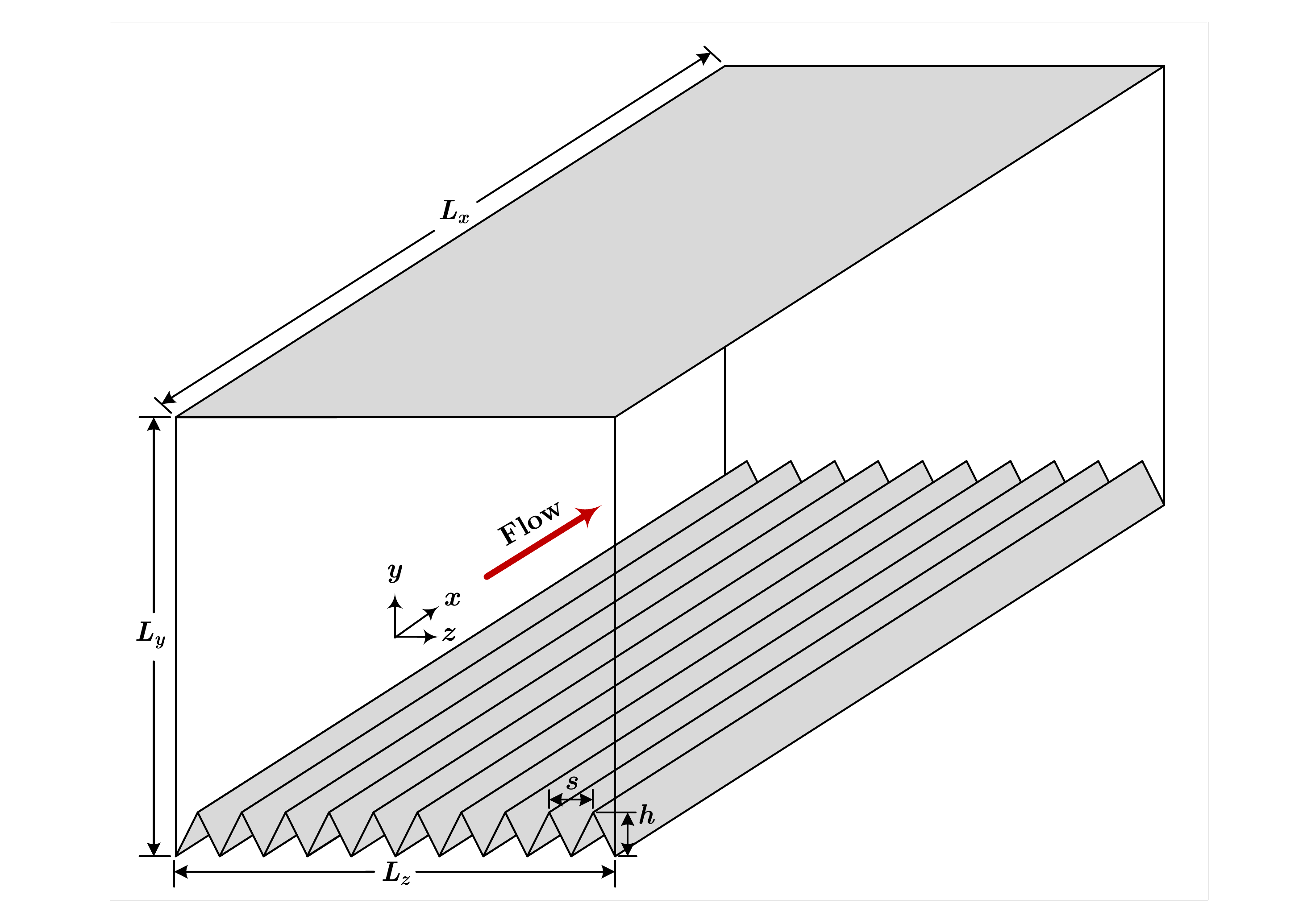}
	\put(72.5,1){\includegraphics[trim = 123mm 83mm 123mm 93mm, clip, width = 1.25in]{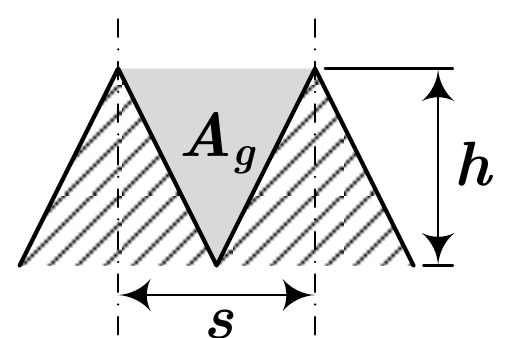}}
	\end{overpic}
	
	\caption{Computational domain of the turbulent channel flow with V-groove riblets mounted on the bottom wall. \label{Channel-V}}
\end{figure*}

	The present study considers a fully-developed channel flow of an incompressible 
fluid. In the Cartesian coordinate system, $x$, $y$ and $z$ denote the streamwise, 
wall-normal and spanwise directions, respectively. The origin of the coordinate axes is 
located at the center of the inlet plane as depicted in Fig.~\ref{Channel-V}. All the 
flow quantities, henceforth, are normalized by outer scales represented by the channel 
half-width $\delta$ and the bulk streamwise velocity $\Um$. Accordingly, the flow Reynolds 
number is defined as $\Rem = \Um\delta/\nu$, where $\nu$ is the kinematic viscosity. The 
governing equations can be written in non-dimensional form as:
%
%=====================================================================================================================================================================================================%
%
\begin{align}
																													 \boldsymbol{\nabla\cdot V} &= 0	\label{eq:ContinuityChan}\\
	\frac{\partial \boldsymbol{V}}{\partial \mathcal{T}} + \boldsymbol{V\cdot \nabla V} &= -\boldsymbol{\nabla} p + \frac{1}{\Rem} \nabla^2\boldsymbol{V} + \mathscr{F}\boldsymbol{e_x}	\label{eq:MomentumChan}
\end{align}
%
%=====================================================================================================================================================================================================%
%
where $\boldsymbol{V}$ = $(u$, $v$, $w)$, $p$ and $\mathcal{T}$ are the non-dimensional 
velocity vector, pressure and time, respectively. $u$, $v$, and $w$ are the respective 
components of $\boldsymbol{V}$ in the streamwise ($x$), wall-normal ($y$), and spanwise 
($z$) directions. $\mathscr{F} > 0$ is a prescribed external force to drive the flow 
through the channel, and $\boldsymbol{e_x}$ is the unit vector in the $x$ direction.
%
%=====================================================================================================================================================================================================%
%
%
%=====================================================================================================================================================================================================%
%

	Periodic boundary conditions are imposed in the $x$ and $z$ directions, while a no-slip 
condition is enforced on the two walls. In the present work, the forcing term $\mathscr{F}$ 
is adjusted dynamically to maintain a constant flow rate (CFR) through the channel. It 
is evaluated based on the deviation of bulk velocity $\Um$ at each time step. Accordingly, 
$\mathscr{F}$ can be perceived as a time-dependent pressure gradient that balances the 
wall shear stress $\tau_w$ in a time-averaged sense. The CFR forcing strategy is chosen 
to allow a direct change of wall shear stress during the course of the simulation.  
Although not shown here, the temporal fluctuations of $\Um$ in all cases are fairly small 
(standard deviation is $< 0.02\%$ of the mean), while the time-averaged values of 
$\mathscr{F}$ and $\tau_w$ differ by less than 0.5\%. The V-groove riblets are mounted 
on the bottom wall to allow a direct comparison of flow properties over two different 
surface geometries under the same level of uncertainty. Note that the channel height 
$L_y = 2\delta$ is measured from the valley of the riblets to the top wall as illustrated 
in Fig.~\ref{Channel-V}. The percentage drag reduction is then computed from the 
difference between the time-averaged drag on the two walls. The drag is derived from 
the stress tensor $\mathsfbi{\sigma} = \nu (\partial u_i/\partial x_j + \partial 
u_j/\partial x_i)$, in which the pressure drag is non-existent because of the streamwise 
homogeneity of riblets.

	Since riblets are created only on one of the walls, the flow is asymmetric about the 
channel centerline which results in different friction velocities on the two walls. In 
actual fact, the friction velocity on the smooth wall tends to increase with the size of 
riblets. Such physical effect may be explained in part by a reduction in the 
\emph{hydraulic diameter} $d_h$ of the channel~\citep{JHNg15}. In a way, this idea can 
account for the higher viscous drag produced by riblets in laminar channel flows, and it 
is in accord with the notion of an increase in the wetted surface area~\citep{Choi91, 
Chu93}. Accordingly, it may be reasonable to say that part of the change in viscous drag 
produced by riblets in internal flows can be attributed to the modification of the 
\emph{hydraulic diameter}. In this respect, the percentage drag reduction computed against 
the baseline channel flow would contain an anomaly because of the inherently different 
$d_h$. On the contrary, computation within a single domain offers the advantage of 
alleviating such problem, because the same effect is exerted on both walls.

	The size of the V-groove riblets is defined using the dimensionless length scale 
$\lgp$ proposed by~\citet{Garcia11}. This length scale is expressed as the square root 
of the groove cross-sectional area $A_g$ (see the schematic diagram alongside 
Fig.~\ref{Channel-V}) measured in wall units, i.e. $\lgp = \sqrt{A_g}\Utau/\nu$. Their 
rationale of using $A_g$ as an indicator of the riblet size is to collectively capture 
the influence of riblet spacing $s$ and height $h$. Note that the definition of $\lgp$ 
in wall units is computed with respect to the surface-averaged friction velocity on the 
riblet wall in this work. Unless otherwise stated, flow quantities stated with an 
overbar, i.e. $\overline{(\cdot)}$, denote the \emph{mean} or \emph{time-averaged} value, 
whereas those written with a prime, i.e. $(\cdot)'$ refer to the \emph{fluctuating} part. 
Root-mean-square (rms) quantities are affixed with a subscript, i.e. $(\cdot)_{rms}$. 
Flow quantities measured in \emph{wall units} or normalized by the inner scales derived 
from $\nu$ and $\Utau$ are indicated with a $+$ superscript, i.e. $(\cdot)^+$, throughout 
this paper.

\subsection{Numerical method and solver}
	
	The solver \emph{nek5000}~\citep{nek5000} employed in this work is based on the spectral 
element spatial discretization scheme~\citep{Patera84}, but uses the Gauss-Lobatto-Legendre 
(GLL) quadrature points within each local, non-overlapping element generated by 
decomposing the global domain~\citep{Deville02}. The spectral element method combines 
the accuracy of spectral methods and the generality of the finite element method. As a 
result, it can treat a wide class of geometrically and physically complex problems while 
maintaining spectral-like accuracy and rate of convergence~\citep{Fischer89}. Furthermore, 
it is advantageous for solving unsteady incompressible flow problem at moderate Reynolds 
number due to its non-dissipative and non-dispersive properties~\citep{Maday89}. In the 
present work, a $\mathbbit{P}_N-\mathbbit{P}_N$ formulation is implemented where the 
velocity and pressure field are represented by a Lagrange polynomial of degree $N$. The 
time integration is performed via a high-order operator-integration-factor splitting (OIFS) 
method~\citep{Maday90} with an implicit linear Stokes solver, coupled with an explicit 
sub-integration of the non-linear advective term~\citep{Fischer02}. Specific formulation 
and implementation details of the solver can be found in~\citep{Fischer94, Fischer97, 
Fischer02}.

\subsection{Numerical setup and configurations}
%
%=====================================================================================================================================================================================================%
%
\begin{table*}[t]
	\caption{The six V-groove riblet configurations considered in the current parametric study. Figure on the right shows the schematic view of the three physical configurations of V-groove riblets under the same scale. \label{tab:ParamGeom}}
	
	\begin{minipage}[c]{0.525\textwidth}
	\centering
		\hfill
			\begin{ruledtabular}
			\begin{tabular*}{2.5in}{@{\extracolsep{\fill} } cl*{4}{c}}
				& \multicolumn{1}{c}{\bf Case} & \multicolumn{1}{c}{$\Rem$} & $s$ 				& $h$ 				&\\
				\hline
				& {\bf I}											 & \multirow{3}{*}{1842}			& $0.2\delta$ & $0.1\delta$ &\\
				& {\bf II}										 &			 											& $0.2\delta$ & $0.2\delta$ &\\
				& {\bf III} 									 &														& $0.4\delta$ & $0.4\delta$ &\\
				\hline
				& {\bf IV} 										 & \multirow{3}{*}{2800}			& $0.2\delta$ & $0.1\delta$ &\\
				& {\bf V} 										 &														& $0.2\delta$ & $0.2\delta$ &\\
				& {\bf VI} 										 &														& $0.4\delta$ & $0.4\delta$ &\\
			\end{tabular*}
			\end{ruledtabular}
		\end{minipage}
		\hfill
		\begin{minipage}[c]{0.45\textwidth}
			\centering
			% trim option's parameter order: left bottom right top
			\includegraphics[trim = 10mm 10mm 10mm 180mm, clip, width = 0.95\textwidth]{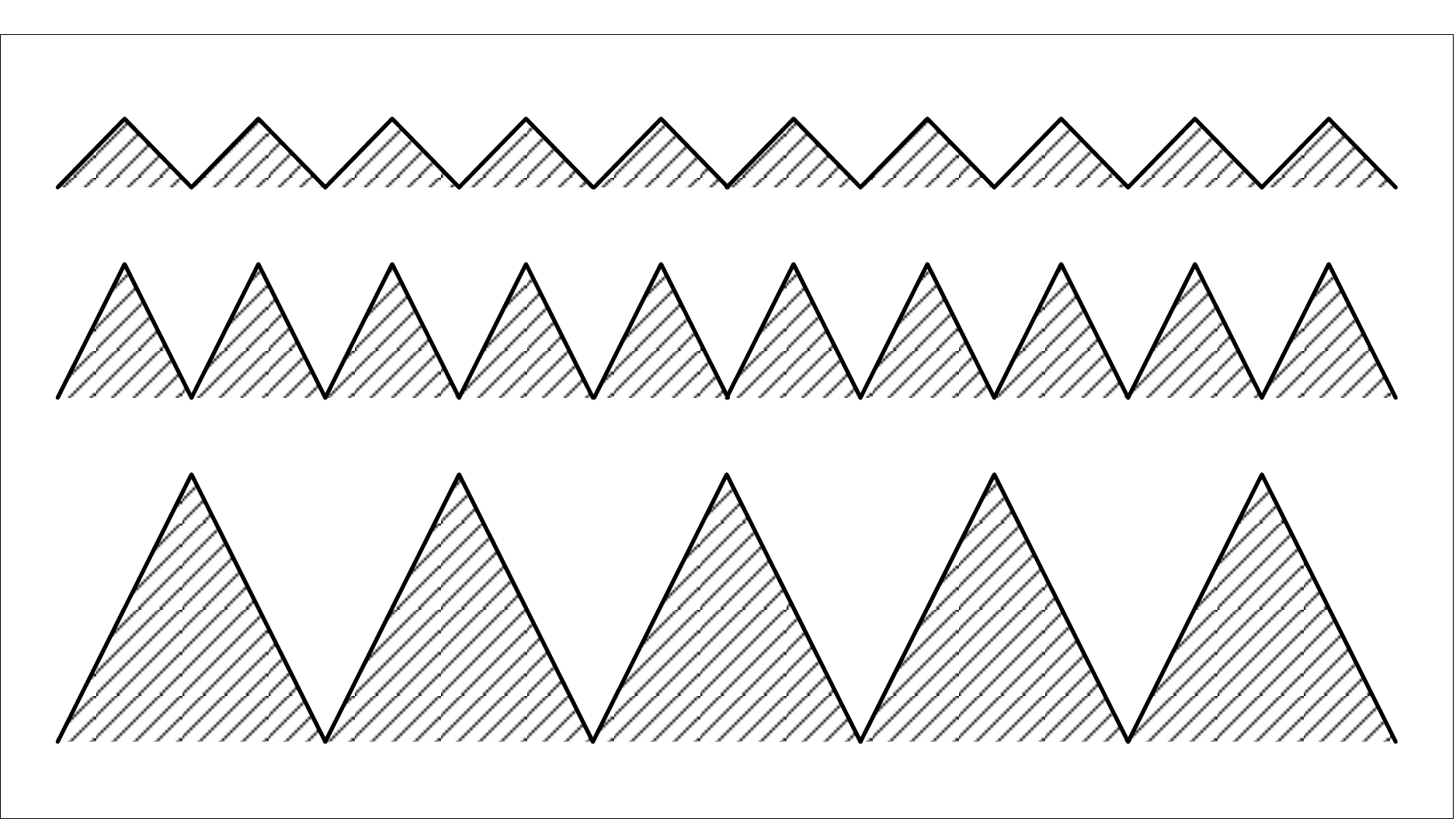}
			%\caption{A schematic view of the V-groove riblet geometries considered in the parametric study.}
		%	\label{ParamSchematic}
		\end{minipage}
\end{table*}
%
%=====================================================================================================================================================================================================%
%

	In the current parametric study, V-groove riblets with three different combinations of 
physical spacing $s$ and height $h$ have been simulated at two bulk Reynolds numbers 
$\Rem =$ 1842 and 2800. Table~\ref{tab:ParamGeom} describes the physical dimensions of 
the six configurations denominated as Cases I to VI. A schematic diagram illustrating 
the relative appearances of the riblets is provided on the side of the table. One of the 
configurations denoted as Case II follows closely the reference DNS of~\citet{Chu93}. 
This flow configuration is chosen as a guideline to assess the current simulation setup 
using the spectral element solver. In Case II, the riblet height and spacing are both 
$0.2\delta$. There are 10 riblets across the span as depicted in Fig.~\ref{Channel-V}. 
Also, the riblets are introduced into the fluid domain explicitly without any form of 
modeling. Although explicitly resolving the riblet profile inevitably slows down the 
convergence of solutions, the improved accuracy of the near-wall flow field is undoubtedly 
crucial to elucidate the nature of turbulence in the presence of riblets.

	Two separate simulations denoted as Cases I and III are also performed to contrast the 
flow fields at the same bulk Reynolds number $\Rem = 1842$ or $\ReT \approx 122$. These 
simulations facilitate the identification of the flow features responsible for the vast 
change in flow behavior. The remaining three configurations, i.e. Cases IV to VI, are 
simulated at $\Rem = 2800$ or $\ReT \approx 180$ to obtain a more representative turbulent 
flow field, and to include the effects of Reynolds number. Although the current work is 
limited to moderate Reynolds number, earlier studies have noted that such computations 
could be adequate for studying flow structures~\citep{Moin98}. \citet{Garcia12} 
have demonstrated that the physical descriptions deduced at $\ReT \approx 180$ are 
essentially correct. Noticeable discrepancies would probably manifest when the 
Reynolds number is far higher, i.e. $\ReT > 10^3$~\citep{Marusic10, Smits11}.

	As in the reference DNS~\citep{Chu93}, the computational domain has dimensions of 
$L_x = 5\delta$, $L_y = 2\delta$ and $L_z = 2\delta$ in the $x$, $y$ and $z$ directions, 
respectively. This translates to $L_x^+ \approx 650$ and $L_y^+ = L_z^+ \approx 260$ 
with respect to the smooth top wall. The current domain extent in the homogeneous 
directions is about two and a half times larger than the \emph{minimal flow unit} ($L_x^+ 
\approx 250 \sim 350$ and $L_z^+ \approx 100$) found by~\citet{Jimenez91}. Since a higher 
mesh resolution is essential to resolve the flow field around riblets, the small domain 
allows the simulations to complete within a reasonable time. The articulation 
by~\citet{Gatti13} on the greater computational efforts expended on simulating the initial 
transient in a larger domain applies to the present work as well. Having said that, an 
asessment on the impact of domain size on the relevant flow properties is presented 
in Appendix~\ref{sec:DomainSize}. The other five configurations are also simulated with 
the same domain size of $5\delta \times 2\delta \times 2\delta$ to maintain consistency.

	For the simulations at $\Rem = 1842$, the domain is partitioned into $21 \times 14 
\times 20$ elements in the $x$, $y$ and $z$ directions, respectively. The total number 
of grid points in each orthogonal direction $x$, $y$ and $z$ is 190, 127 and 181, 
respectively after including the GLL nodes within each element generated based on a 
$9^{\text{\tiny th}}$ order Lagrangian polynomial. On the other hand, two additional 
layer of elements are added in the wall-normal direction for the simulations at $\Rem 
= 2800$, increasing the number of grid points to 145. In Case II, the average resolution 
in the streamwise and spanwise directions are respectively $\Delta x^+ \approx 3.4$ and 
$\Delta z^+ \approx 1.4$ with reference to the friction velocity on the smooth wall. The 
resolutions become $\Delta x^+ \approx 2.3$ and $\Delta z^+ \approx 1.0$ when computed 
using the surface-averaged friction velocity on the riblet wall. In the wall-normal 
direction, a hyperbolic sine function is used to distribute the element size such that 
more grid points are clustered near the two walls. Accordingly, the first element adjacent 
to the riblet wall which comprises of 10 grid points is below $y^+ = 5$, and the first 
grid point away from the riblet wall is at $y^+ \approx 0.12$. On the other hand, the 
average wall-normal resolution near the channel centerline is $\Delta y^+ \approx 5$.
%
%=====================================================================================================================================================================================================%
%
\begin{figure*}[t!]
	\begin{minipage}[b]{0.49\textwidth}
		\centering
		\subfigure[][]
		{
			% trim option's parameter order: left bottom right top
			\includegraphics[trim = 25mm 22.5mm 25mm 7.5mm, clip, width = 0.95\textwidth]{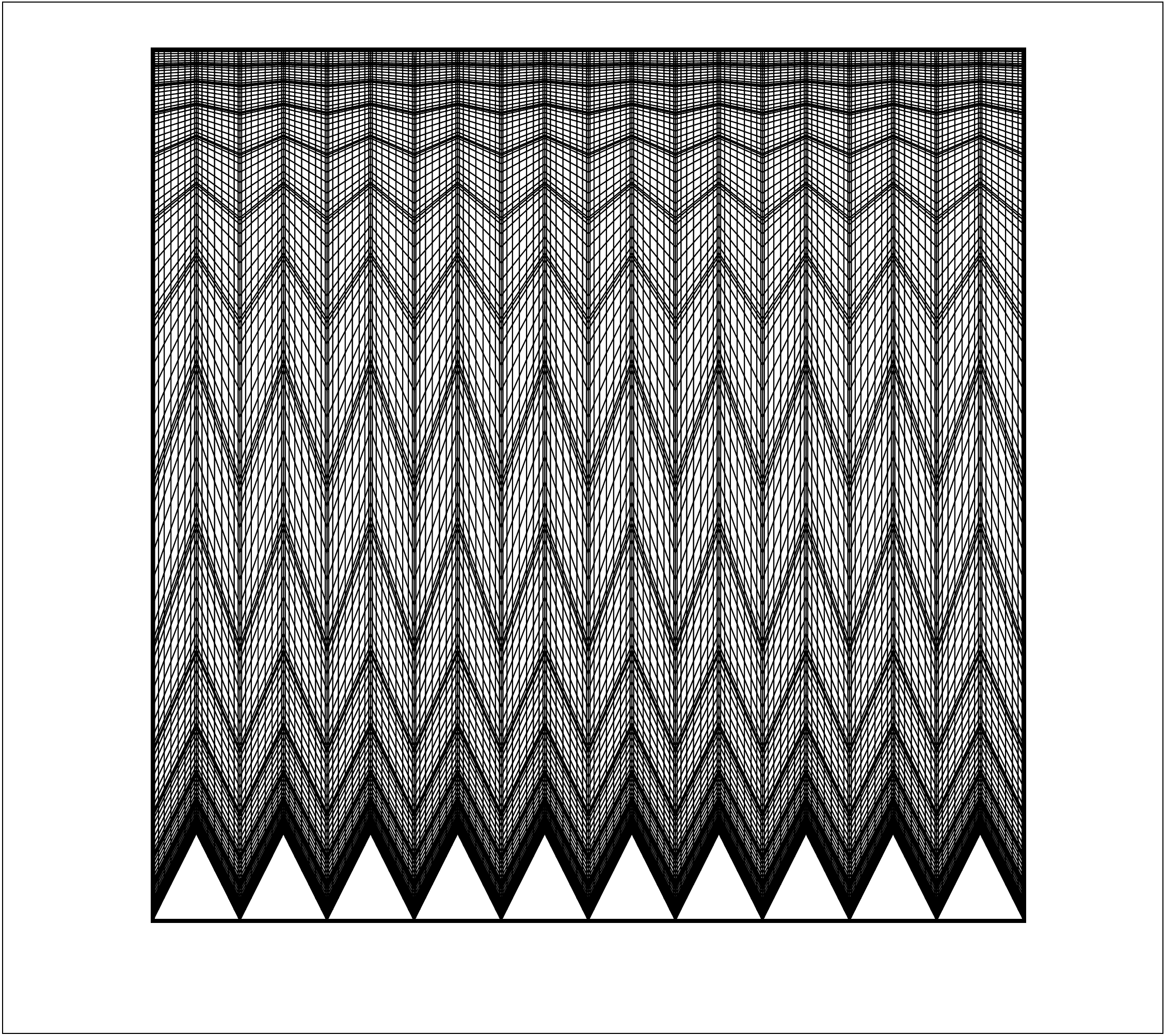}
			\label{Channel-V-Mesh}
		}
	\end{minipage}
	\hfill
	\begin{minipage}[b]{0.49\textwidth}
		\centering
		\subfigure[][]
		{
			% trim option's parameter order: left bottom right top
			\includegraphics[trim = 25mm 22.5mm 25mm 7.5mm, clip, width = 0.95\textwidth]{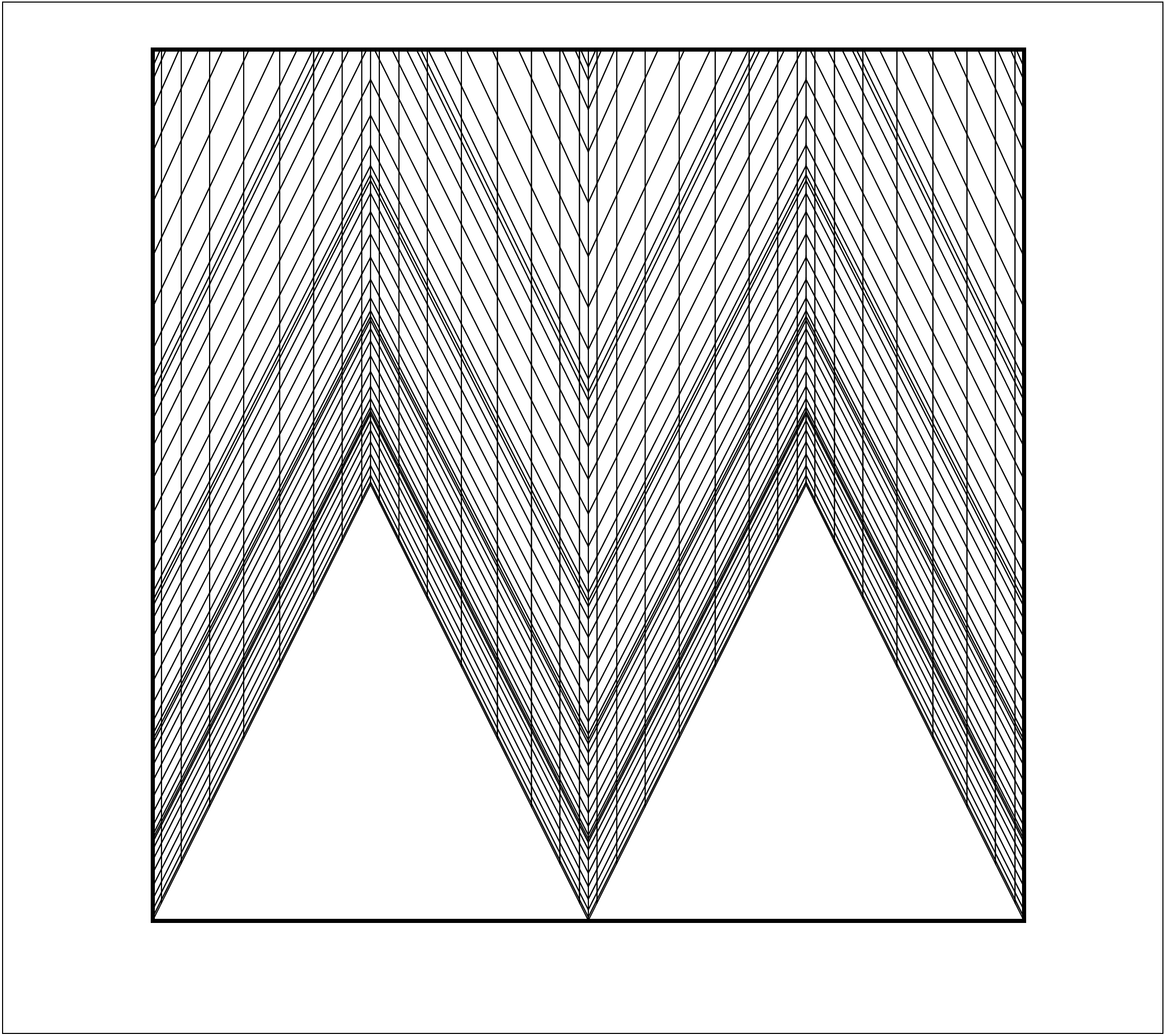}
			\label{Channel-V-Mesh-Riblet}
		}
	\end{minipage}
	
	\caption
		{ 
			\subref{Channel-V-Mesh}				 Spectral element mesh for the simulation of Case II $(s = 0.2\delta$, $h = 0.2\delta)$, and
			\subref{Channel-V-Mesh-Riblet} Close-up view of the mesh near the V-groove riblets.
		}
\end{figure*}
%
%=====================================================================================================================================================================================================%
%

	 The mesh resolutions in all configurations, especially near the riblets, are made 
sufficiently high to avoid under-resolved simulation that can give rise to undesirable 
wiggles and over- or under-predicted profile of streamwise velocity fluctuations $u'$, 
see section 9.3.1 in the book written by~\citet{Karniadakis05}. In particular, the mesh 
resolution is designed to be higher than the plane turbulent channel flow to accommodate 
the possible reduction in scales when riblets are introduced into the flow domain. 
This coupled with the use of a $9^{\text{\tiny th}}$ order spectral element discretization 
helps to enhance the accuracy of the simulations and to diminish dispersion 
errors~\citep{Fischer02}. Figure~\ref{Channel-V-Mesh} depicts the generated spectral 
element mesh for Case II, while Fig.~\ref{Channel-V-Mesh-Riblet} shows a magnified view 
of the same mesh in the proximity of the riblets. Table~\ref{tab:SetupSize} presents the 
computational mesh setups of the six riblet configurations.
%
%=====================================================================================================================================================================================================%
%
\begin{table*}[t]
	\caption{Computational mesh setups of the V-groove riblet configurations. The values in wall units are computed based on the surface-averaged friction velocity at the riblet wall. \label{tab:SetupSize}}
	
		\centering
		\begin{ruledtabular}
		\begin{tabular*}{0.95\textwidth}{@{\extracolsep{\fill} } cl*{10}{c}}
			& {\bf Case} & $N_x$ 								& $N_y$ 							 & $N_z$ 								& $\Delta x_{min}^+$ & $\Delta x_{max}^+$ & $\Delta y_{min}^+$ & $\Delta y_{max}^+$ & $\Delta z_{min}^+$ & $\Delta z_{max}^+$ &\\
			\hline
			& {\bf I} 	 & \multirow{3}{*}{190} & \multirow{3}{*}{127} & \multirow{3}{*}{181} & 0.98 							 & 4.02 							& 0.15 							 & 6.44 							& 0.41 							 & 1.69 								 &\\
			& {\bf II} 	 & 											& 										 & 											& 0.82 							 & 3.38 							& 0.12 							 & 5.25 							& 0.35 							 & 1.42 								 &\\
			& {\bf III}	 & 											& 										 & 											& 1.03 							 & 4.23 							& 0.13 							 & 5.97 			 				& 0.43 							 & 1.78 								 &\\
			\hline
			& {\bf IV} 	 & \multirow{3}{*}{190} & \multirow{3}{*}{145} & \multirow{3}{*}{181} & 1.46 							 & 6.01 							& 0.19 							 & 8.64 							& 0.61 							 & 2.53 								 &\\
			& {\bf V} 	 & 											& 										 & 											& 1.29 							 & 5.29 							& 0.16 							 & 7.26 							& 0.54 							 & 2.22 								 &\\
			& {\bf VI}	 & 											& 										 & 											& 1.56 							 & 6.42 							& 0.17 							 & 7.91 							& 0.66 							 & 2.70 								 &\\
		\end{tabular*}
		\end{ruledtabular}
\end{table*}
%
%=====================================================================================================================================================================================================%
%

	The simulations of all six configurations are carried out for a non-dimensional period 
$T\Um/\delta = 750$ using the baseline turbulent channel flow solution at $\ReT \approx 
180$ as the initial condition. The non-dimensional time step size $\Delta t\Um/\delta$ 
is 0.005 in all six configurations. At $\Rem = 2800$, this time step in wall unit, i.e. 
$\Delta t\Utau^2/\nu \approx 0.06$, is much smaller than the value recommended 
by~\citet{Choi93} $(\Delta t^+ \approx 0.4)$ to accurately predict turbulence statistics 
of a plane channel flow. As the solver employs a characteristics-based time stepping 
method, i.e. the OIFS procedure, a less restrictive Courant or CFL number up to 4 
is allowable~\citep{Pironneau82, Maday90}. In the present set of simulations, the CFL 
number varies from 1 to 3 while the temporal discretization scheme attains a third-order 
accuracy. The average computation time of the six configurations is about 10 seconds per 
time step when executed over 48 parallel processes.

\subsection{Impact of domain size}
\label{sec:DomainSize}

	In simulations of fully-developed turbulent channel flows with periodic boundary 
conditions, the domain size plays a pivotal role in determining the extent of boundary 
layer that possess ``healthy'' turbulence~\cite{Flores10, Lozano-Duran14}. Hence, a 
preliminary study is carried out to assess the impact of domain truncation on flow 
statistics. It begins with the baseline turbulent channel flow, followed by one of the 
V-groove riblet configurations. The assessment concerning the baseline configuration is 
achieved by comparing the simulations with domain sizes of $5\delta \times 2\delta \times 
2\delta$ and $4\upi\delta \times 2\delta \times 2\upi\delta$ with similar mesh resolutions 
at $\Rem = 1842$ and 2800.
%
%=====================================================================================================================================================================================================%
%
\begin{figure*}[t]
	\begin{minipage}[c]{0.495\textwidth}
		\centering
		\subfigure[][]
		{
			% trim option's parameter order: left bottom right top
			\includegraphics[trim = 30mm 2.5mm 37.5mm 15mm, clip, width = 0.95\textwidth]{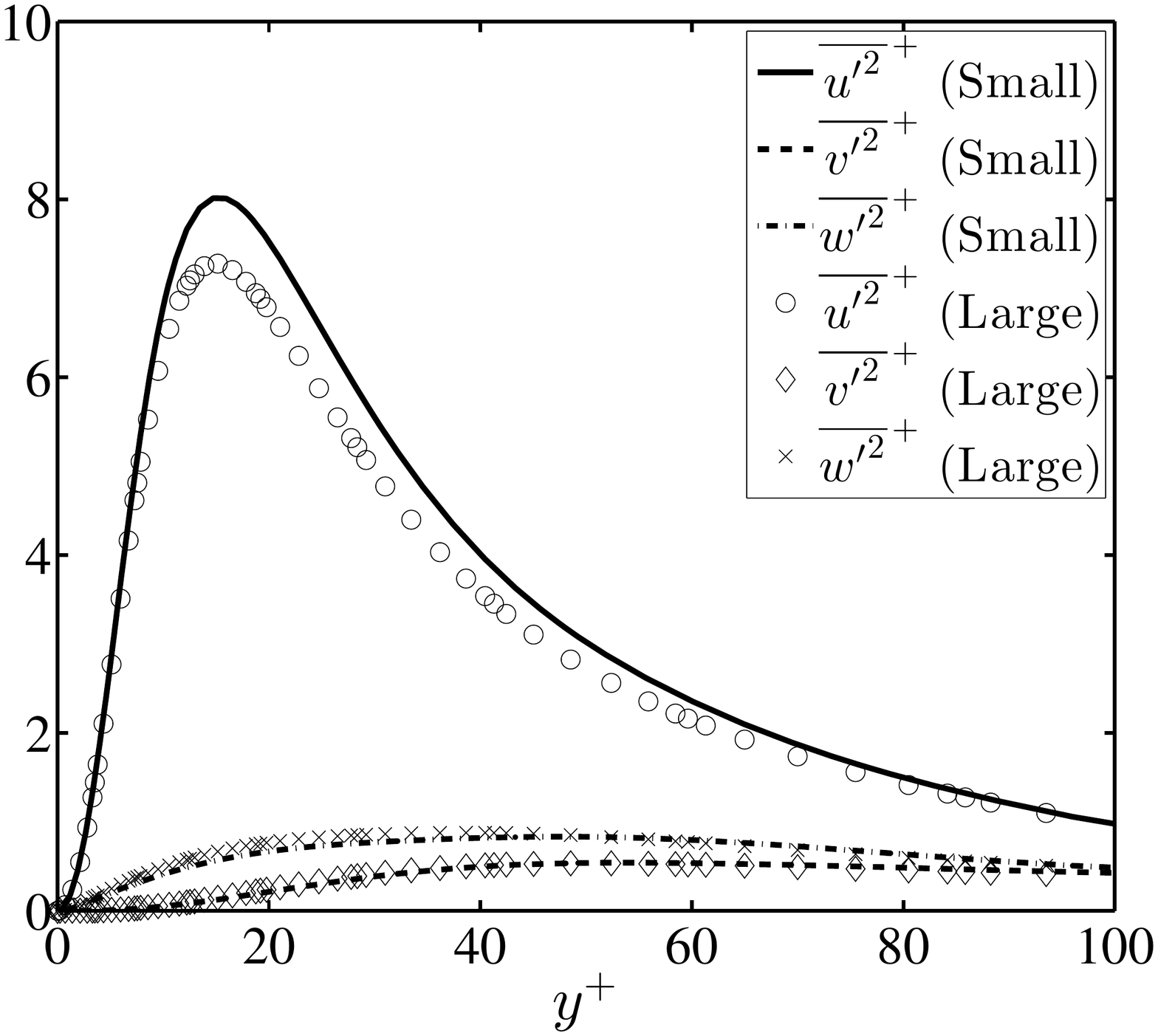}
			\label{UVWmsS1842DomainSize}
		}
	\end{minipage}
	\hfill
	\begin{minipage}[c]{0.495\textwidth}
		\centering
		\subfigure[][]
		{
			% trim option's parameter order: left bottom right top
			\includegraphics[trim = 30mm 2.5mm 37.5mm 15mm, clip, width = 0.95\textwidth]{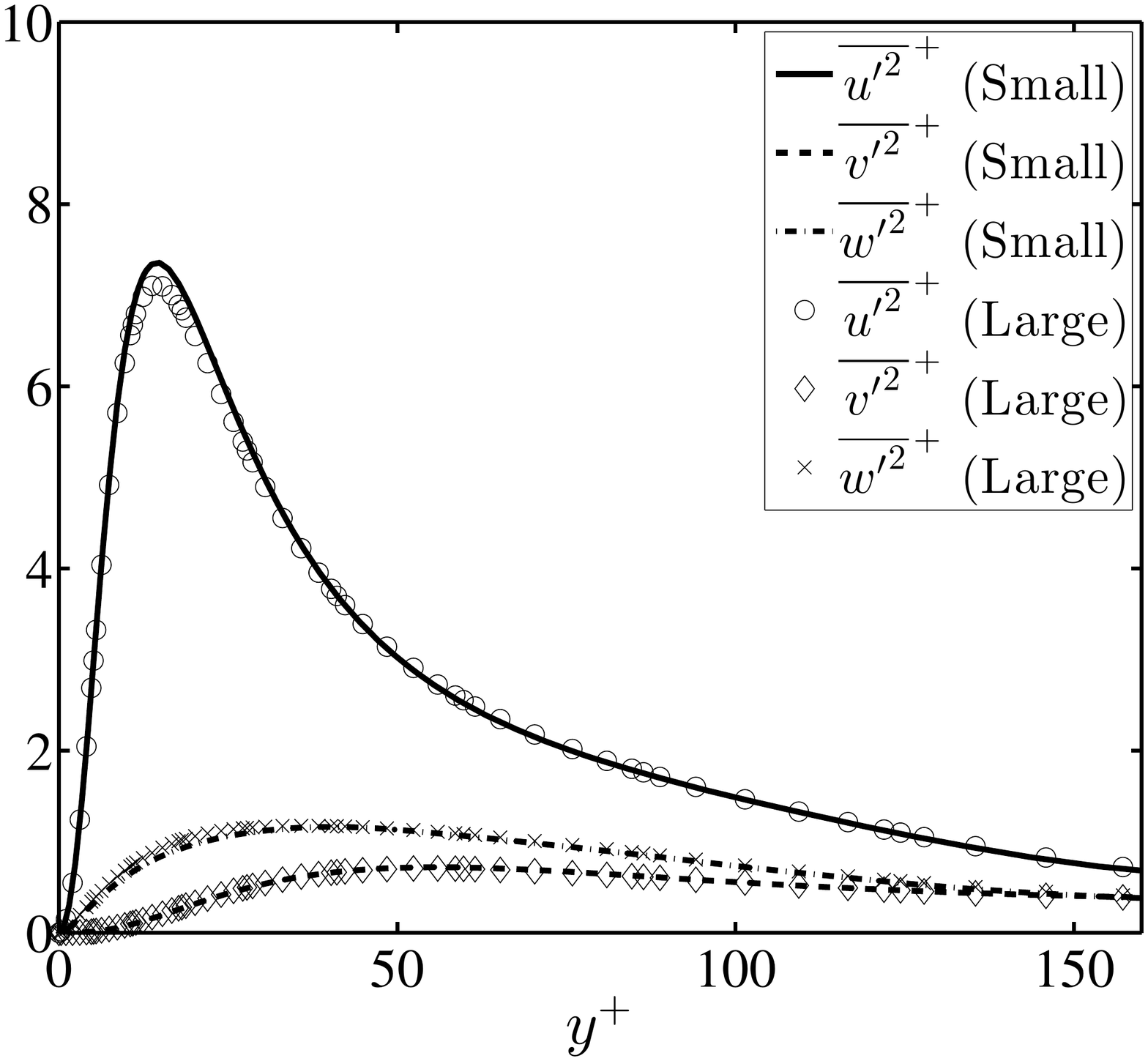}
			\label{UVWmsS2800DomainSize}
		}
	\end{minipage}
	
	\caption
		{
			Comparison of the mean-square velocity fluctuations profiles plotted in wall units acquired from simulations of the baseline turbulent channel flow with different domain sizes at:
			\subref{UVWmsS1842DomainSize} $\Rem = 1842$, and 
			\subref{UVWmsS2800DomainSize} $\Rem = 2800$.
			\label{UVWmsSDomainSize}
		}
\end{figure*}
%
%=====================================================================================================================================================================================================%
%

	Figure~\ref{UVWmsSDomainSize} demonstrates that the effect of a smaller domain is 
noticeable mainly on the statistics of the streamwise velocity fluctuations $u'$, 
especially at $\Rem = 1842$ because of the smaller domain size in wall units, i.e. 
$L_x^+ \approx 650$ and $L_z^+ \approx 260$. The higher peak of $u'$ is in line with the 
observation that the two-point correlation of $u'$ (not shown here) does not decay to zero 
at one-half of the streamwise domain extent. On the other hand, the smaller domain size 
does not result in considerable disparity at $\Rem = 180$, because it is relatively large 
when translated to wall units, i.e. $L_x^+ \approx 930$ and $L_z^+ \approx 370$.
%
%=====================================================================================================================================================================================================%
%
\begin{figure*}[t]
	\begin{minipage}[c]{0.495\textwidth}
		\centering
		\subfigure[][]
		{
			% trim option's parameter order: left bottom right top
			\includegraphics[trim = 30mm 2.5mm 37.5mm 15mm, clip, width = 0.95\textwidth]{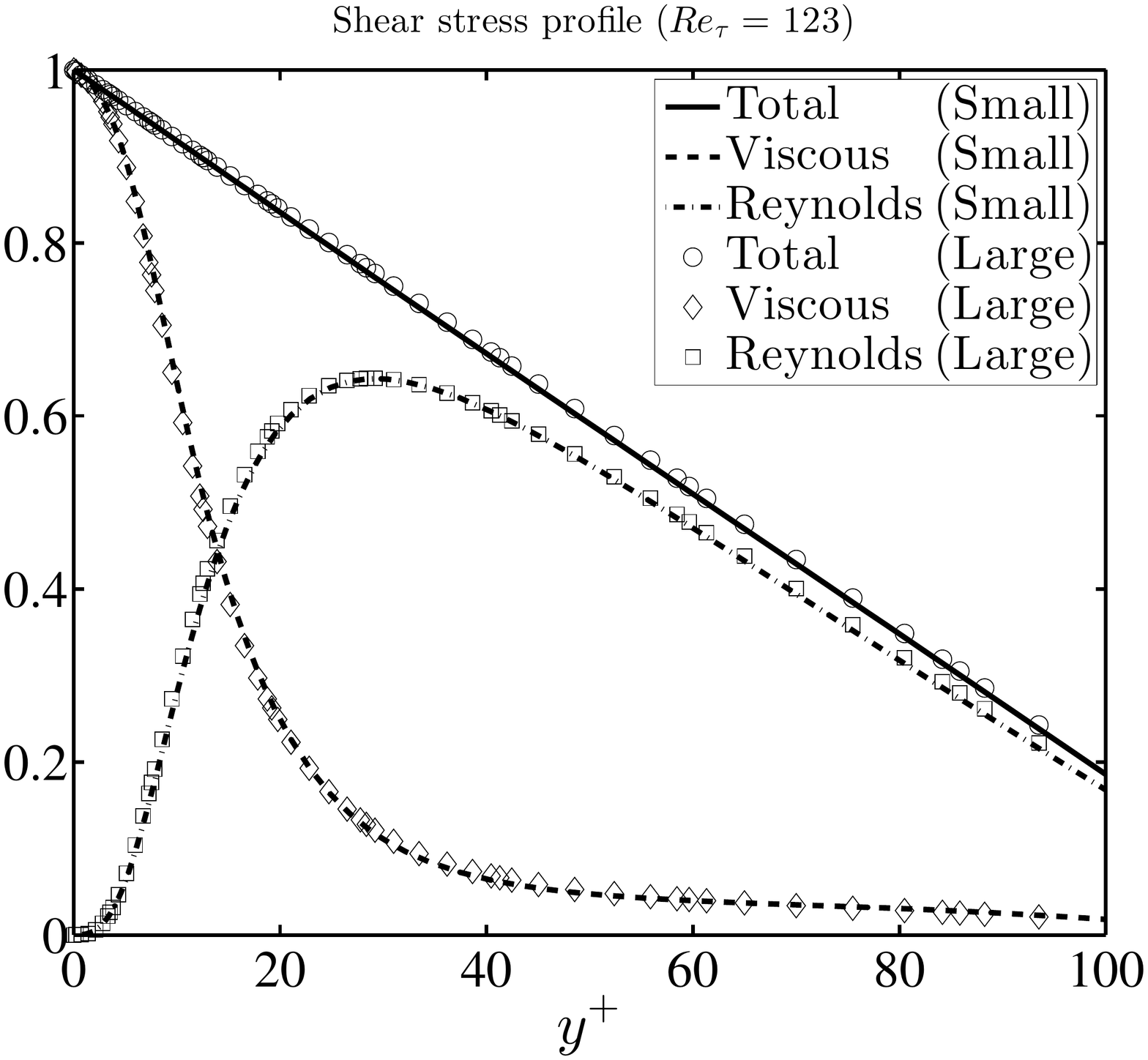}
			\label{StressS1842DomainSize}
		}
	\end{minipage}
	\hfill
	\begin{minipage}[c]{0.495\textwidth}
		\centering
		\subfigure[][]
		{
			% trim option's parameter order: left bottom right top
			\includegraphics[trim = 30mm 2.5mm 37.5mm 15mm, clip, width = 0.95\textwidth]{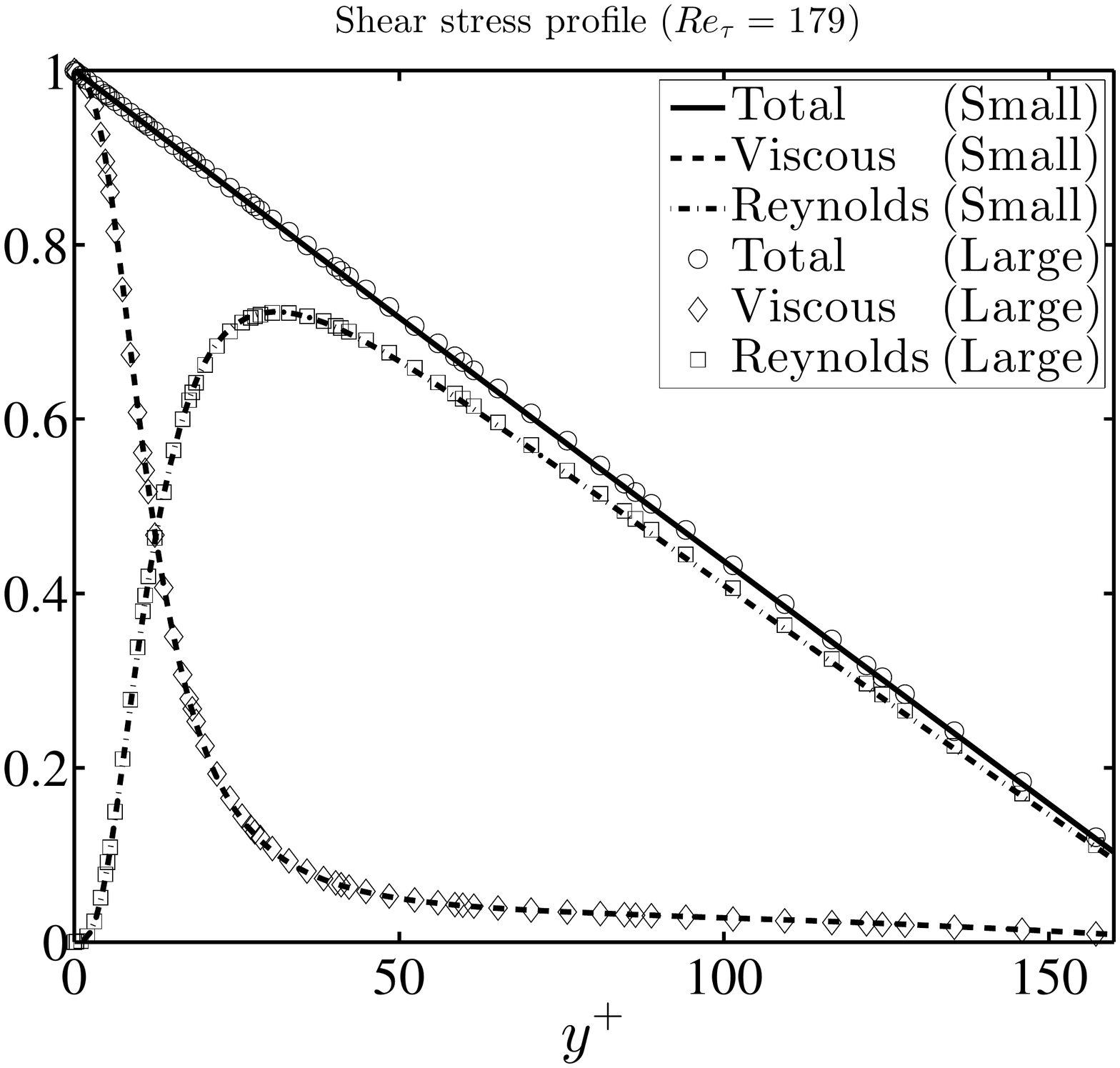}
			\label{StressS2800DomainSize}
		}
	\end{minipage}
	
	\caption
		{
			Comparison of the time-averaged shear stress profiles plotted in wall units acquired from simulations of the baseline turbulent channel flow with different domain sizes at:
			\subref{StressS1842DomainSize} $\Rem = 1842$, and 
			\subref{StressS2800DomainSize} $\Rem = 2800$.
			\label{StressSDomainSize}
		}
\end{figure*}
%
%=====================================================================================================================================================================================================%
%

\begin{figure*}[t]
	\begin{minipage}[c]{0.495\textwidth}
		\centering
		\subfigure[][]
		{
			% trim option's parameter order: left bottom right top
			\includegraphics[trim = 30mm 2.5mm 37.5mm 15mm, clip, width = 0.95\textwidth]{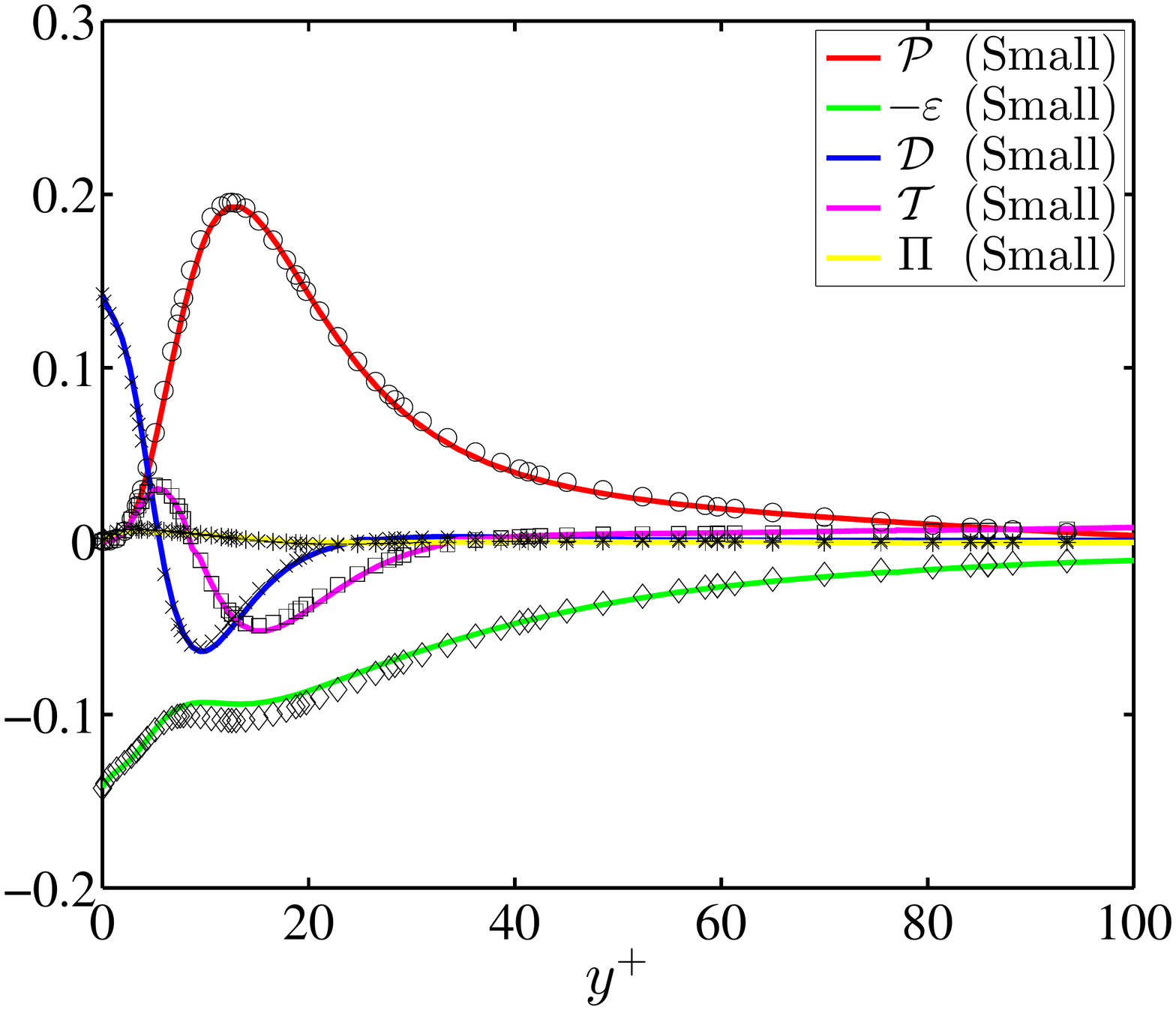}
			\label{TkekS1842DomainSize}
		}
	\end{minipage}
	\hfill
	\begin{minipage}[c]{0.495\textwidth}
		\centering
		\subfigure[][]
		{
			% trim option's parameter order: left bottom right top
			\includegraphics[trim = 30mm 2.5mm 37.5mm 15mm, clip, width = 0.95\textwidth]{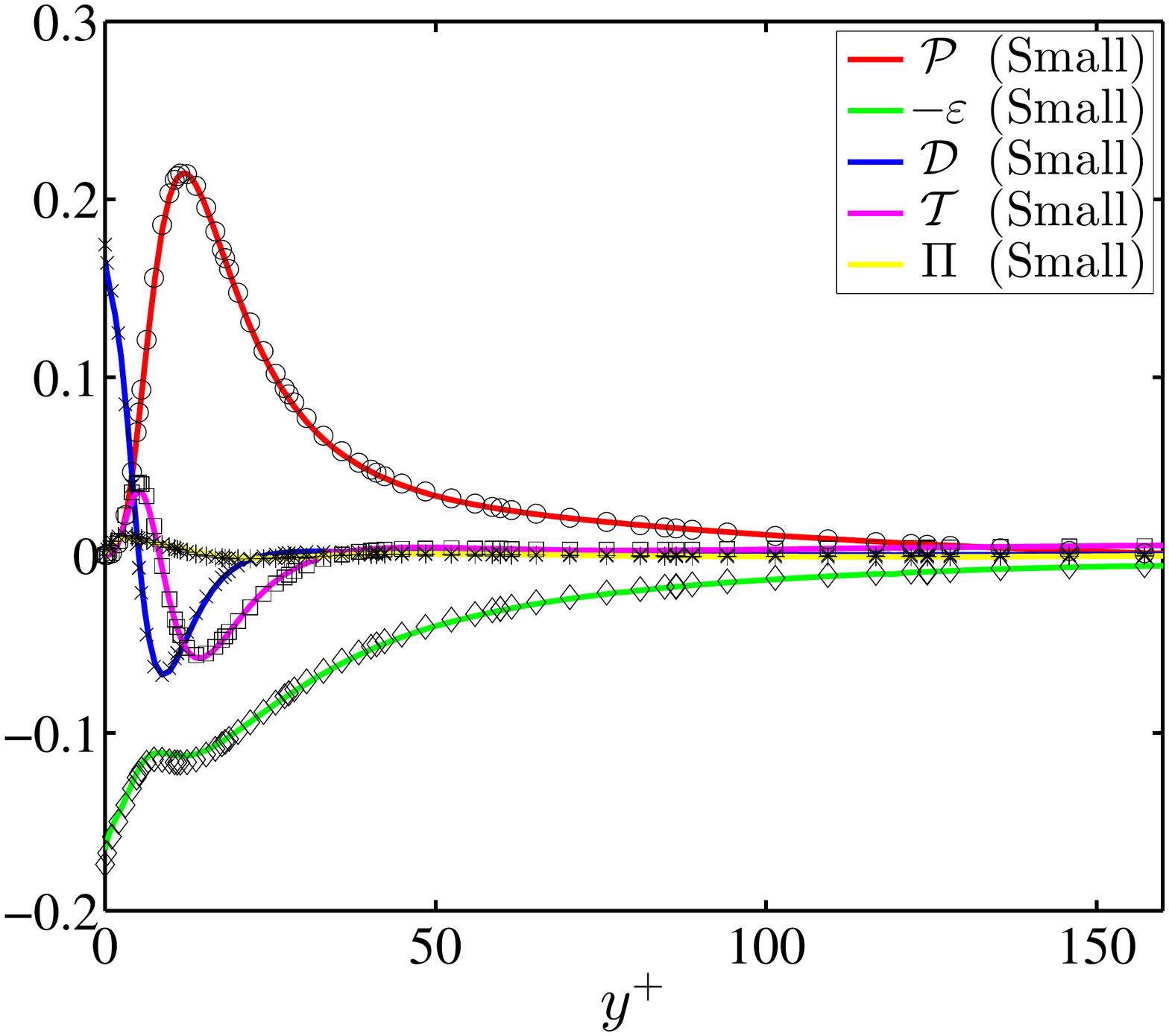}
			\label{TkekS2800DomainSize}
		}
	\end{minipage}
	
	\caption
		{
			Comparison of the budget of turbulence kinetic energy plotted in wall units acquired from simulations of the baseline turbulent channel flow with different domain sizes at:
			\subref{TkekS1842DomainSize} $\Rem = 1842$, and 
			\subref{TkekS2800DomainSize} $\Rem = 2800$.
			The markers ($\circ$,$\smalldiamond$,$\times$,$\smallsquare$,\textasteriskcentered) represent the corresponding profiles acquired from the larger domain configuration.
			\label{TkekSDomainSize}
		}
\end{figure*}

	In a statistically stationary and fully-developed turbulent channel flow, the total 
shear stress comprises of the combined contribution by the mean viscous stress and the 
Reynolds stress is linear along the wall-normal direction. In Fig.~\ref{StressSDomainSize}, 
there is no noticeable discrepancy between the two sets of shear stress profiles. The close 
agreement corroborates the finding by~\citet{Gatti13} that the variation in the skin 
friction coefficient $C_f$ at $\ReT \approx 200$ is more or less unaffected by the domain 
size when $6 < L_xL_z/\delta^2 < 60$, see figure 3 in their paper. In fact, they 
demonstrated that the uncertainty is almost canceled out when the drag reduction is 
computed under the same simulation condition. Another important metric to assess the 
implication of domain size is the budget of turbulence kinetic energy $k$ which 
describes the energy transfer process instituted by the turbulent motions. The transport 
equation of $k$, encompassing the five terms $\mathcal{P}_k$, $\mathcal{D}_k$, 
$\mathcal{T}_k$, $\Pi_k$, and $\varepsilon_k$, is equivalent to equation (1) given 
by~\citet{Mansour88}. Once again, Fig.~\ref{TkekSDomainSize} illustrates that the 
reduced domain size does not affect the turbulence processes considerably.
%
%=====================================================================================================================================================================================================%
%
%
%=====================================================================================================================================================================================================%
%

\begin{figure*}[t]
	\begin{minipage}[c]{0.495\textwidth}
		\centering
		\subfigure[][]
		{
			% trim option's parameter order: left bottom right top
			\includegraphics[trim = 30mm 2.5mm 37.5mm 15mm, clip, width = 0.95\textwidth]{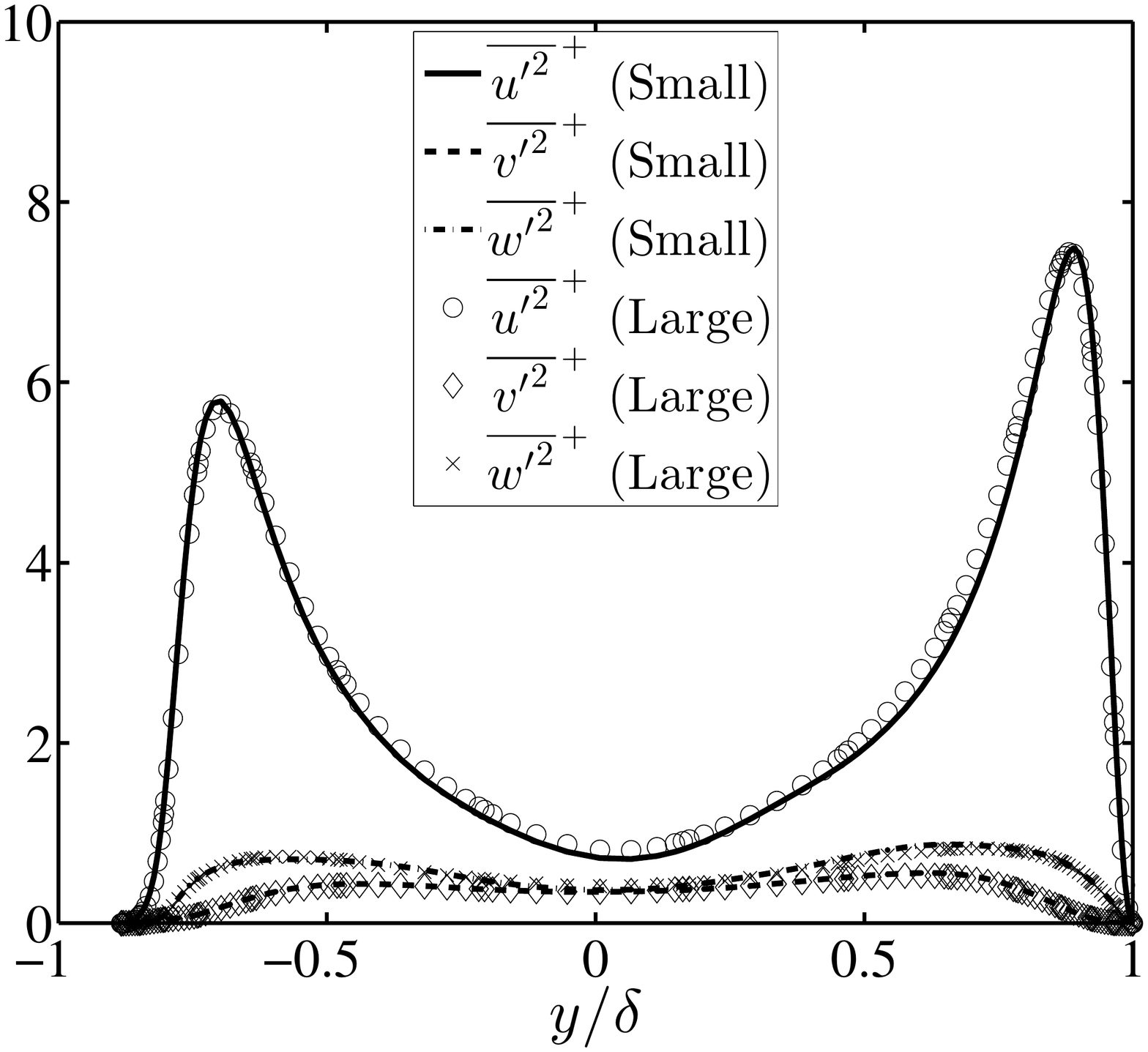}
			\label{UVWmsRDomainSize}
		}
	\end{minipage}
	\hfill
	\begin{minipage}[c]{0.495\textwidth}
		\centering
		\subfigure[][]
		{
			% trim option's parameter order: left bottom right top
			\includegraphics[trim = 30mm 2.5mm 37.5mm 15mm, clip, width = 0.95\textwidth]{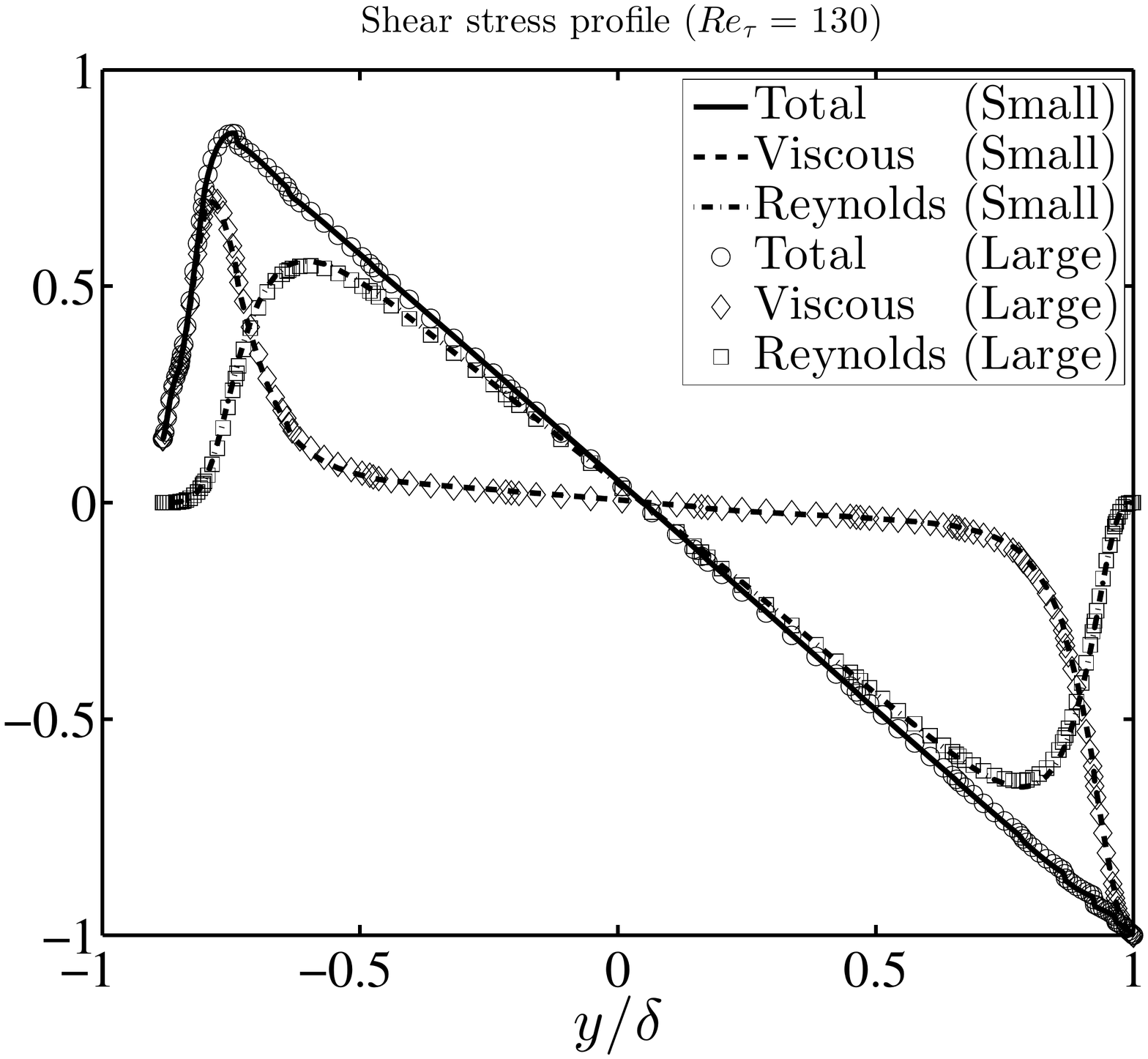}
			\label{StressRDomainSize}
		}
	\end{minipage}
	
	\caption
	{
		Impact of domain size on the statistics of Case II ($\lgp \approx 12$) across the channel:
		\subref{UVWmsRDomainSize}	 Mean-square velocity fluctuations, and 
		\subref{StressRDomainSize} Time-averaged shear stresses. 
		Note that the flow quantities are scaled in wall units using the friction velocity of the top (smooth) wall.
		\label{CaseIIDomainSize}
	}
\end{figure*}

	Apart from the baseline configuration, the riblet configuration denoted as Case II is 
also compared with another simulation performed in a domain that is one and a half times 
longer in the streamwise $(x)$ and spanwise $(z)$ directions. Figure~\ref{CaseIIDomainSize} 
reveals no significant discrepancy between the two sets of simulation, probably because 
the stabilizing effect of riblets compensates, to some extent, the higher intermittency 
in the flow evolution. Judging from these assessments, the findings derived from the 
present simulations are expected to be marginally affected by the domain size. Although 
there might be more noticeable discrepancy if compare with an even larger domain 
simulation, the issue is not pursued further because it would deviate from the scope of 
the present work, apart from being too costly.
%
%=====================================================================================================================================================================================================%
%
%
%=====================================================================================================================================================================================================%
%

\section{Results and Discussions}
\label{sec:Result}

	The time averaging of flow quantities is performed within a non-dimensional time period 
$T\Um/\delta$ of 450 which corresponds to about 100 eddy-turnover time units. Furthermore, 
these statistical profiles have been spatial-averaged in the streamwise direction, and 
also ``piecewise-averaged'' in the spanwise direction that collapses the domain into 
one-half of a V-groove riblet. In the following discussions, the statistics are plotted 
at three spanwise positions, namely ``tip'', ``mid-point'' and ``valley'' as illustrated 
schematically in Fig.~\ref{UavgBenchmark}.

\subsection{Benchmarking study}
\label{sec:Benchmark}
%
%=====================================================================================================================================================================================================%
%
\begin{figure*}[t]
	\begin{minipage}[c]{0.495\textwidth}
		\centering
		\subfigure[][]
		{
			% Insert 'grid, tics=10' in the square bracket to show the grid in 10% intervals.
			% trim option's parameter order: left bottom right top
			\begin{overpic}[trim = 0.5mm 50mm 10mm 51mm, clip, width = 0.95\textwidth]{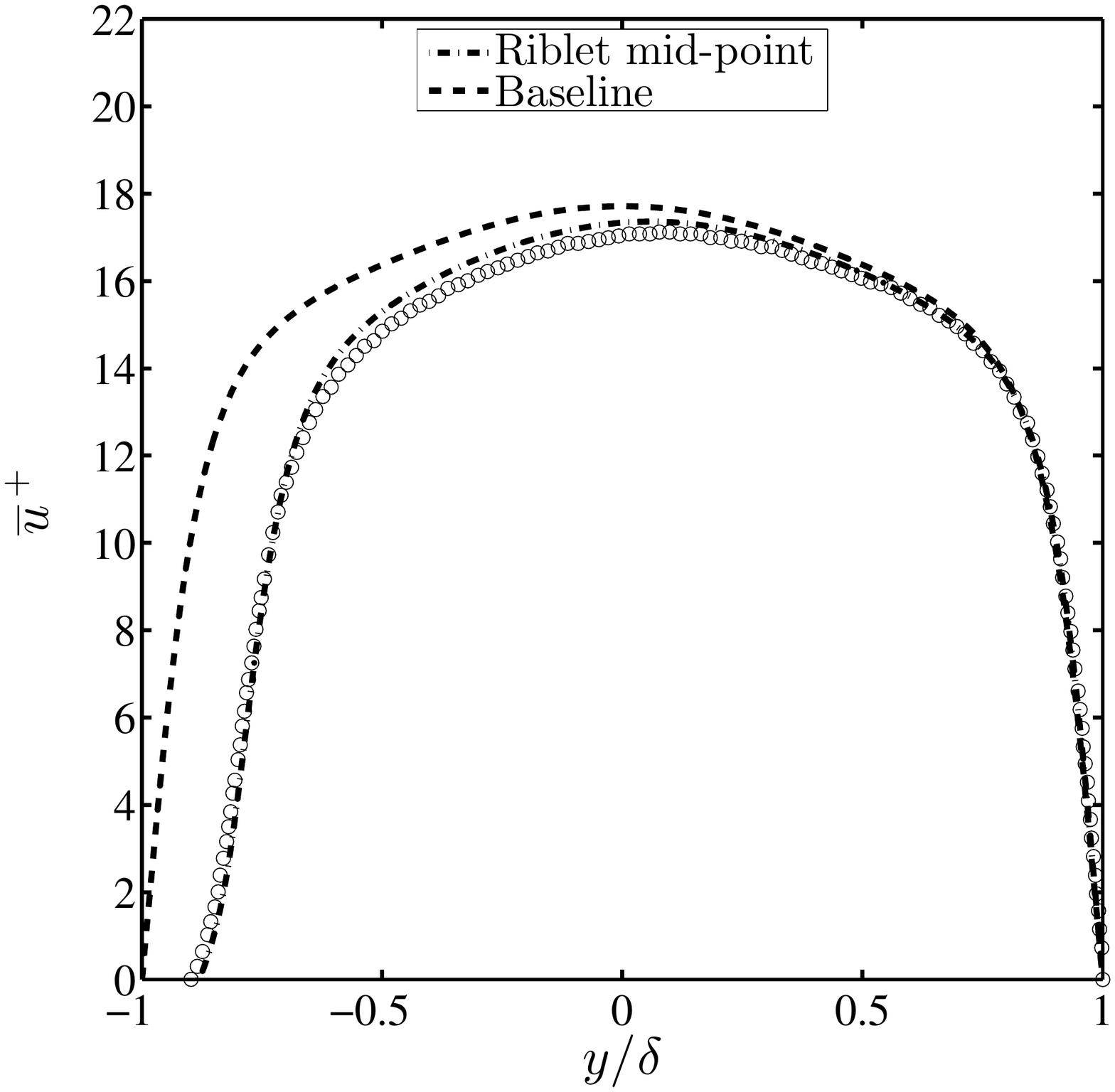}
			\put(37.5,15){\includegraphics[trim = 120mm 67.5mm 105mm 93mm, clip, width = 1.25in]{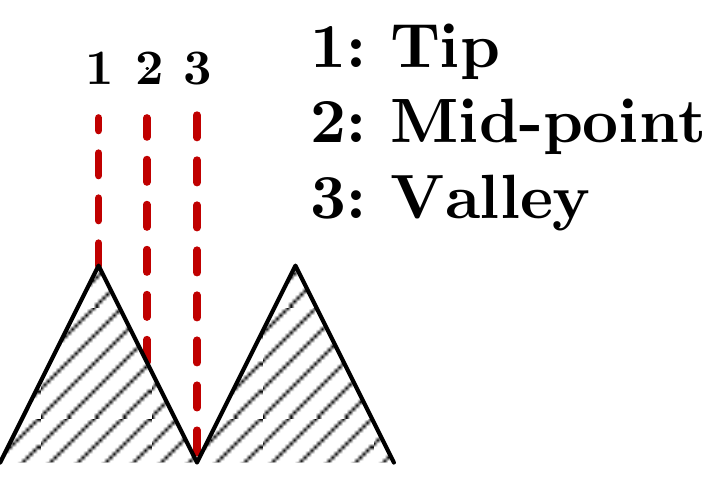}}
			\end{overpic}
			\label{UavgBenchmark}
		}
	\end{minipage}
	\hfill
	\begin{minipage}[c]{0.495\textwidth}
		\centering
		\subfigure[][]
		{
			% trim option's parameter order: left bottom right top
			\includegraphics[trim = 0.5mm 50mm 10mm 51mm, clip, width = 0.95\textwidth]{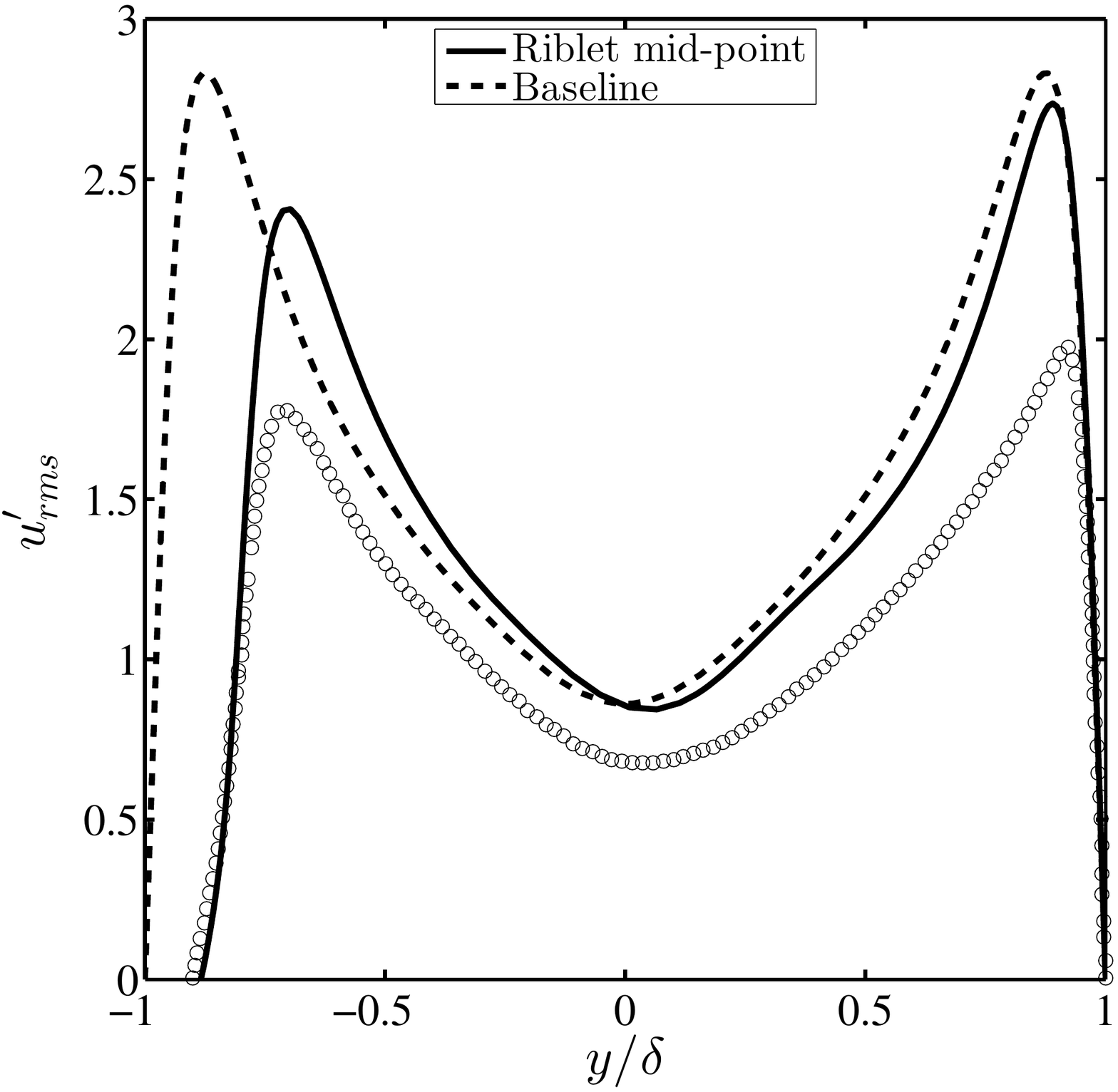}
			\label{UrmsBenchmark}
		}
	\end{minipage}
	
	\begin{minipage}[c]{0.495\textwidth}
		\centering
		\subfigure[][]
		{
			% trim option's parameter order: left bottom right top
			\includegraphics[trim = 0.5mm 50mm 10mm 51mm, clip, width = 0.95\textwidth]{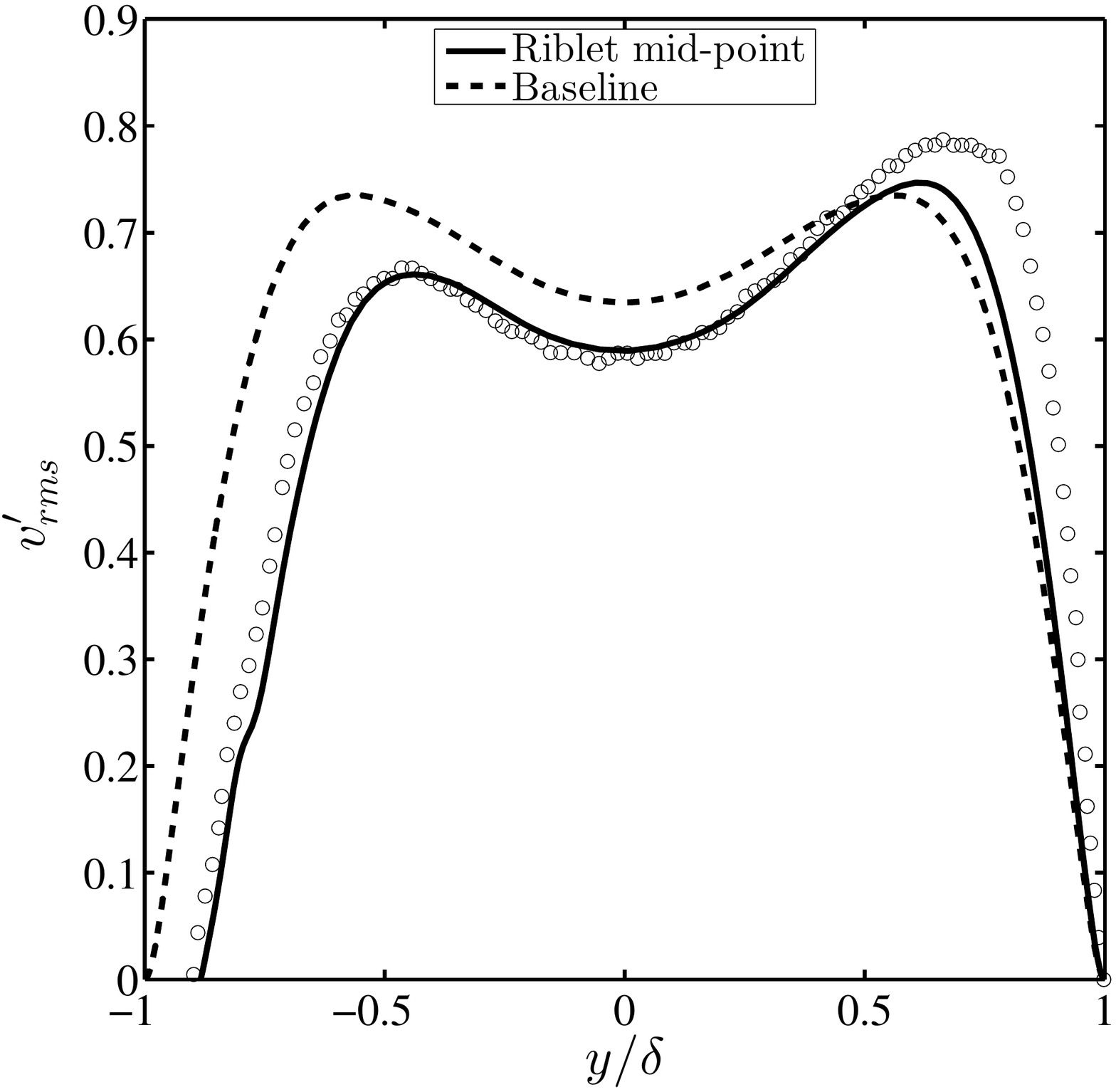}
			\label{VrmsBenchmark}
		}
	\end{minipage}
	\hfill
	\begin{minipage}[c]{0.495\textwidth}
		\centering
		\subfigure[][]
		{
			% trim option's parameter order: left bottom right top
			\includegraphics[trim = 0.5mm 50mm 10mm 51mm, clip, width = 0.95\textwidth]{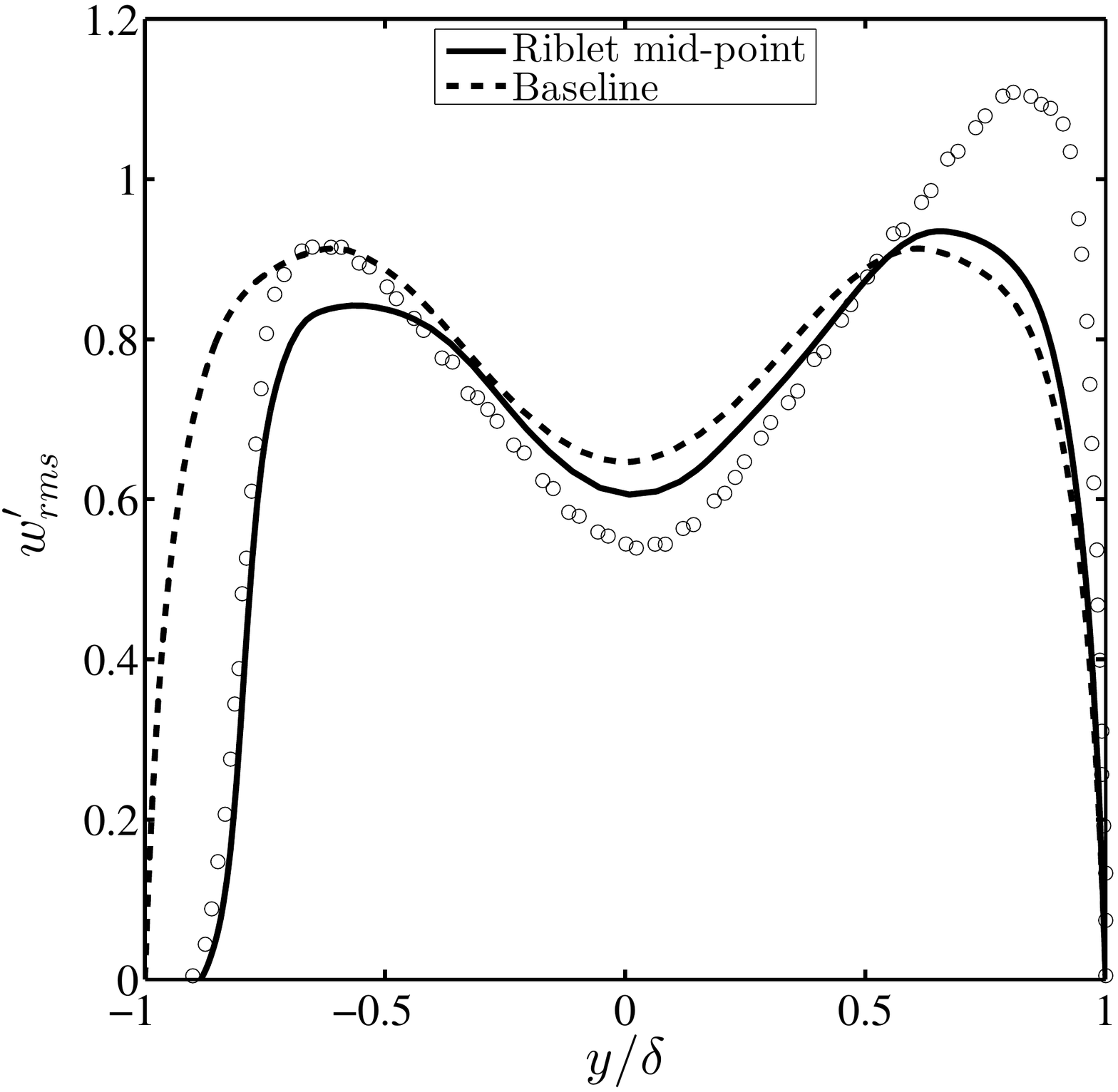}
			\label{WrmsBenchmark}
		}
	\end{minipage}
	
	For caption see the next page.
\end{figure*}
%
%=====================================================================================================================================================================================================%
%

	Case II in this parametric study represents a benchmarking configuration adapted from the 
work of~\citet{Chu93}, since their numerical framework was formulated in part using the 
spectral element method. The resulting riblet spacing and height in wall units are 
respectively $\Sp = 17.2$ and $\Hp = 17.2$, based on the surface-averaged friction 
velocity of the riblet wall. It is worthwhile to mention that the reference profiles 
reproduced in the present paper could contain a small degree of error arising from the 
digitization process.
%
%=====================================================================================================================================================================================================%
%
\begin{figure*}[t]
	\begin{minipage}[c]{0.495\textwidth}
		\centering
		\subfigure[][]
		{
			% trim option's parameter order: left bottom right top
			\includegraphics[trim = 0.5mm 50mm 10mm 51mm, clip, width = 0.95\textwidth]{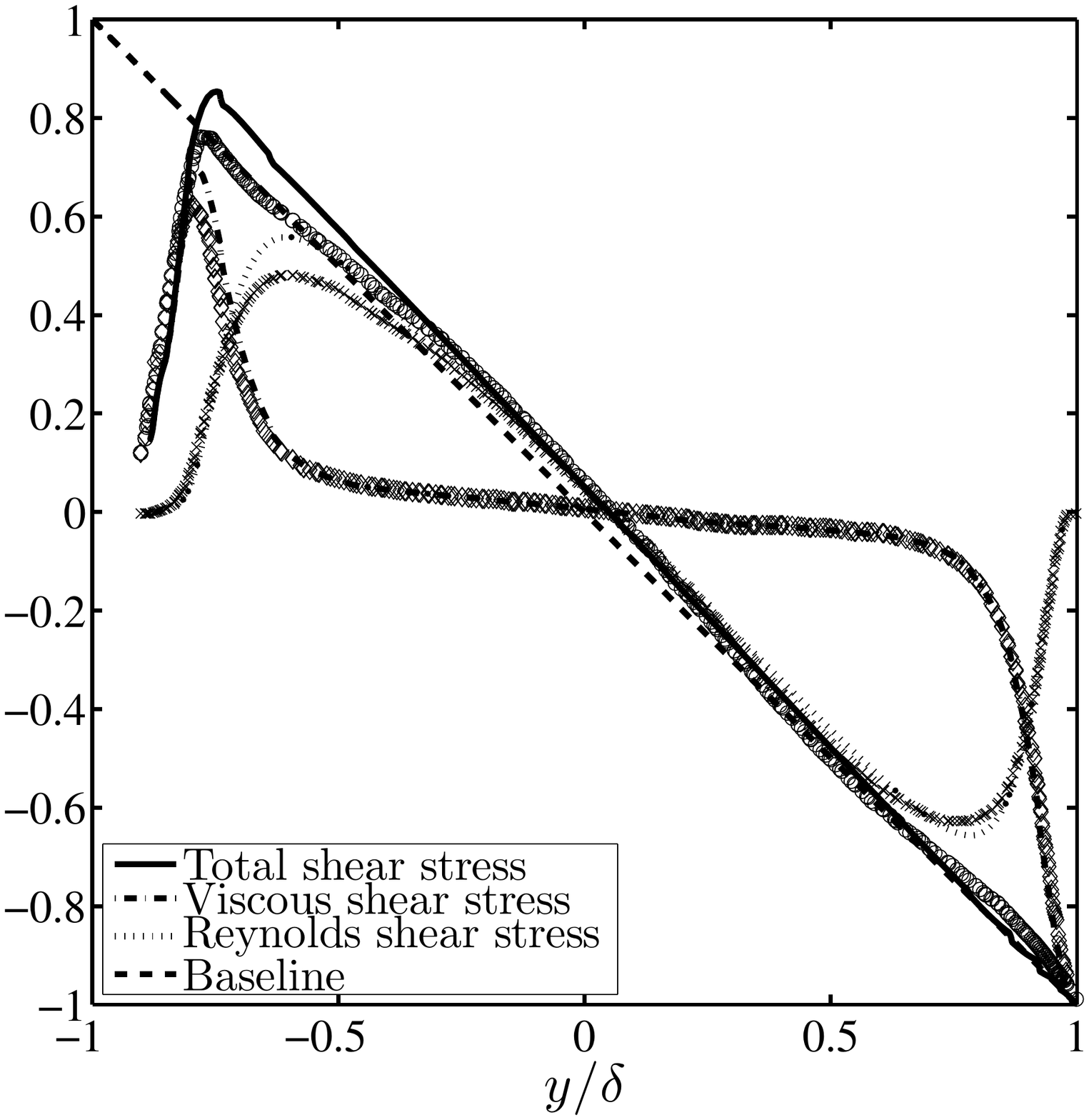}
			\label{StressBenchmark}
		}
	\end{minipage}
	
	\caption
		{
			Benchmarking of statistics collected from Case II $(\lgp \approx 12)$ at the riblet mid-point:
			\subref{UavgBenchmark}	 Mean streamwise velocity profile, 
			\subref{UrmsBenchmark}	 Root-mean-square streamwise velocity fluctuations, 
			\subref{VrmsBenchmark}	 Root-mean-square wall-normal velocity fluctuations, 
			\subref{WrmsBenchmark}	 Root-mean-square spanwise velocity fluctuations, and
			\subref{StressBenchmark} Shear stress distribution.
			The profiles are scaled in wall units based on the friction velocity of the top (smooth) wall. The markers ($\circ$,$\smalldiamond$,$\times$) represent the digitized profiles reported by~\citet{Chu93}, while the baseline profiles refer to those acquired from the configuration with two plane walls.
			\label{CaseIIBenchmark}
		}
\end{figure*}
%
%=====================================================================================================================================================================================================%
%
\begin{figure*}[t]
	\centering
		% trim option's parameter order: left bottom right top
		\includegraphics[trim = 10mm 5mm 10mm 16mm, clip, width = 0.65\textwidth]{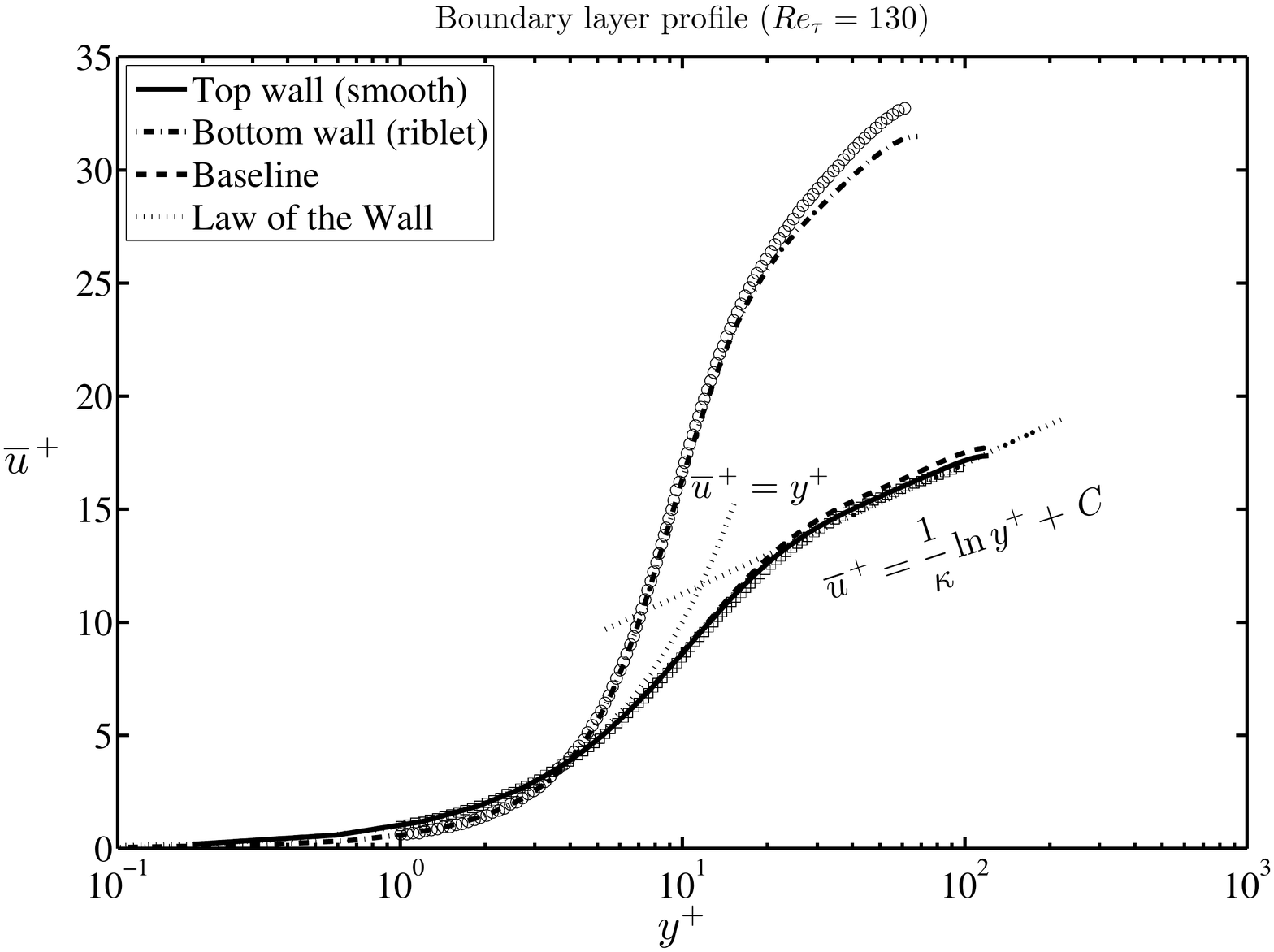}
		
		\caption{Boundary layer profile from Case II $(\lgp \approx 12)$ at the riblet mid-point normalized using the \emph{local} friction velocity $(\kappa = 0.4$ and $C = 5.5)$. The marker $(\circ, \smallsquare)$ represents the digitized profiles reported by~\citet{Chu93}, while the baseline profiles refer to those acquired from the configuration with two smooth walls. \label{CaseIIBLBenchmark}}
\end{figure*}
%
%=====================================================================================================================================================================================================%
%

	Figure~\ref{CaseIIBenchmark} depicts the profiles of mean streamwise velocity, 
root-mean-square velocity fluctuations and shear stress distribution across the channel 
at a spanwise location above the riblet mid-point. Note that all profiles are normalized 
in wall units using the friction velocity of the top (smooth) wall for the ease of 
comparison. Figure~\ref{CaseIIBLBenchmark} shows the boundary layer profile above the 
riblet mid-point. In this figure, the von K{\'a}rm{\'a}n constant $\kappa$ and the 
constant $C$ each has a value of 0.4~\citep{Jimenez07, Lumley01} and 5.5~\citep{KMM87} 
respectively in the log-law equation $\overline{u}^+ = \ln{y^+}/\kappa + C$ describing 
the logarithmic layer. Note that the profiles on the riblet wall are computed using the 
\emph{local} friction velocity for comparison with the reference profiles. The statistics 
acquired from the baseline plane turbulent channel flow at $\Rem = 1842$ with a same 
domain size of $5\delta \times 2\delta \times 2\delta$ is also included in the two figures 
for comparison. Although not shown here, the baseline profiles ($\ReT \approx 123$) 
are comparable with those acquired from a simulation at $\ReT \approx 110$~\cite{Iwamoto02}.

	The profiles for Case II illustrated in figures~\ref{CaseIIBenchmark} 
and~\ref{CaseIIBLBenchmark} are generally in good agreement with those reported 
by~\citet{Chu93}. However, the profiles of streamwise velocity fluctuations $u'$ 
between Case II and the reference presented in Fig.~\ref{UrmsBenchmark} contain rather 
significant discrepancy. The peak of the reference profile near the top wall ($y/\delta 
= 1$) has a lower magnitude than the profiles for Case II and also the baseline 
configuration. On the other hand, the peak magnitudes of $v'_{rms}$ and $w'_{rms}$ near 
the top wall reported by~\citet{Chu93} are somewhat higher than the present profiles, and 
only the profile of $v'_{rms}$ near the riblet wall shows reasonable agreement. 
Some possible explanations for the discrepancies could be the different mesh resolutions 
and length of time-averaging. \citet{Chu93} collected the statistics within 300 
non-dimensional time units from a medium resolution mesh which has cross-flow grid 
spacings of $\Delta y_{min}^+ = 0.42$ and $\Delta z_{min}^+ = 0.43$ near the riblet wall. 
A study on the effect of mesh resolution in the spectral element framework has shown that 
the profile of $u'_{rms}$ is under-predicted because of under-resolution in the 
wall-normal direction~\citep[p. 475]{Karniadakis05}. On a separate note, simulations of 
the baseline configuration at $\ReT \approx 110$ and 150~\citep{Iwamoto02} revealed that 
the peak magnitude of $u'_{rms}$ is about 2.6. In this respect, the peak magnitude of 
$u'_{rms}$ in Case II near the top wall is on a par with the data in the literature, 
same goes for those associated with $v'_{rms}$ and $w'_{rms}$.
%
%=====================================================================================================================================================================================================%
%
\begin{table*}[t]
	\caption{Comparison of flow properties near the smooth (top) and riblet (bottom) walls between the present simulation of Case II and the earlier DNS by {\protect\NoHyper\citet{Chu93}\protect\endNoHyper}. The riblet spacing and height are $s = 0.2\delta$ and $h = 0.2\delta$ respectively. \label{tab:PropertiesBenchmark}}
		
		\centering
		\begin{ruledtabular}
		\begin{threeparttable}		
			\begin{tabular*}{0.98\textwidth}{@{\extracolsep{\fill} } l*{4}{d}}
																																							 & \multicolumn{2}{c}{\bf Reference simulation} 								 		 & \multicolumn{2}{c}{\bf Present simulation}\\
																																							 & \multicolumn{1}{c}{Smooth wall} & \multicolumn{1}{c}{Riblet wall} & \multicolumn{1}{c}{Smooth wall} & \multicolumn{1}{c}{Riblet wall}\\
				\hline																																			
				$\hspace{2mm} \ReT$ 																									 & 131														 &	128\tnote{*}						 			 & 130 													 	 & 128\tnote{*}\\[3pt]
				$\hspace{2mm} \dsty \frac{U_c}{\Um}$ 		 															 & \multicolumn{2}{c}{1.22} 			 																	 & \multicolumn{2}{c}{1.22}\\[9pt]
				$\hspace{2mm} \dsty \frac{U_c}{\Utau}$ 																 & 17.08 													 &	17.54\tnote{*}							 	 & 17.25 												 	 & 17.50\tnote{*}\\[9pt]
				$\hspace{2mm} \delta^*$ 																							 & 0.140 												 	 &	0.205							 						 & 0.144 												 	 & 0.200\\
				$\hspace{2mm} \theta$ 																								 & 0.079 													 &	0.089					 								 & 0.082 												 	 & 0.085\\
				$\hspace{2mm} H$ 	 																										 & 1.77 							 						 &	2.30					 							 	 & 1.75													 	 & 2.34\\
				$\hspace{2mm} \dsty G = \frac{U_c}{\Utau}\left(\frac{H - 1}{H}\right)$ & 7.45 													 & 14.85				 									 & 7.36       										 & 14.98\\[9pt]
				$\hspace{2mm} \dsty J = \frac{U_c - \Um}{\Utau}$											 & 3.08 													 & 4.74				 										 & 3.10 													 & 4.70\\[6pt]
			\end{tabular*}
			
			\begin{tablenotes}
				\item[*] \scriptsize{Based on the \emph{equivalent} friction velocity averaged over the entire riblet wall.}
			\end{tablenotes}
		\end{threeparttable}
	\end{ruledtabular}
\end{table*}
%
%=====================================================================================================================================================================================================%
%

	The comparison of shear stress distributions in Fig.~\ref{StressBenchmark} also shows 
some discrepancies, especially near the riblet wall. The peaks of the profiles of Reynolds 
stress $-\overline{u'v'}^+$ have the largest disparity most likely because of the lower 
magnitude of $u'$ reported by the reference. On the other hand, a small discrepancy of 
the peak magnitudes is observed on the profiles of viscous shear stress 
$\partial\overline{u}^+/\partial y^+$. In all, these discrepancies lead to a difference 
in the peak magnitudes of total shear stress near the riblet wall between the present and 
the reference simulations. Table~\ref{tab:PropertiesBenchmark} summarizes and compares the 
flow properties between the current simulation of V-groove riblets of Case II and the 
reference~\cite{Chu93}. These properties include centerline velocity $U_c$, friction 
velocity $\Utau$, displacement thickness $\delta^*$, momentum thickness $\theta$, shape 
factor $H$, Clauser shape parameter $G$, and velocity defect ratio $J$. Once again, the 
decent agreement between the two sets of data ascertains the current simulation setup.
		
\subsection{Variation of drag reduction performance with the size of riblets}
\label{sec:ScalingDRSize}
%
%=====================================================================================================================================================================================================%
%	
\begin{table}[t]
	\caption{Drag reduction performance of V-groove riblets in the channel. The values in wall units are computed based on the surface-averaged friction velocity at the riblet (bottom) wall. $\ReT$ is the Reynolds number based on the friction velocity on the top (smooth) wall. \label{tab:RibletPerformance}}
		
		\centering
		\begin{ruledtabular}
		\begin{tabular*}{0.75\textwidth}{@{\extracolsep{\fill} } cl*{6}{c}dc}
			& {\bf Case} & $s$ 				 & $h$ 				 & $\Sp$ & $\Hp$ & $\lgp$ & $\ReT$ & \multicolumn{1}{c}{$\Delta${\bf Drag} (\%)} &\\
			\hline
			& {\bf I}		 & $0.2\delta$ & $0.1\delta$ & 20.4	 & 10.2	 & 10.21  & 125 	 & -5.08 														 					 &\\
			&	{\bf II}	 & $0.2\delta$ & $0.2\delta$ & 17.2	 & 17.2	 & 12.15  & 130 	 & -4.59														 					 &\\
			& {\bf III}	 & $0.4\delta$ & $0.4\delta$ & 43.0	 & 43.0	 & 30.41  & 137		 & +35.81 													 					 &\\
			\hline
			& {\bf IV} 	 & $0.2\delta$ & $0.1\delta$ & 30.6	 & 15.3	 & 15.28  & 181 	 & +0.79 																			 &\\
			& {\bf V} 	 & $0.2\delta$ & $0.2\delta$ & 26.9	 & 26.9	 & 19.10  & 186		 & +13.86 																		 &\\
			& {\bf VI} 	 & $0.4\delta$ & $0.4\delta$ & 65.2	 & 65.2	 & 46.15  & 196		 & +54.20 																		 &\\
		\end{tabular*}
		\end{ruledtabular}
\end{table}
%
%=====================================================================================================================================================================================================%
%

	Table~\ref{tab:RibletPerformance} tabulates the drag reduction performance of all six 
V-groove riblet configurations. In the same table, the corresponding values of 
$\ReT$, $\Sp$, $\Hp$ and $\lgp$ are also included. Since the smaller computational domain 
inevitably leads to a higher degree of intermittency in the flow evolution, the percentage 
drag reduction is calculated by taking the mean value within a reasonable time span towards 
the end of statistical averaging. In the same way, an uncertainty of $\pm 1\%$ has been 
estimated with a 95\% confidence level. In Case II, the skin friction drag reduction 
achieved is approximately 4.6\%, lower than the approximately 6\% reduction reported by 
the reference DNS. Nonetheless, the experimental results summarized by~\cite{Bechert97} 
shows 3\% to 5\% skin friction drag reduction for an identical riblet geometry. Therefore, 
the current percentage drag reduction falls within the scatter of experimental data. 
Likewise, \citet{Garcia11Thesis} reported a mean drag reduction of 4.32\% after 
reproducing the drag-reducing configuration simulated by~\citet{Choi93}, but with about 
18 times longer length of averaging. A separate simulation of Case II with 50\% longer 
domain extent in the two wall-parallel directions produces 3.6\% drag reduction and 
reduces the standard deviation from 0.47\% to 0.36\%. In view of this, the impact of 
changing the domain size alone is within the statistical uncertainty. Although the 
present work has to bear with the slightly larger uncertainty, it is not anticipated 
to affect the findings or conclusions markedly since the analysis of the six riblet 
configurations is carried out under similar settings.

	Apart from Case II, Case I also yields a drag reduction of 5.1\%. However, this 
configuration with $\Sp \approx 20$ and $h/s = 0.5$ has been shown experimentally 
by~\citet{Bechert97} to reduce slightly lesser drag than Case II. The discrepancy is 
again likely to be caused by the statistical uncertainty. Moreover, the fact that the 
$\lgp$ values of Cases I and II are rather similar, and close to the optimum value, i.e. 
$\lgp \approx 11$~\cite{Garcia11} means their performance are difficult to distinguish. 
On the contrary, the remaining four cases are all drag-increasing configurations. It is 
observed that the V-groove riblets produce viscous drag reduction only when $\lgp 
\lesssim 15$. This condition is slighly more restrictive than for the thin blade 
configuration which reduces drag if $\lgp \lesssim 17$~\citep{Garcia11Thesis}.
%
%=====================================================================================================================================================================================================%
%
\begin{figure}[t!]
	\begin{minipage}[c]{0.495\textwidth}
		\centering
		\subfigure[][]
		{
			% trim option's parameter order: left bottom right top
			\includegraphics[trim = 47mm 4.5mm 47mm 15mm, clip, width = 0.95\textwidth]{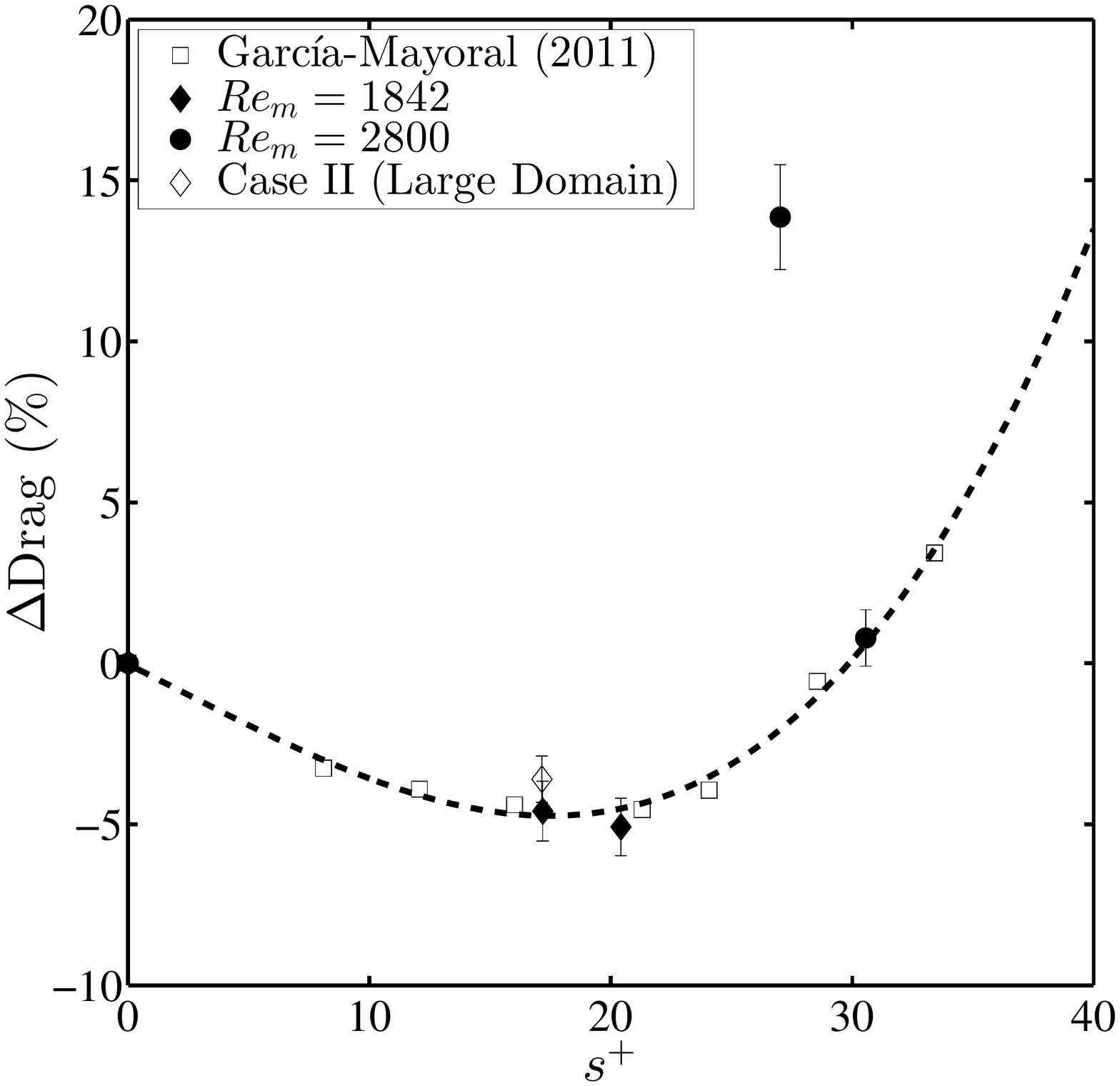}
			\label{DRsplus}
		}
	\end{minipage}
	\hfill
	\begin{minipage}[c]{0.495\textwidth}
		\centering
		\subfigure[][]
		{
			% trim option's parameter order: left bottom right top
			\includegraphics[trim = 47mm 4.5mm 47mm 15mm, clip, width = 0.95\textwidth]{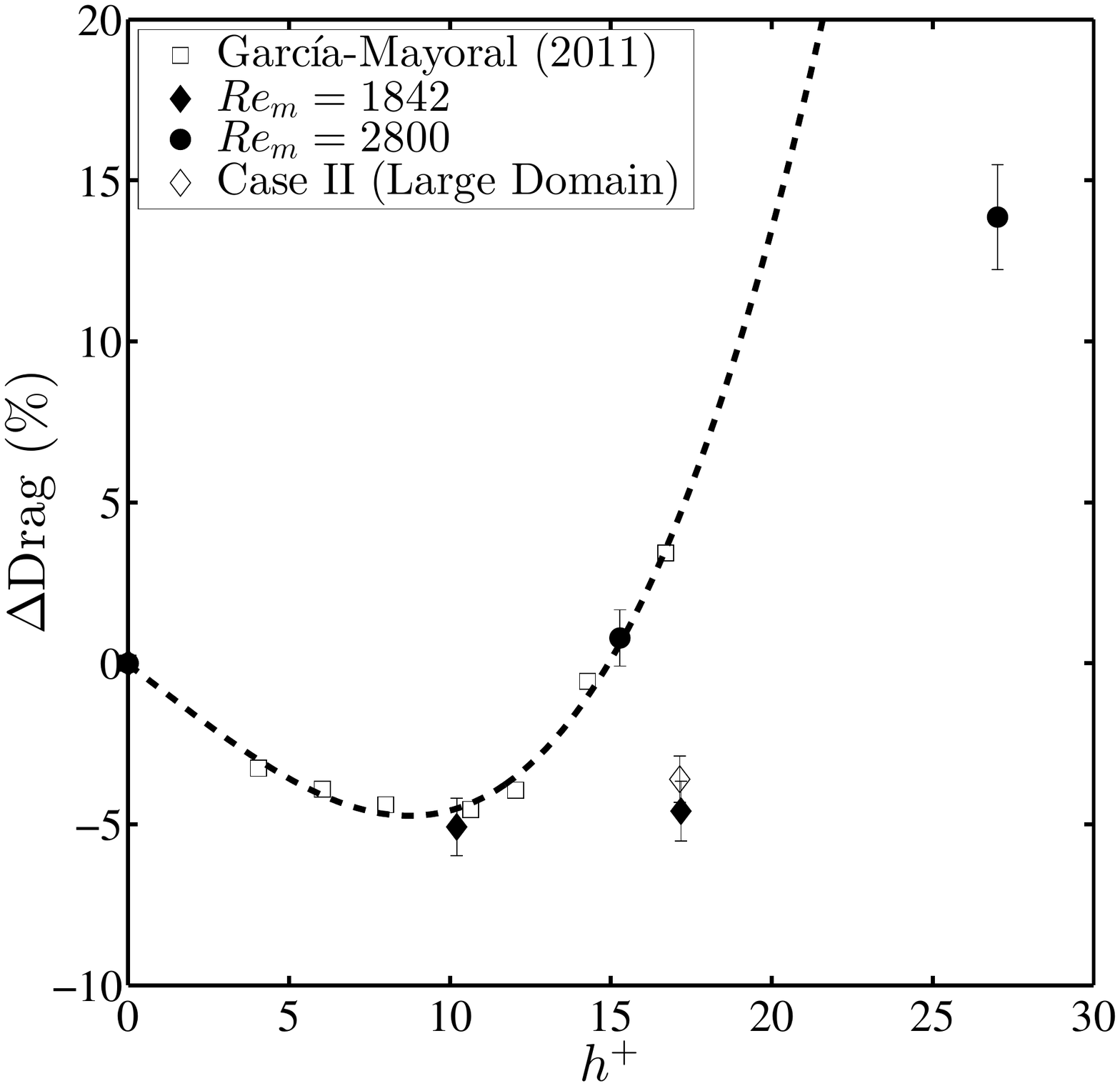}
			\label{DRhplus}
		}
	\end{minipage}
	
	\begin{minipage}[c]{0.495\textwidth}
		\centering
		\subfigure[][]
		{
			% trim option's parameter order: left bottom right top
			\includegraphics[trim = 47mm 4.5mm 47mm 15mm, clip, width = 0.95\textwidth]{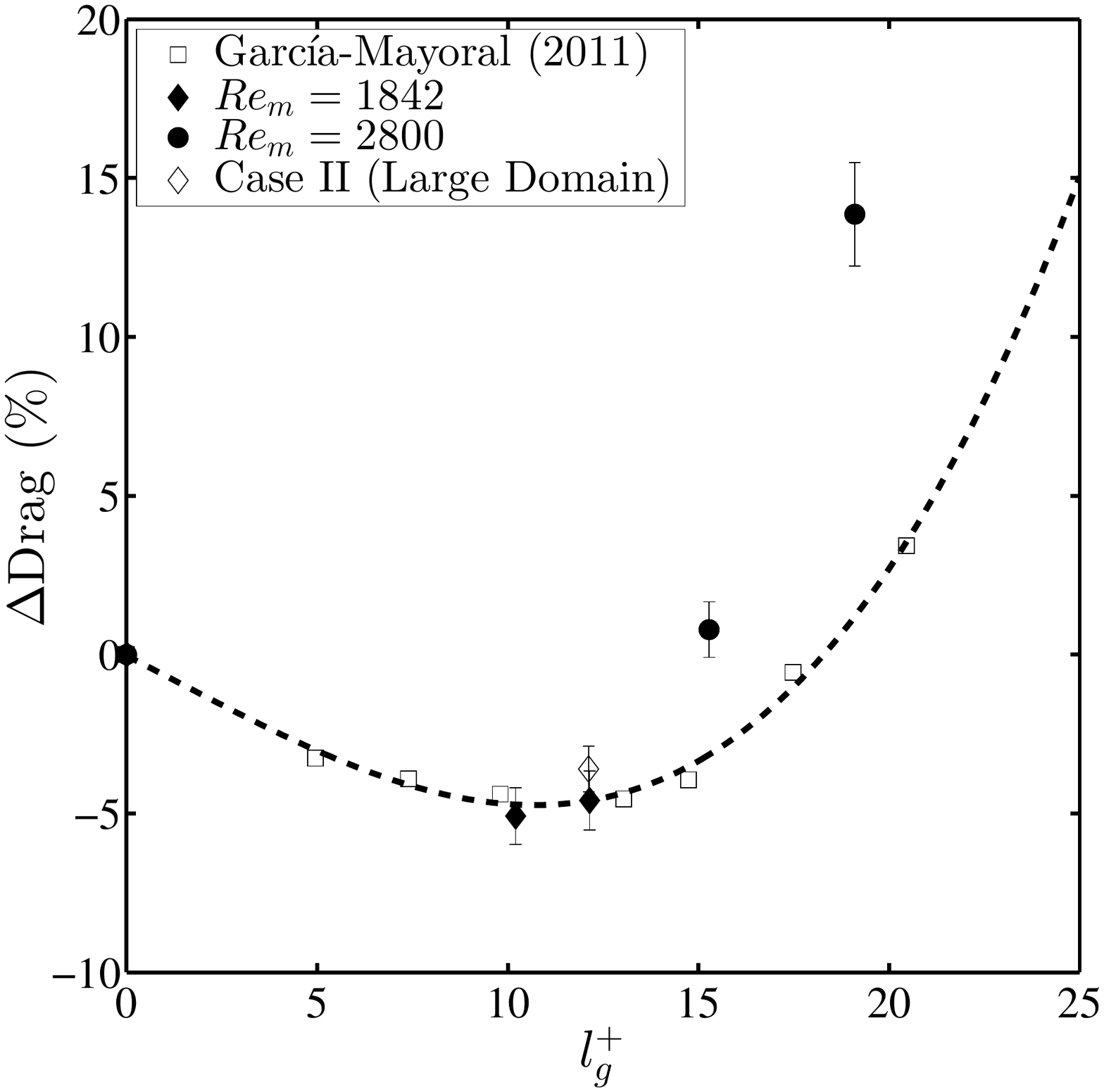}
			\label{DRlgplus}
		}
	\end{minipage}
		
		\caption
		{
			The scaling of drag reduction curves in terms of 
			\subref{DRsplus}  $\Sp$, 
			\subref{DRhplus}  $\Hp$, and 
			\subref{DRlgplus} $\lgp$.
			Note that the surface-averaged friction velocity on the riblet wall is used to compute the length scale in wall units. The profile for the thin blade riblets reported by~\citet{Garcia11Thesis} is included. The dash line is a curve-fitted profile of the reference data. The marker ($\smalldiamond$) is the corresponding value of Case II simulated using a 50\% larger domain size. Error bars represent a 95\% confidence level.
		}
\end{figure}
%
%=====================================================================================================================================================================================================%
% 

	Figure~\ref{DRlgplus} illustrates the percentage skin friction drag reduction achieved 
by the six V-groove riblet configurations against $\lgp$. Note that only those 
configurations with $\lgp < 25$ are shown in the figure, and the error bars represent a 
95\% confidence level. Also, the baseline turbulent channel flow is assigned a value of 
$\lgp = 0$. The drag reduction curve of the thin blade riblets reported 
by~\citet{Garcia11Thesis} is included to demonstrate the ``universality'' of the length 
scale $l_g^+$. By comparing Fig.~\ref{DRlgplus} with either Fig.~\ref{DRsplus} 
or~\ref{DRhplus}, it is evident that $\lgp$ mitigates the dependence on the ratio of $h/s$. 
Thus, it offers a better collapse of the data points belonging to the six V-groove 
configurations. The greatest appeal of $\lgp$ is the monotonic variation of the percentage 
drag reduction, as opposed to the ``zig-zag'' trend when using either $\Sp$ or $\Hp$ (see 
also Table~\ref{tab:RibletPerformance}). In view of this, there is a strong indication 
that the corresponding flow fields and statistics should also scale monotonically with 
$\lgp$.

	In Fig.~\ref{DRlgplus}, the agreement with the reference drag reduction curve of thin 
blade riblets is decent for the cases which have $\lgp \approx 10$ and 12, notwithstanding 
the uncertainty. Although the discrepancy widens with the increase of $\lgp$, 
\citet{Garcia11Thesis} has shown that the data scattering is still significant. One 
plausible explanation could be the rising importance of inertial effects in the groove 
due to a greater exposure to the overlying turbulence structures residing in the buffer 
or logarithmic layers. Alternatively, the breakdown of ``universality'' may imply that 
the influence of shape-dependent mechanisms starts to manifest as the local flow 
transitions from the \emph{viscous} to the \emph{k-roughness} regime. In this respect, it 
is conjectured that the precise groove geometry becomes an influential factor that dictates 
the near-wall turbulent motions. One shortcoming of the present work is the lack of data to 
confirm the fitting in the \emph{viscous regime}, i.e. when $\lgp < 10$, because smaller 
riblets would require even higher mesh resolutions. Nevertheless, the physics of flow in 
the \emph{viscous regime} has been addressed extensively by~\citet{Bechert89, Bechert90}, 
\citet{Luchini91}, and~\citet{Gruneberger11}. Besides, determining the role of riblet 
shape in causing the divergent of drag reduction curves when the size of riblets is near 
or beyond the performance optimum at $\lgp \approx 11$ would be more illuminating.

\subsection{Variation of turbulence statistics with the size of riblets}
\label{sec:ScalingStatSize}
%
%=====================================================================================================================================================================================================%
%
\begin{figure*}[p]
	\centering
	\subfigure[][]
	{
		% trim option's parameter order: left bottom right top
		\includegraphics[trim = 10mm 27.5mm 10mm 37mm, clip, width = 0.525\textwidth]{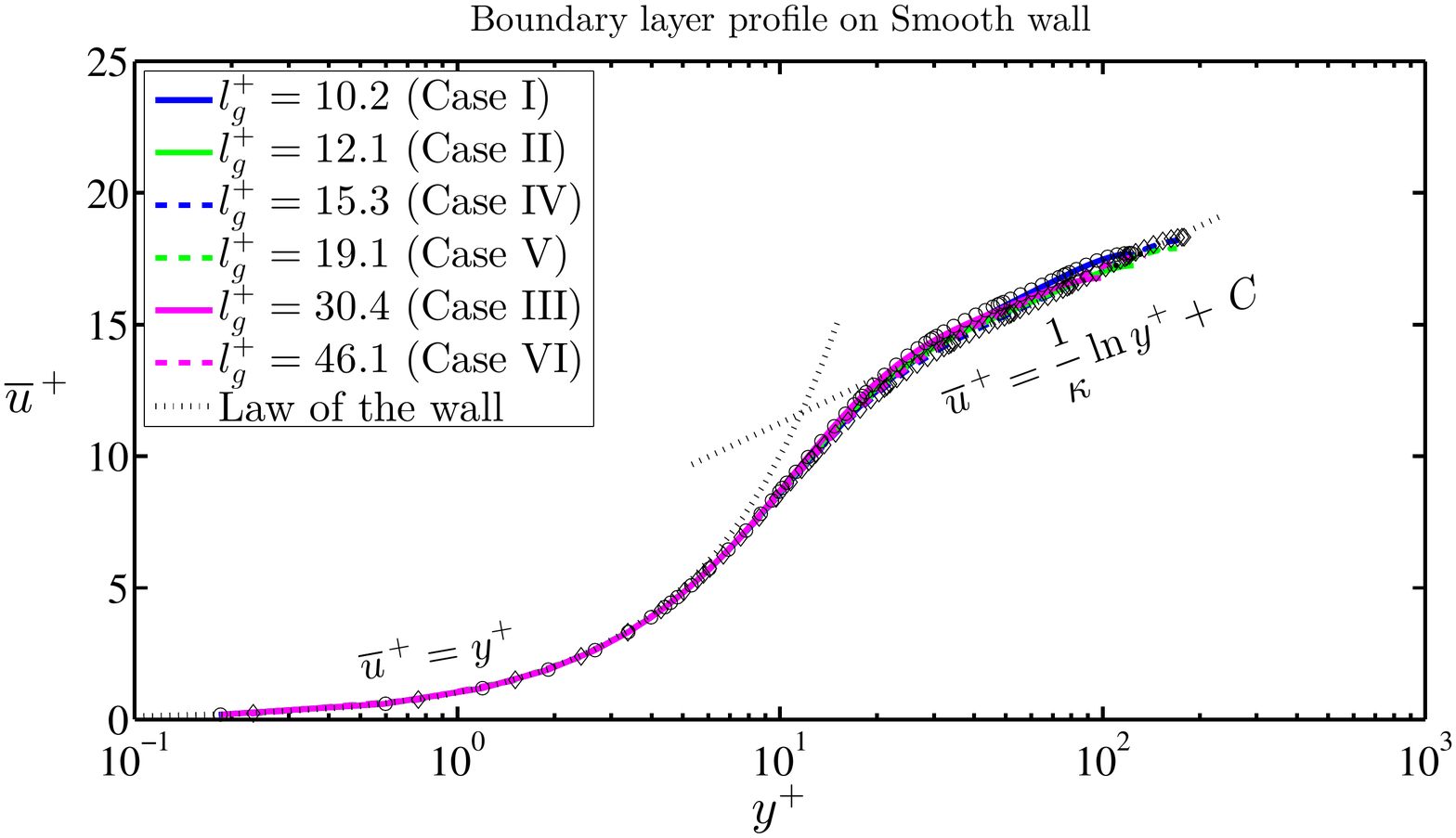}
		\label{BLProfileSmooth}
	}
	
	\centering
	\subfigure[][]
	{
		% trim option's parameter order: left bottom right top
		\includegraphics[trim = 10mm 5mm 10mm 15.5mm, clip, width = 0.525\textwidth]{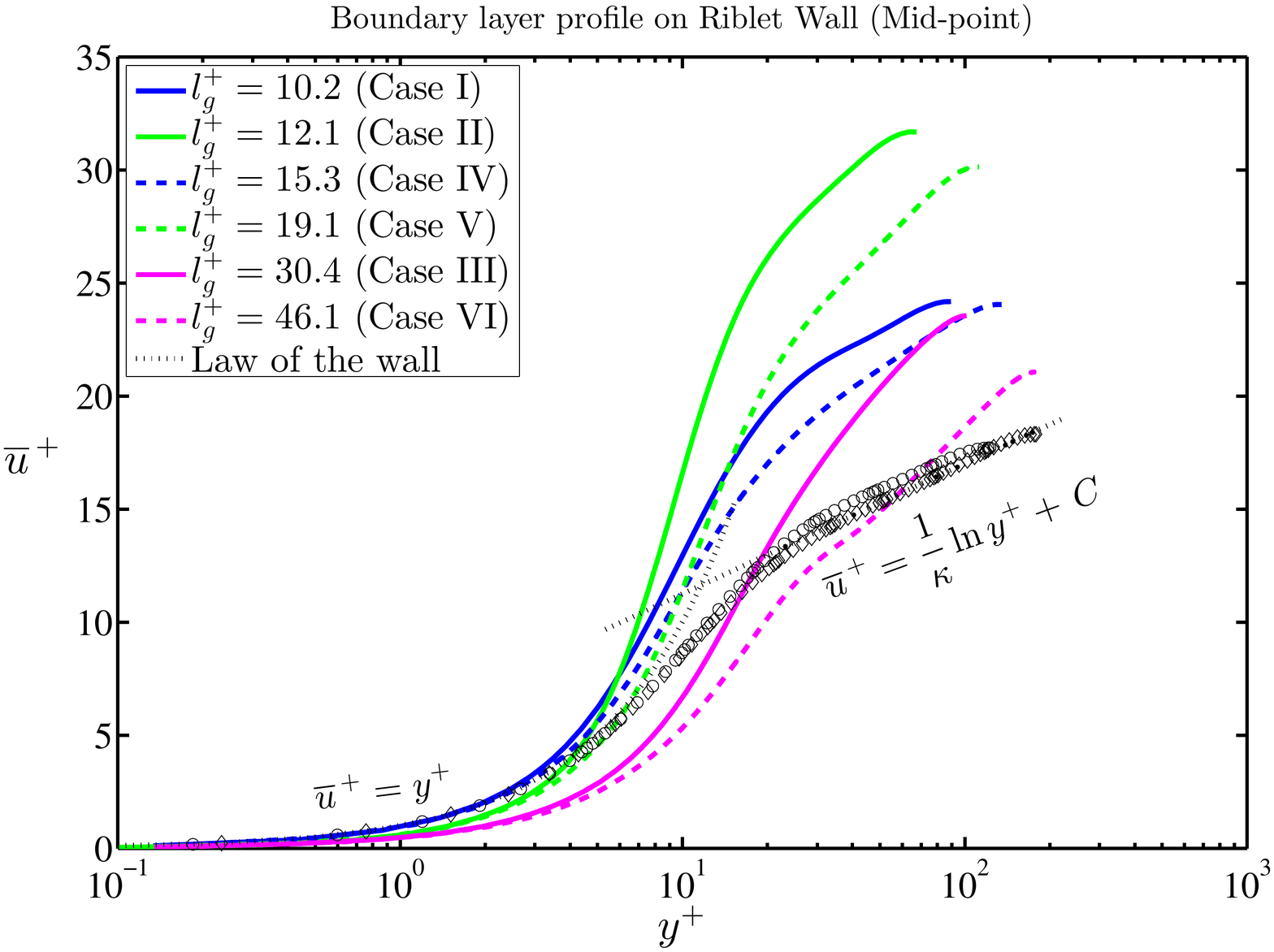}
		\label{BLProfileRiblet}
	}
	
	\centering
	\subfigure[][]
	{
		% trim option's parameter order: left bottom right top
		\includegraphics[trim = 10mm 27.5mm 10mm 37mm, clip, width = 0.525\textwidth]{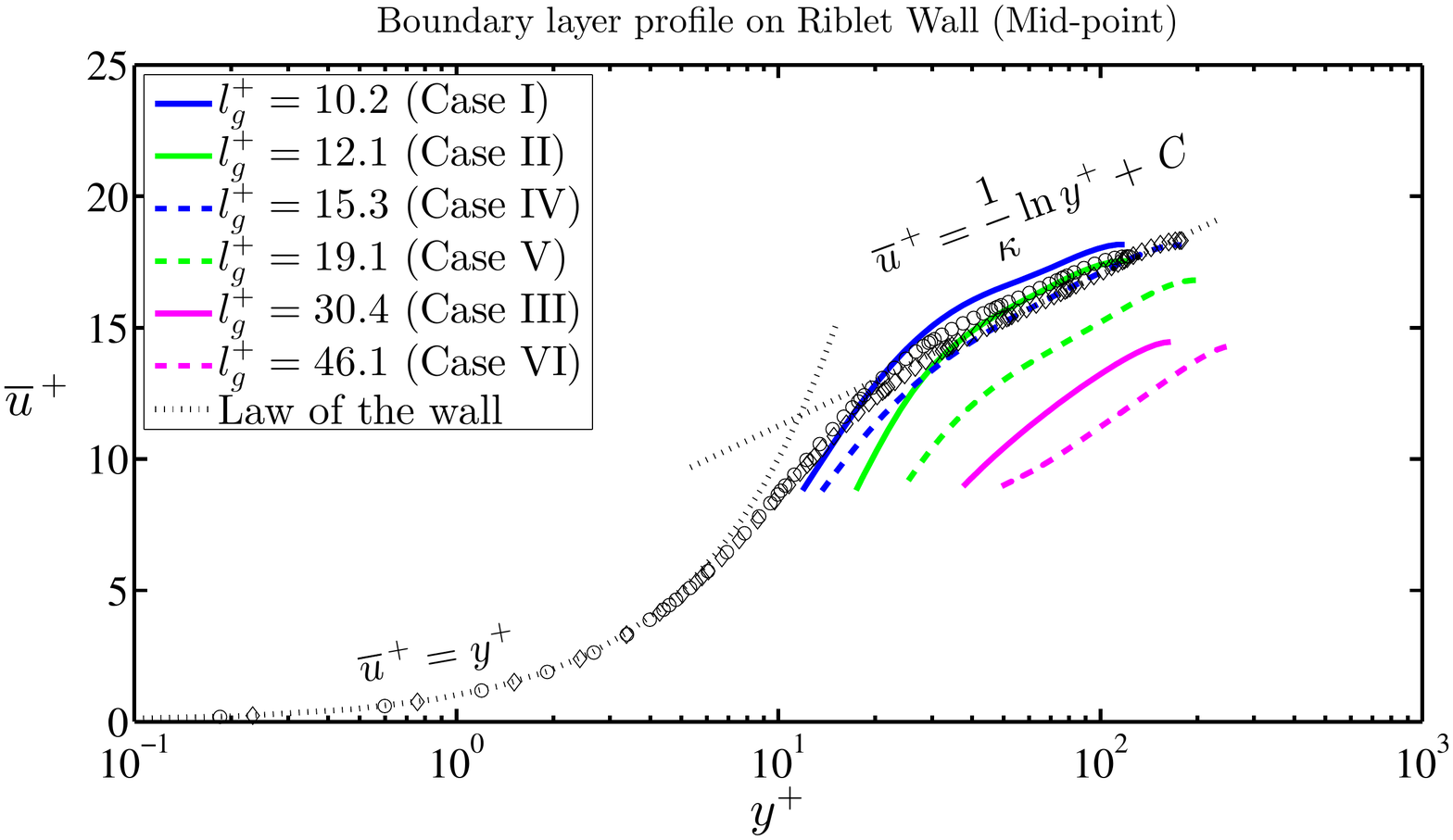}
		\label{BLProfileRibletEqui}
	}
	
	\caption[Boundary layer profiles of the six V-groove riblet configurations.]
		{
			Boundary layer profiles of the six V-groove riblet configurations $(\kappa = 0.4$ and $C = 5.5)$:
			\subref{BLProfileSmooth} 		 On the smooth top wall
			\subref{BLProfileRiblet} 		 Above the riblet mid-point normalized using the \emph{local} friction velocity, and
			\subref{BLProfileRibletEqui} Above the riblet mid-point normalized using the \emph{equivalent} friction velocity.
			Profiles drawn in solid and dashed lines represent configurations at $\Rem = 1842$ and 2800, respectively. Likewise, the markers ($\circ$) and ($\smalldiamond$) represent the baseline profiles acquired from the configuration with two smooth walls at $\Rem = 1842$ and 2800, respectively.
			\label{BLProfile}
		}
\end{figure*}
%
%=====================================================================================================================================================================================================%
%

	Figure~\ref{BLProfile} depicts the boundary layer profiles of all six configurations at 
a spanwise location above the riblet mid-point. Note that the profiles on the riblet wall 
are computed using the \emph{local} friction velocity. Although the V-groove riblets 
protrude significantly into the flow, the boundary layer profile on the top smooth wall 
is hardly affected as seen in Fig.~\ref{BLProfileSmooth}. In contrast, the boundary 
layer profile on the riblet surface plotted in wall units using the \emph{local} friction 
velocity is modified considerably as shown in Fig.~\ref{BLProfileRiblet}. In Cases III 
and VI with $s = h = 0.4\delta$, the inner layer is altered significantly and the log-layer 
is no longer established with the same slope. On the other hand, significant deviation from 
the law of the wall is mainly observed on the log-layer for the remaining four riblet 
configurations with $s = 0.2\delta$. As for the inner layer, the effect is marginal 
especially for Cases I and III with $h = 0.1\delta$.

	In Fig.~\ref{BLProfileRiblet}, the shifting of the boundary layer profile does not 
exhibit a clear variation trend with $\lgp$. Instead, the shifting appears to be dependent 
on the physical dimensions of riblets and/or the bulk Reynolds number $\Rem$. With the 
exception of Cases III and VI, the log-layer profile at the same $\Rem$ is shifted upwards 
with a steeper slope following an increase of the riblet height measured in outer units. 
In view of this, the threshold value of $\lgp$ above which the boundary layer is altered 
considerably seems to fall in the range $20 < \lgp < 30$. Therefore, it is likely that the 
significant modification of near-wall flow field in these two cases is associated with a 
different flow feature or mechanism. Interestingly, the riblets in both Cases III and VI 
have a spacing (see table~\ref{tab:RibletPerformance}) that is larger than the average 
diameter of the inherent near-wall streamwise vortices (around 30 wall units). This 
suggests that the new feature or mechanism could be related to the postulation 
by~\citet{Choi93}.

	Figure~\ref{BLProfileRibletEqui} shows another view of the log-layer profiles on 
the riblet wall scaled using an \emph{equivalent} friction velocity $\Utau^*$. 
$\Utau^*$ is defined as the friction velocity of a plane wall which experiences 
the same amount of drag as on a riblet wall having an equivalent projected surface 
area. $\Utau^*$ can also be perceived as the friction velocity averaged over the entire 
riblet wall. This idea is analogical to computing the friction velocity at the ``apparent 
origin'' (see the discussions by~\citet{McLean13}, p.219-222). Under this alternative 
normalization, it is clear that the shifting of the log-layer has a direct relation with 
the drag-reducing ability of the V-groove riblets, and also scales monotonically with 
$\lgp$. An upward shift is observed in Cases I and II with $\lgp < 15$, while a downward 
shift is observed on the remaining four drag-increasing configurations in which $\lgp > 
15$. The upward shift in the log-layer or an increase in the intercept $C$ in the log-law 
equation is a common feature in drag-reduced flows, and it can be perceived as a thickening 
of the viscous sublayer~\citep{Walsh90B, Choi93}.
%
%=====================================================================================================================================================================================================%
%
\begin{figure*}[p]
	\begin{minipage}[c]{0.495\textwidth}
		\centering
		\subfigure[][]
		{
			% trim option's parameter order: left bottom right top
			\includegraphics[trim = 0.5mm 50mm 10mm 51mm, clip, width = 0.95\textwidth]{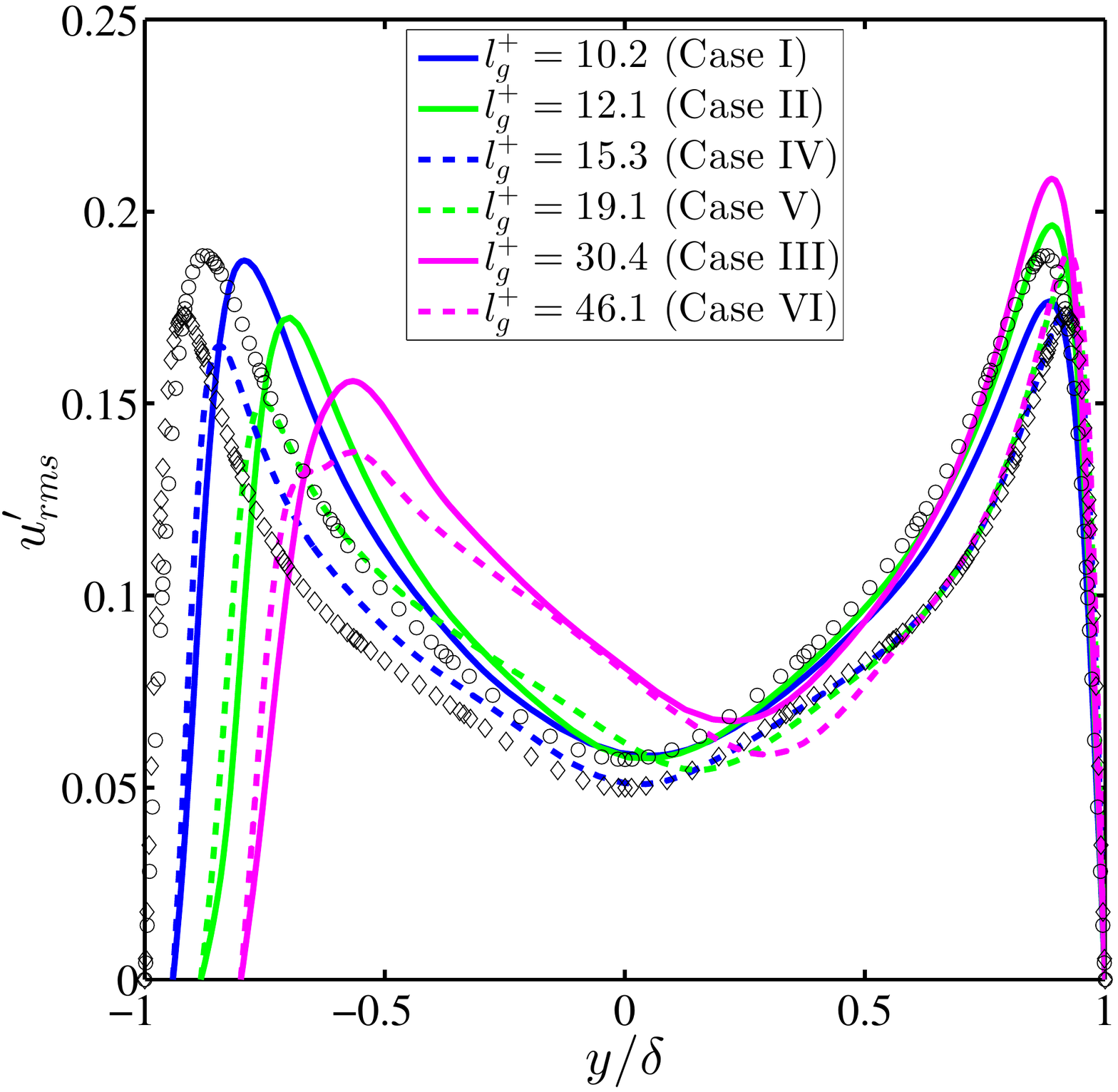}
			\label{UrmslgplusMid}
		}
	\end{minipage}
	\hfill
	\begin{minipage}[c]{0.495\textwidth}
		\centering
		\subfigure[][]
		{
			% Insert 'grid, tics=10' in the square bracket to show the grid in 10% intervals.
			% trim option's parameter order: left bottom right top
			\begin{overpic}[trim = 0.5mm 50mm 10mm 51mm, clip, width = 0.95\textwidth]{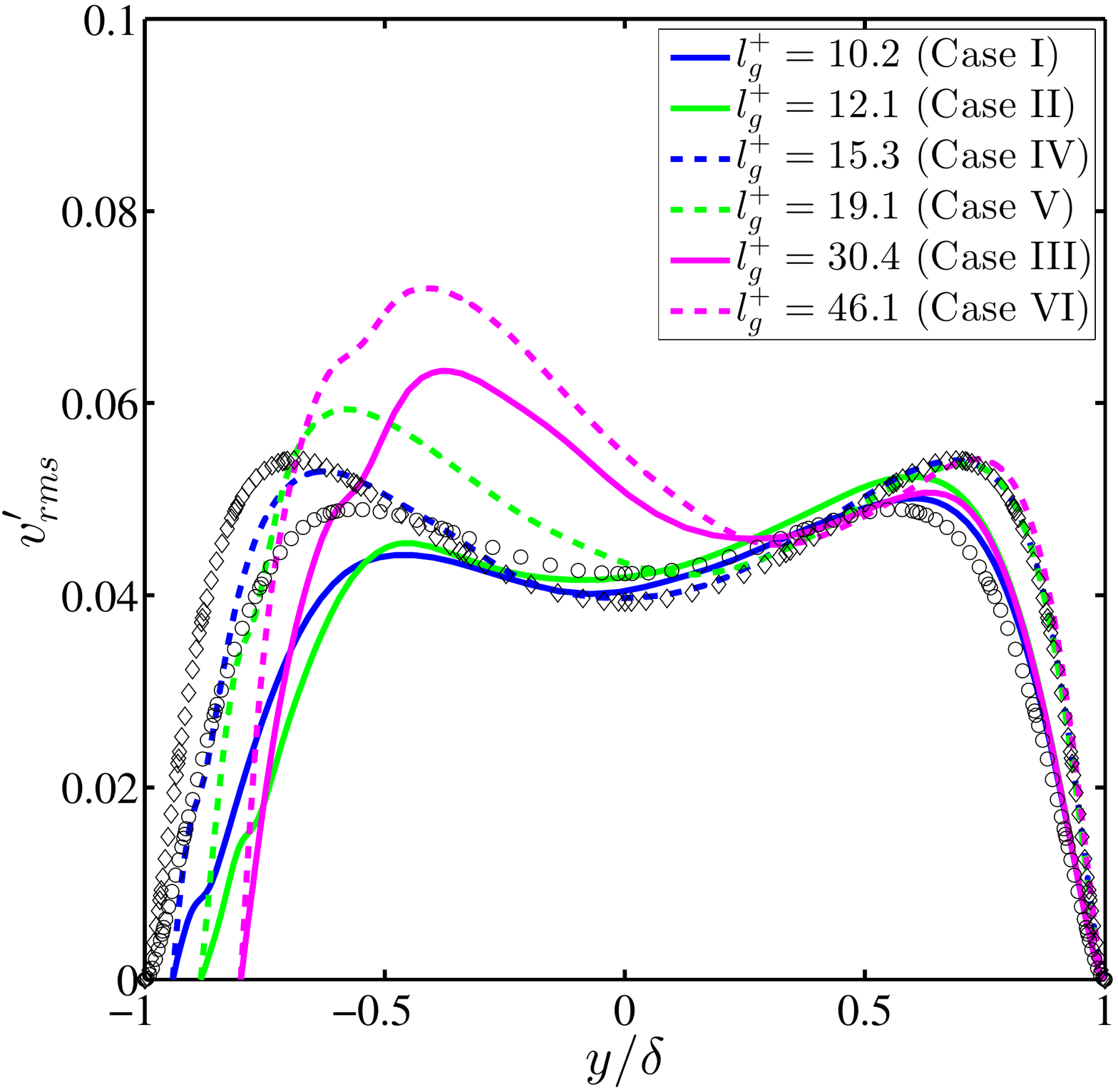}
			\put(8.25,65.5){\parbox{0.2\linewidth}{\linespread{1}\raggedleft\scriptsize hump\\ $\searrow$}}
			\end{overpic}
			\label{VrmslgplusMid}
		}
	\end{minipage}
	
	\begin{minipage}[c]{0.495\textwidth}
		\centering
		\subfigure[][]
		{
			% trim option's parameter order: left bottom right top
			\includegraphics[trim = 0.5mm 50mm 10mm 51mm, clip, width = 0.95\textwidth]{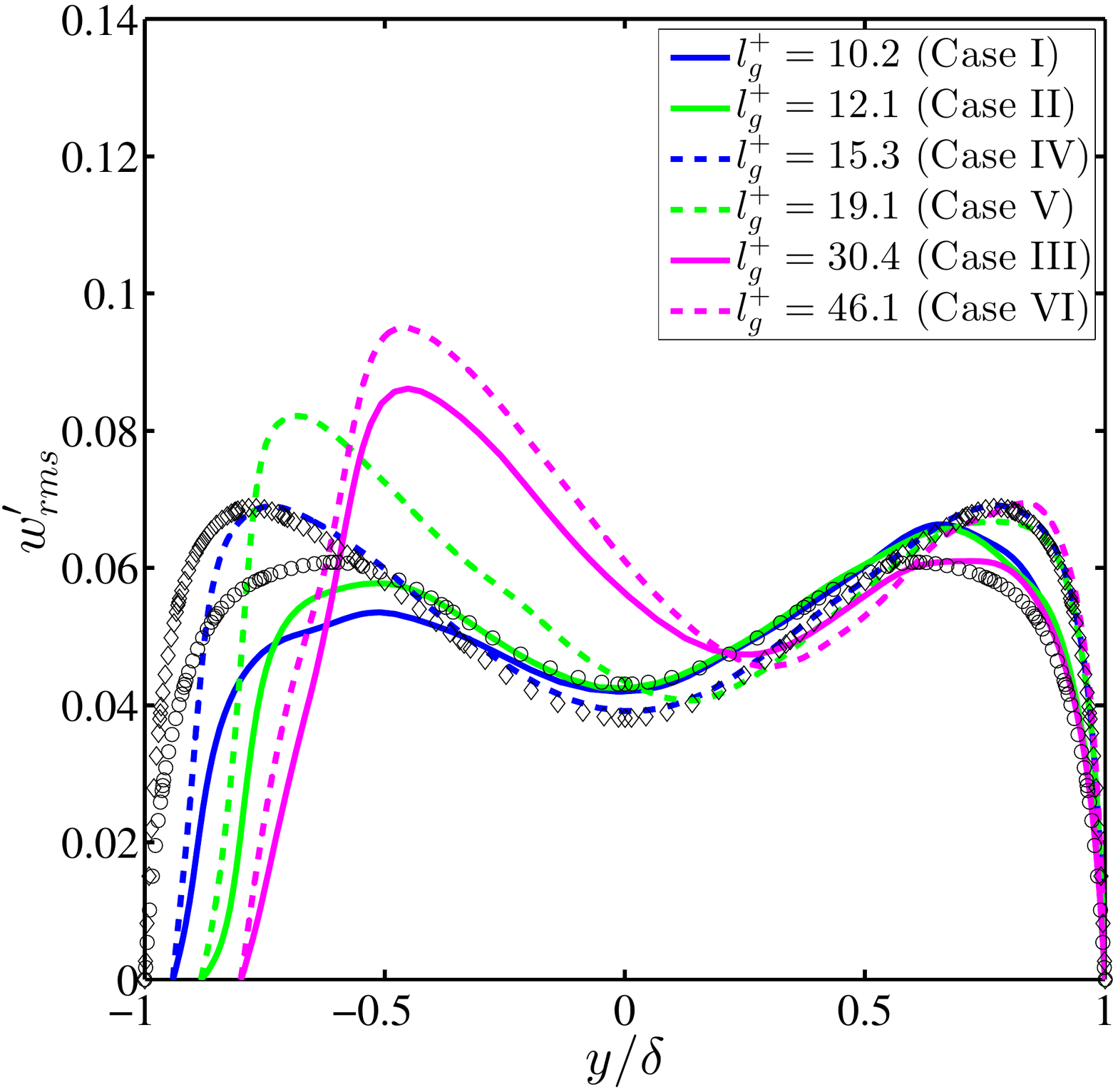}
			\label{WrmslgplusMid}
		}
	\end{minipage}
	\hfill
	\begin{minipage}[c]{0.495\textwidth}
		\centering
		\subfigure[][]
		{
			% Insert 'grid, tics=10' in the square bracket to show the grid in 10% intervals.
			% trim option's parameter order: left bottom right top
			\begin{overpic}[trim = 0.5mm 50mm 10mm 51mm, clip, width = 0.95\textwidth]{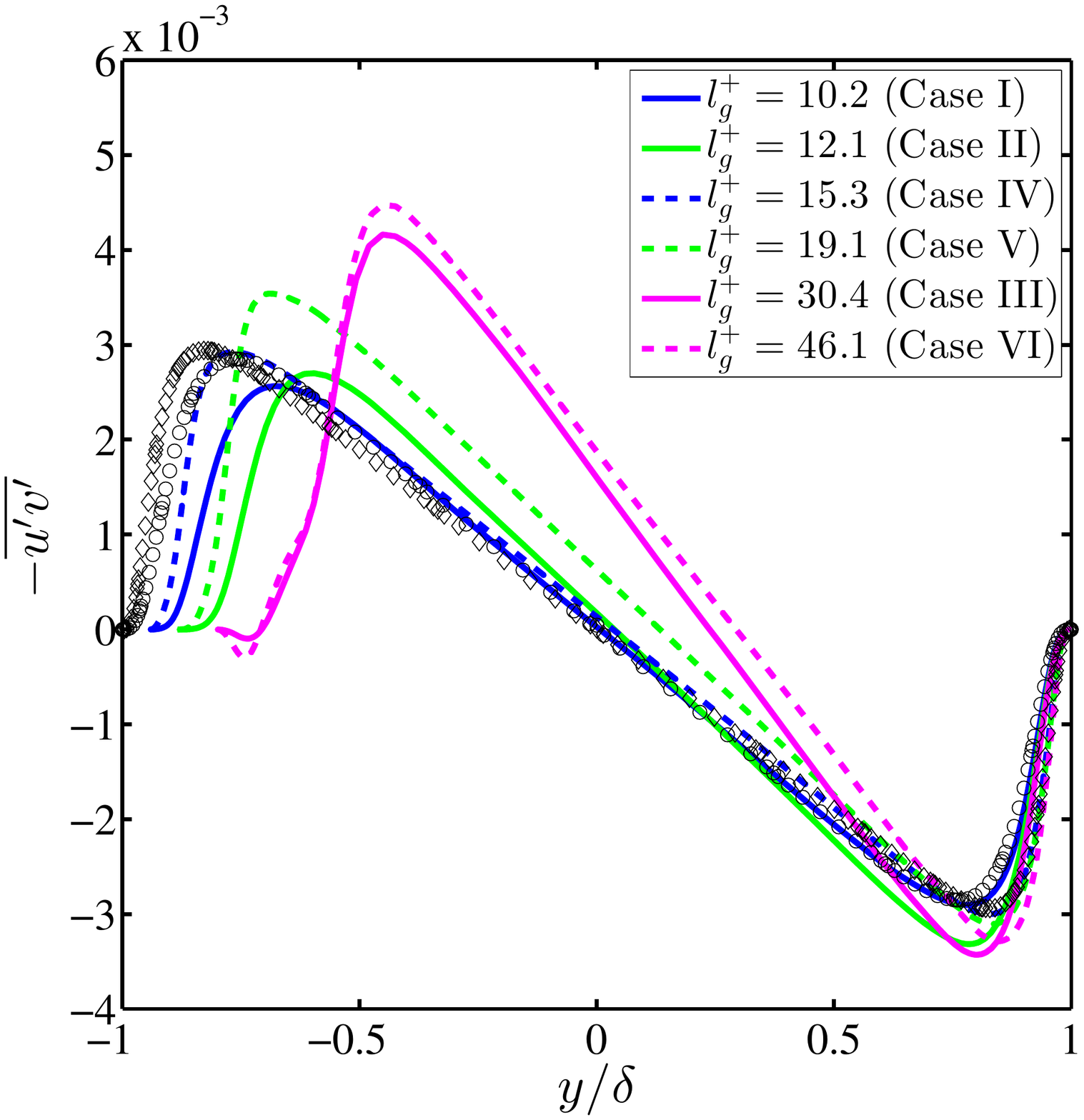}
			\put(12.75,35.25){\parbox{0.2\linewidth}{\linespread{1}\centering\scriptsize $\uparrow$\\ hump}}
			\end{overpic}
			\label{UVmslgplusMid}
		}
	\end{minipage}
	
	\caption
		{
			Profiles of turbulent fluctuations of all six V-groove riblet configurations across the channel above the riblet mid-point:
			\subref{UrmslgplusMid} $u'_{rms}$, 
			\subref{VrmslgplusMid} $v'_{rms}$, 
			\subref{WrmslgplusMid} $w'_{rms}$, and
			\subref{UVmslgplusMid} $-\overline{u'v'}$.
			Profiles drawn in solid and dashed lines represent configurations at $\Rem = 1842$ and 2800, respectively. Likewise, the markers ($\circ$) and ($\smalldiamond$) represent the baseline profiles acquired from the configuration with two smooth walls at $\Rem = 1842$ and 2800, respectively. Note that all profiles are scaled in outer units using the bulk velocity $\Um$.
			\label{ReyStresslgplusMid}
		}
\end{figure*}
%
%=====================================================================================================================================================================================================%
%

	Figure~\ref{ReyStresslgplusMid} depicts the wall-normal distribution of turbulent 
fluctuations across above the riblet mid-point. The profiles for the baseline 
configurations with two plane walls at $\Rem = 1842$ and 2800 simulated with the same 
domain size are also included to illustrate the effects of riblets. Note that all the 
profiles are scaled in outer units using the bulk streamwise velocity $\Um$. The rationale 
of not scaling in wall units is because the friction velocities are affected by the 
different cross-sectional area across the six riblet configurations. Therefore, adhering 
to the conventional scaling in wall units could yield ambiguous and even misleading 
comparison, unlike using $\Um$ that is kept the same across all configurations. 
Furthermore, it is well-known that the profiles of velocity fluctuations do not collapse 
universally at low Reynolds numbers even when scaled in wall 
units~\citep{Spalart88, Moser99}.

	In general, the V-groove riblets are able to reduce the peak of streamwise velocity 
fluctuations $u'$ as compared to the respective baseline configuration. Interestingly, 
Fig.~\ref{UrmslgplusMid} shows that the reduction is proportional to $\lgp$. Furthermore, 
configurations that have $h/s = 1$ are more effective in reducing the peak of $u'$, 
whereas cases I and III with $h/s = 0.5$ have a peak near the riblet wall that is 
comparable in magnitude to the baseline configuration. The reduction of $u'$ may be 
attributed to the ability of riblets in aligning the near-wall flow structures thereby 
manipulating the bursting process~\citep{Tardu93}. In particular, impeding the lateral 
motions of near-wall longitudinal vortices leads to premature bursts which in turn give 
rise to a lower production of turbulence energy~\cite{Choi01}. On the other hand, the 
difference of peaks between the top and bottom walls appears to correlate with the flow 
asymmetry about the channel centerline, which is observed to be proportional to the riblet 
height $\Hp$. Judging from these observations, the impact of riblets on $u'$ may not 
directly reflect their drag-reducing ability. Instead, the drag reduction performance of 
V-groove riblets is strongly correlated with the cross-flow velocity fluctuations as 
illustrated in figures~\ref{VrmslgplusMid} and~\ref{WrmslgplusMid}.

	Both Cases I and II which produce a viscous drag reduction exhibit peaks of $v'$ and $w'$ 
that are discernibly lower than the baseline configuration, and also the corresponding 
peaks near the top smooth wall. On the contrary, the rest of the drag-increasing cases 
typically have a higher peaks of $v'$ and $w'$ near the riblet wall. In this 
respect, a reduction in viscous drag can only be attained when turbulent activities in 
the cross-flow directions are disrupted. Similarly, a direct relation with the drag 
reduction performance can be observed from the Reynolds stress profiles shown in 
Fig.~\ref{UVmslgplusMid}. This is in line with the conclusions of~\citet{Fukagata02} 
and~\citet{Marusic07} that manipulating the Reynolds stress $-\overline{u'v'}$ is the key 
to achieve effective control of wall turbulence and substantial viscous drag reduction. 
More astonishingly, one can see that the peaks of the cross-flow velocity fluctuations and 
Reynolds stress near the riblet wall also scale monotonically with $\lgp$. 
%
%=====================================================================================================================================================================================================%
%
\begin{figure}[t]
	\begin{minipage}[c]{0.495\textwidth}
		\centering
		\subfigure[][]
		{
			% trim option's parameter order: left bottom right top
			\includegraphics[trim = 45mm 4.5mm 47mm 15mm, clip, width = 0.95\textwidth]{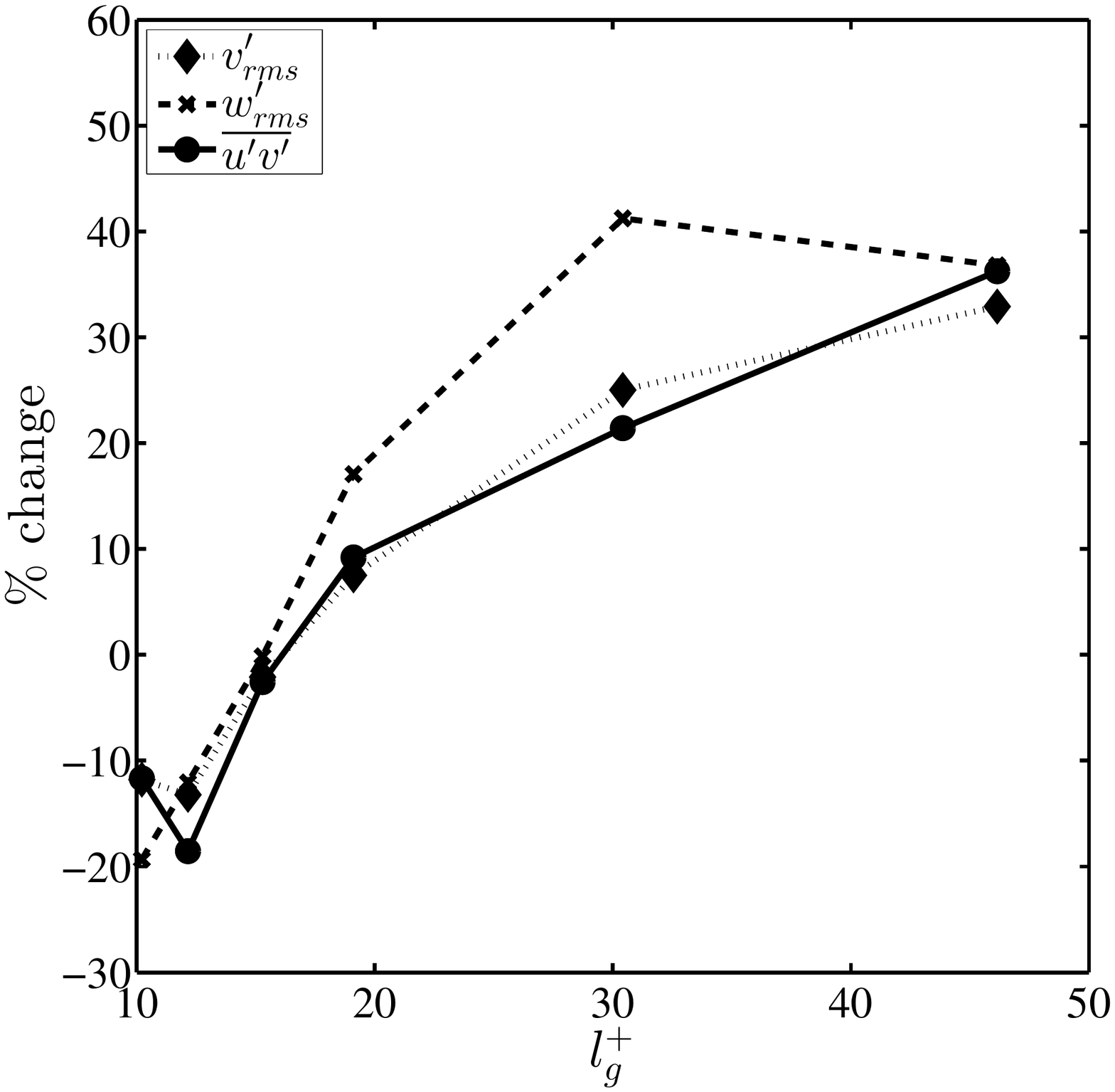}
			\label{RMSPercentSmooth}
		}
	\end{minipage}
	\hfill
	\begin{minipage}[c]{0.495\textwidth}
		\centering
		\subfigure[][]
		{
			% trim option's parameter order: left bottom right top
			\includegraphics[trim = 45mm 4.5mm 47mm 15mm, clip, width = 0.95\textwidth]{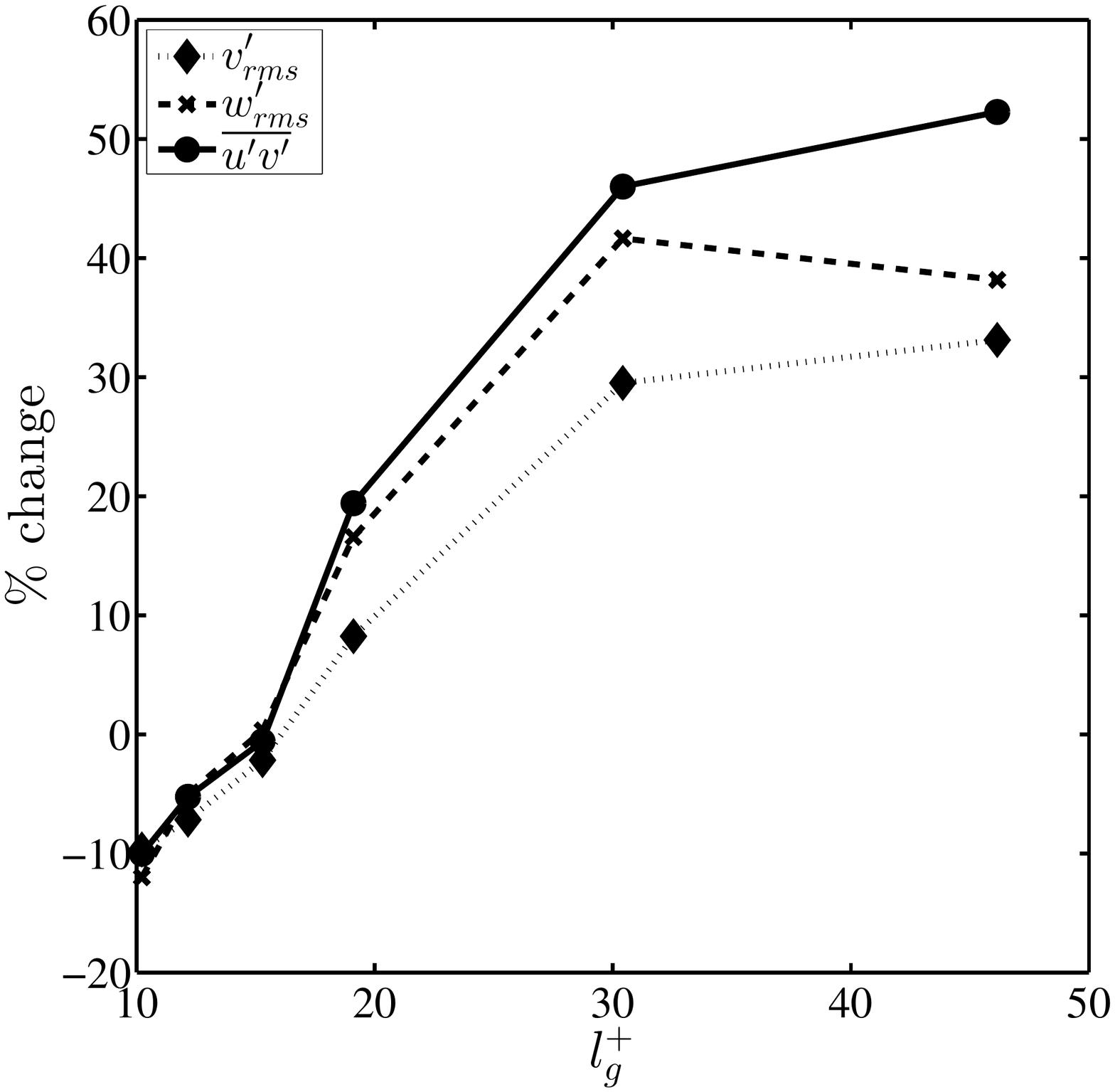}
			\label{RMSPercentBaseline}
		}
	\end{minipage}
		
	\caption
		{
			Percentage change of the peak magnitudes of $v'_{rms}$, $w'_{rms}$ and $-\overline{u'v'}$ of the six riblet configurations above the riblet mid-point as a function of $\lgp$: 
		\subref{RMSPercentSmooth}		Computed with respect to the top (smooth) wall, and 
		\subref{RMSPercentBaseline} Computed with respect to the baseline configuration.
		\label{RMSPercentlgplus}
		}
\end{figure}
%
%=====================================================================================================================================================================================================%
%

	Figure~\ref{RMSPercentlgplus} shows the percentage change in the peak magnitudes of 
$v'_{rms}$, $w'_{rms}$ and $-\overline{u'v'}$ above the riblet mid-point plotted against 
$\lgp$. In Fig.~\ref{RMSPercentSmooth}, the percentage change is computed with respect to 
the top smooth wall and it shows a positive correlation with $\lgp$ in general. The same 
trend is observed in Fig.~\ref{RMSPercentBaseline} when computed with respect to the 
baseline configuration. A small deviation from the trend is mainly observed in 
Fig.~\ref{RMSPercentlgplus} on the profiles of $v'_{rms}$ and $-\overline{u'v'}$ at $\lgp 
\approx 10$ and 12. The reason could be because the performance and statistics are rather 
difficult to distinguish given the close proximity in riblet size. Moreover, these two 
configurations also have different ratios of $h/s$ which means they experience different 
degree of flow asymmetry about the centerline.

	Interestingly, the profiles in Fig.~\ref{RMSPercentBaseline} exhibit a somewhat linear 
trend and a rather decent collapse when $10 < \lgp < 16$, i.e. Cases I, II and IV. This 
trend seems to imply that the rise in Reynolds stress goes hand in hand with the 
augmentation of cross-flow velocity fluctuations $v'$ and $w'$. In this regard, the 
changes of $v'$, $w'$ and $-\overline{u'v'}$ may be caused by similar mechanism. As 
$\lgp$ gets larger than 16 (or $\Sp > 30$), the profiles start to diverge and the Reynolds 
stress increases at a greater magnitude, followed by the spanwise and then the wall-normal 
velocity fluctuations. Such behavior indicates that there are additional mechanisms 
associated with the cross-flow that come into play as the groove size becomes larger. 
Moreover, the fact that the peak magnitude of $u'$ is reduced while those of $v'$ and $w'$ 
are augmented suggests that the increasing reach of riblets in the bulk flow induces a 
transfer of energy from the streamwise to the cross-flow components of velocity 
fluctuations. Previously, \citet{Choi93} (figure 18 in their paper) showed that V-groove 
riblets at $\Sp = 40$ produced a considerable spanwise variation of the peaks of turbulence 
kinetic energy $k$, in which the maximum occurred at the riblet tip. Hence, it may not be 
surprising that larger riblets exhibit additional flow features or mechanisms centering 
about the tip.
%
%=====================================================================================================================================================================================================%
%
\begin{figure*}[p]
	\begin{minipage}[c]{0.495\textwidth}
		\centering
		\subfigure[][]
		{
			% Insert 'grid, tics=10' in the square bracket to show the grid in 10% intervals.
			% trim option's parameter order: left bottom right top
			\begin{overpic}[trim = 0.5mm 50mm 10mm 51mm, clip, width = 0.95\textwidth]{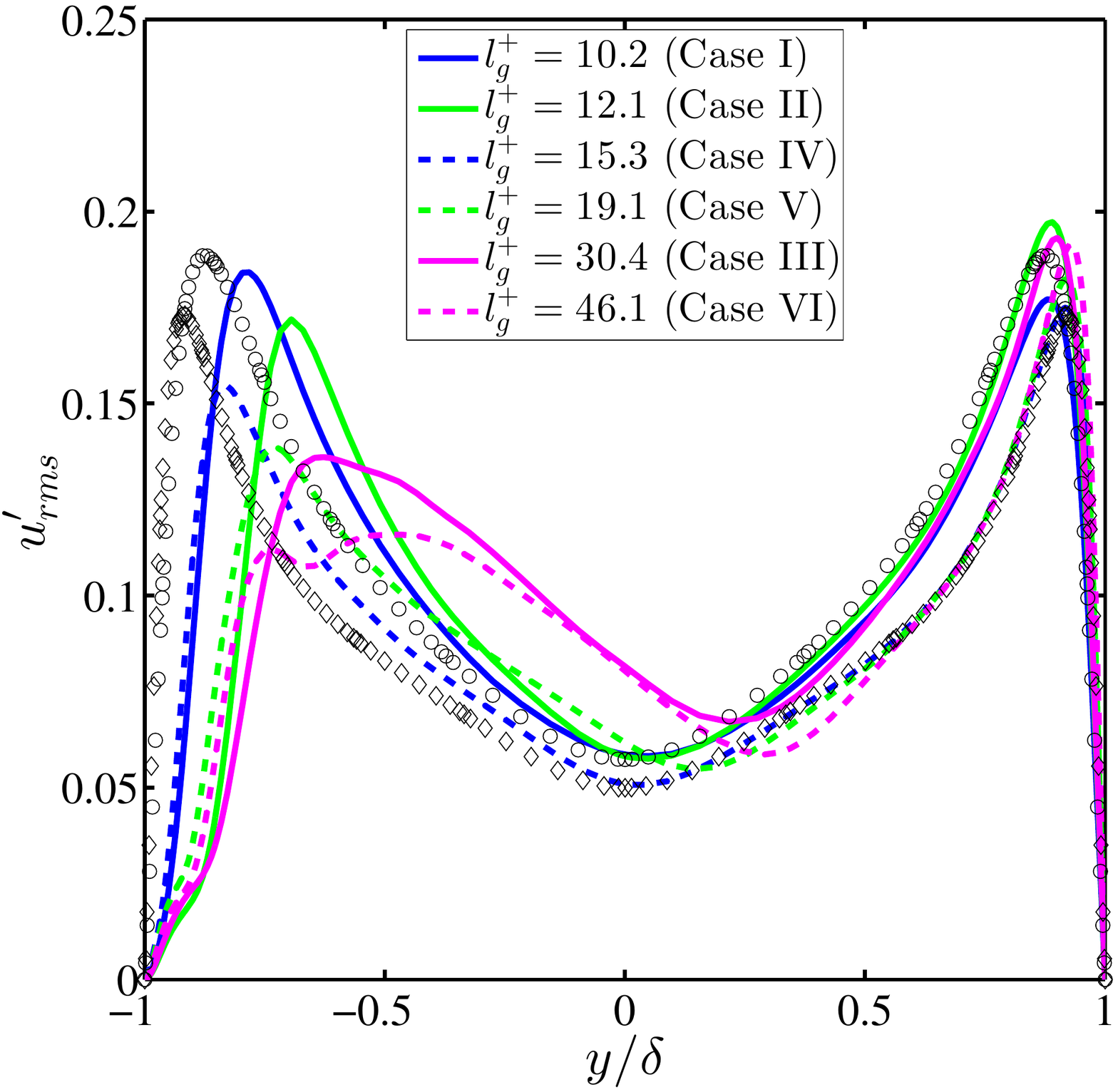}
			\put(9,15){\parbox{0.2\linewidth}{\linespread{1}\raggedleft\scriptsize $\leftarrow$ hump}}
			\put(21.75,37.5){\parbox{0.2\linewidth}{\linespread{1}\raggedright\scriptsize $\Bigg\uparrow$\\ hump}}
			\end{overpic}
			\label{UrmslgplusVal}
		}
	\end{minipage}
	\hfill
	\begin{minipage}[c]{0.495\textwidth}
		\centering
		\subfigure[][]
		{
			% trim option's parameter order: left bottom right top
			\includegraphics[trim = 0.5mm 50mm 10mm 51mm, clip, width = 0.95\textwidth]{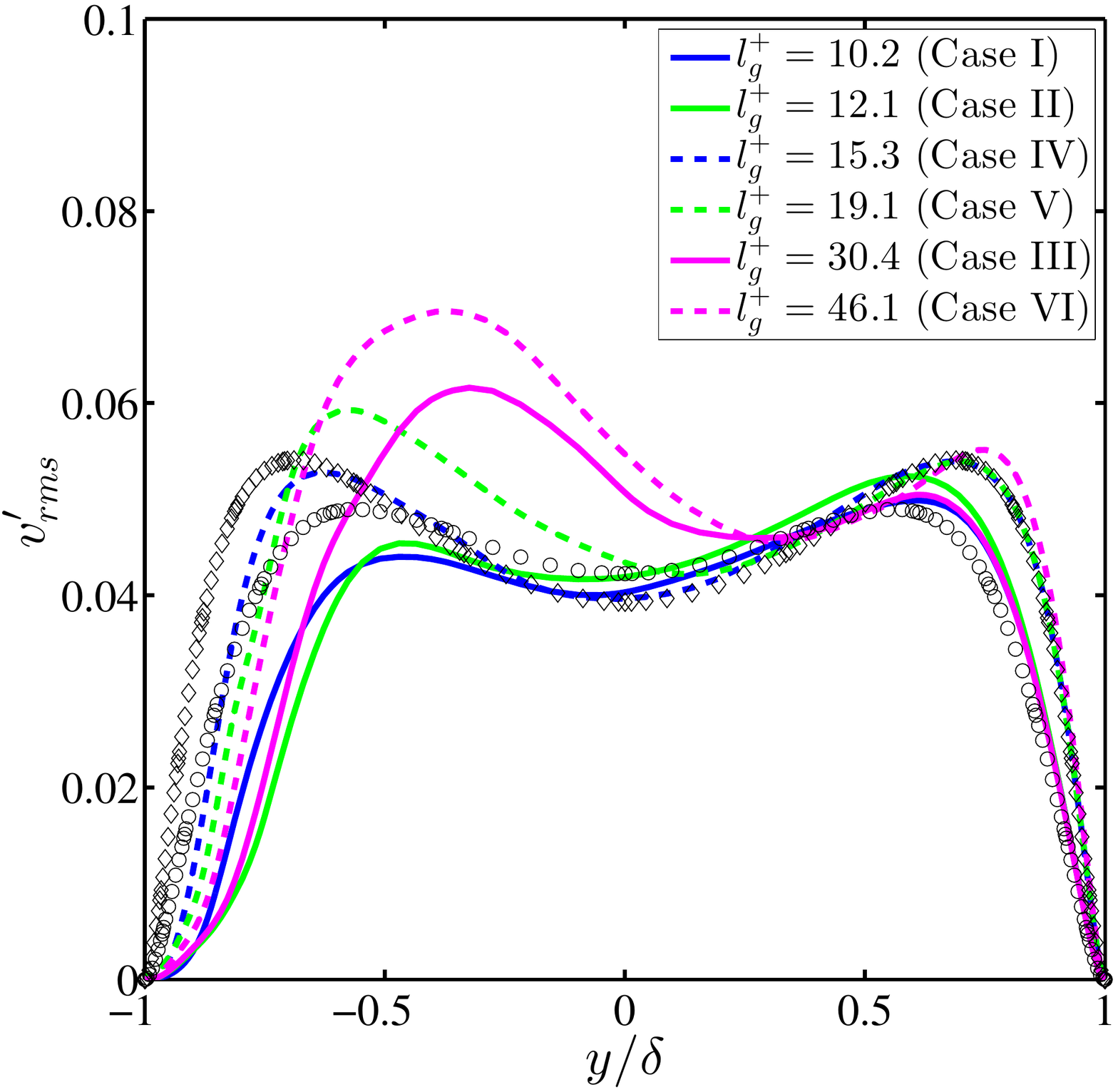}
			\label{VrmslgplusVal}
		}
	\end{minipage}
	
	\begin{minipage}[c]{0.495\textwidth}
		\centering
		\subfigure[][]
		{
			% trim option's parameter order: left bottom right top
			\begin{overpic}[trim = 0.5mm 50mm 10mm 51mm, clip, width = 0.95\textwidth]{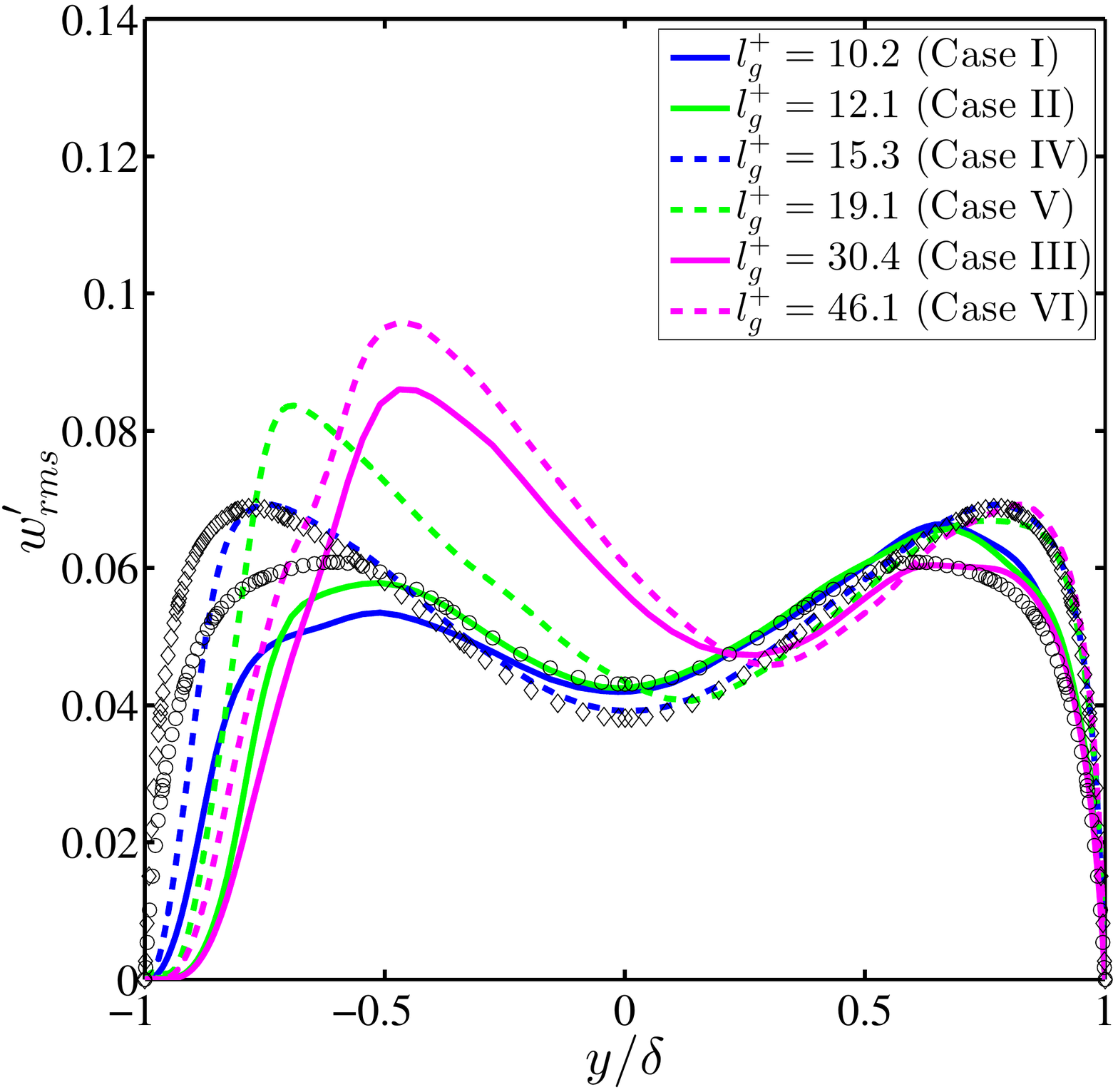}
				\put(23.5,25){\parbox{0.2\linewidth}{\linespread{1}\raggedright\scriptsize $\Bigg\uparrow$\\ slight hump}}
			\end{overpic}
			\label{WrmslgplusVal}
		}
	\end{minipage}
	\hfill
	\begin{minipage}[c]{0.495\textwidth}
		\centering
		\subfigure[][]
		{
			% Insert 'grid, tics=10' in the square bracket to show the grid in 10% intervals.
			% trim option's parameter order: left bottom right top
			\begin{overpic}[trim = 0.5mm 50mm 10mm 51mm, clip, width = 0.95\textwidth]{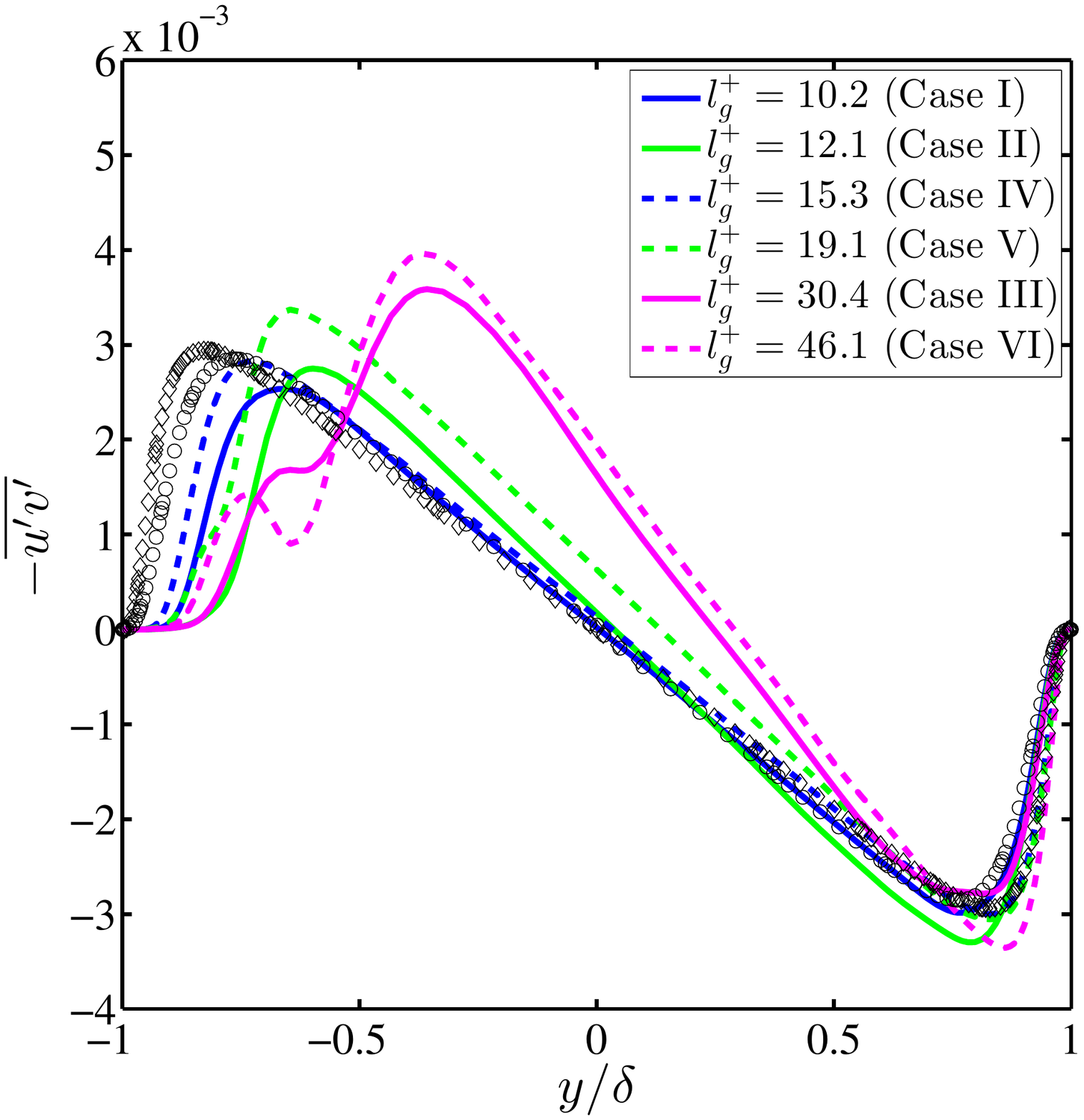}
			\put(24.5,47){\parbox{0.2\linewidth}{\linespread{1}\raggedright\scriptsize $\uparrow$\\ hump}}
			\end{overpic}
			\label{UVmslgplusVal}
		}
	\end{minipage}
	
	\caption
		{
			Profiles of turbulent fluctuations of all six V-groove riblet configurations across the channel above the riblet valley:
			\subref{UrmslgplusVal} $u'_{rms}$, 
			\subref{VrmslgplusVal} $v'_{rms}$, 
			\subref{WrmslgplusVal} $w'_{rms}$, and
			\subref{UVmslgplusVal} $-\overline{u'v'}$.
			Profiles drawn in solid and dashed lines represent configurations at $\Rem = 1842$ and 2800, respectively. Likewise, the markers ($\circ$) and ($\smalldiamond$) represent the baseline profiles acquired from the configuration with two smooth walls at $\Rem = 1842$ and 2800, respectively. Note that all profiles are scaled in outer units using the bulk velocity $\Um$.
			\label{ReyStresslgplusVal}
		}
\end{figure*}
%
%=====================================================================================================================================================================================================%
%

	One interesting observation in Fig.~\ref{VrmslgplusMid} is that the profiles of 
$v'_{rms}$ of all the six riblet configurations feature a small hump at a level around 
the riblet tips. This seems to be a sign of turbulence activities happening close to the 
tips since no discernible hump is found on the corresponding profiles above the riblet 
valley, see Fig.~\ref{VrmslgplusVal}. The appearance of hump may be attributed to the 
shedding of vortices due to the sharp tip of V-groove riblets as speculated 
by~\citet{Goldstein95}. Judging from the present results, the appearance of hump seems to 
be a ubiquitous feature in flows over V-groove riblets that is independent of the riblet 
size. The hump feature on the profile of $v'_{rms}$ could also be perceived as a 
testament to the ability of riblets in aligning the flow structures, or in localizing the 
overlying streamwise vortices above the riblet grooves as reported by~\citet{Garcia11}.

	Although no hump is formed on the profiles of $v'_{rms}$ above the riblet valley, 
humps are found on the profiles of $u'_{rms}$ depicted in Fig.~\ref{UrmslgplusVal}. 
However, the humps are formed in regions within the groove, and only for riblet 
configurations with $h/s = 1$, i.e. except Cases I and IV. Similar kind of hump feature 
on the profiles of $u'_{rms}$ above the riblet valley has been reported by~\citet{Chu93}, 
while~\citet{Choi93} and~\citet{Goldstein95} did not observe any formation of humps from 
their simulations of V-groove riblets. Nonetheless, \citet{Goldstein95} speculated that 
such feature is related to the shedding of vortices by the pointed riblet tip, but they 
concluded that the presence and importance of the hump remains an open question. A 
forthcoming study will show that such phenomenon occurring in the groove can be explained 
by looking at the flow topology on the riblet wall, and that its nature is dictated by 
the groove shape.

	Apart from the humps found on the profiles of $u'_{rms}$ above the riblet valley, 
Fig.~\ref{UVmslgplusVal} reveals that humps also appear on the profiles of 
$-\overline{u'v'}$ belonging to Cases V, III and VI with $\lgp \approx 19$, 30 and 46, 
respectively. The hump is formed more prominently as $\lgp$ increases. In particular, 
the hump in Case VI is formed deeper in the groove than in Case III despite they share 
the same physical riblet height $h = 0.4\delta$. A similar hump is also observed in 
Fig.~\ref{UrmslgplusVal} on the profile of $u'_{rms}$ for Case VI at roughly the same 
wall-normal distance away from the riblet valley. On the other hand, a peculiar 
feature in Fig.~\ref{UVmslgplusMid} is the small hump on the profile of Case VI at 
$y/\delta \approx -0.7$ which indicates a negative contribution to the Reynolds stress. 
Similarly, a small dip is also observed on the corresponding profile of Case III. In 
other words, when $\lgp$ gets sufficiently large, the flow field in some part the groove 
can become slightly dominated by $Q1$ and $Q3$ motions as opposed to the typical $Q2$ 
(\emph{ejection}) and $Q4$ (\emph{sweep}) motions based on the idea of \emph{quadrant 
analysis}~\citep{Wallace16}. The above discussions point to the fact that the flow field 
in the vicinity of V-groove riblets becomes increasingly complex when the their size 
gets sufficiently large, i.e. when $\lgp > 15$ based on the present set of configurations. 
This condition also corresponds with the divergent of profiles in 
Fig.~\ref{RMSPercentBaseline}.
%
%=====================================================================================================================================================================================================%
%
\begin{figure*}[p]
	\begin{minipage}[c]{0.495\textwidth}
		\centering
		\subfigure[][]
		{
			% trim option's parameter order: left bottom right top
			\includegraphics[trim = 0.5mm 50mm 10mm 51mm, clip, width = 0.95\textwidth]{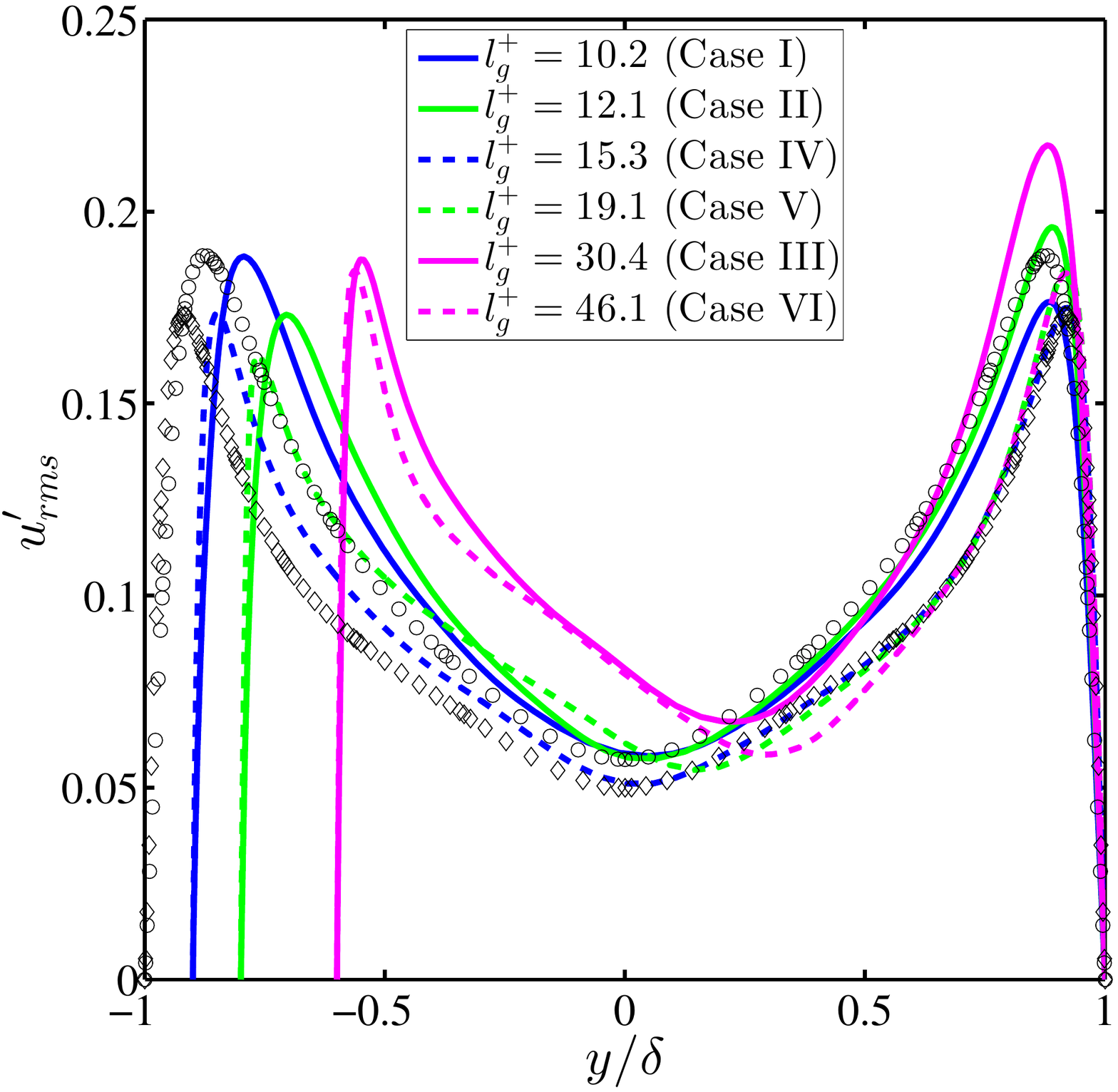}
			\label{UrmslgplusTip}
		}
	\end{minipage}
	\hfill
	\begin{minipage}[c]{0.495\textwidth}
		\centering
		\subfigure[][]
		{
			% trim option's parameter order: left bottom right top
			\includegraphics[trim = 0.5mm 50mm 10mm 51mm, clip, width = 0.95\textwidth]{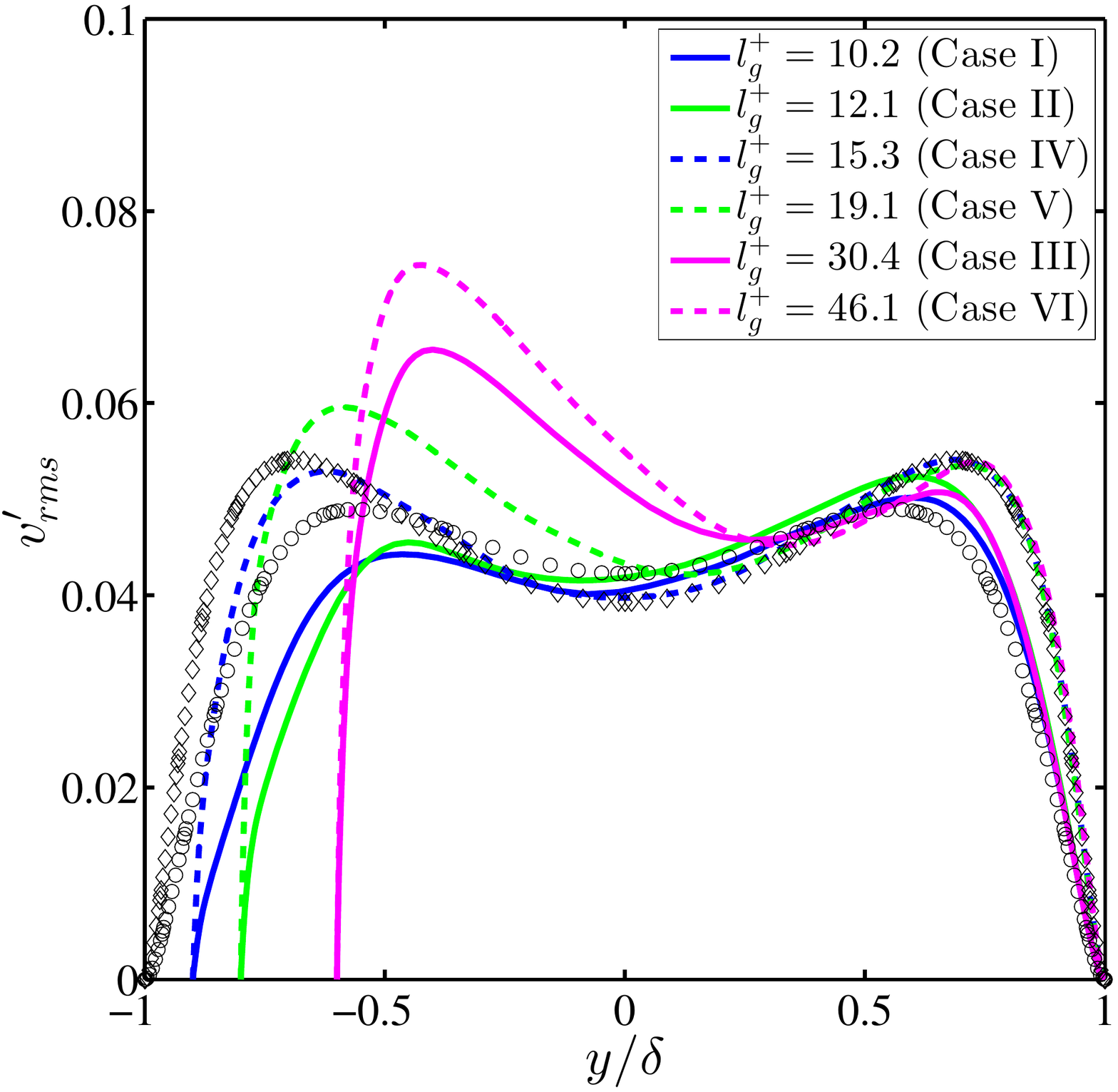}
			\label{VrmslgplusTip}
		}
	\end{minipage}
	
	\begin{minipage}[c]{0.495\textwidth}
		\centering
		\subfigure[][]
		{
			% Insert 'grid, tics=10' in the square bracket to show the grid in 10% intervals.
			% trim option's parameter order: left bottom right top
			\begin{overpic}[trim = 0.5mm 50mm 10mm 51mm, clip, width = 0.95\textwidth]{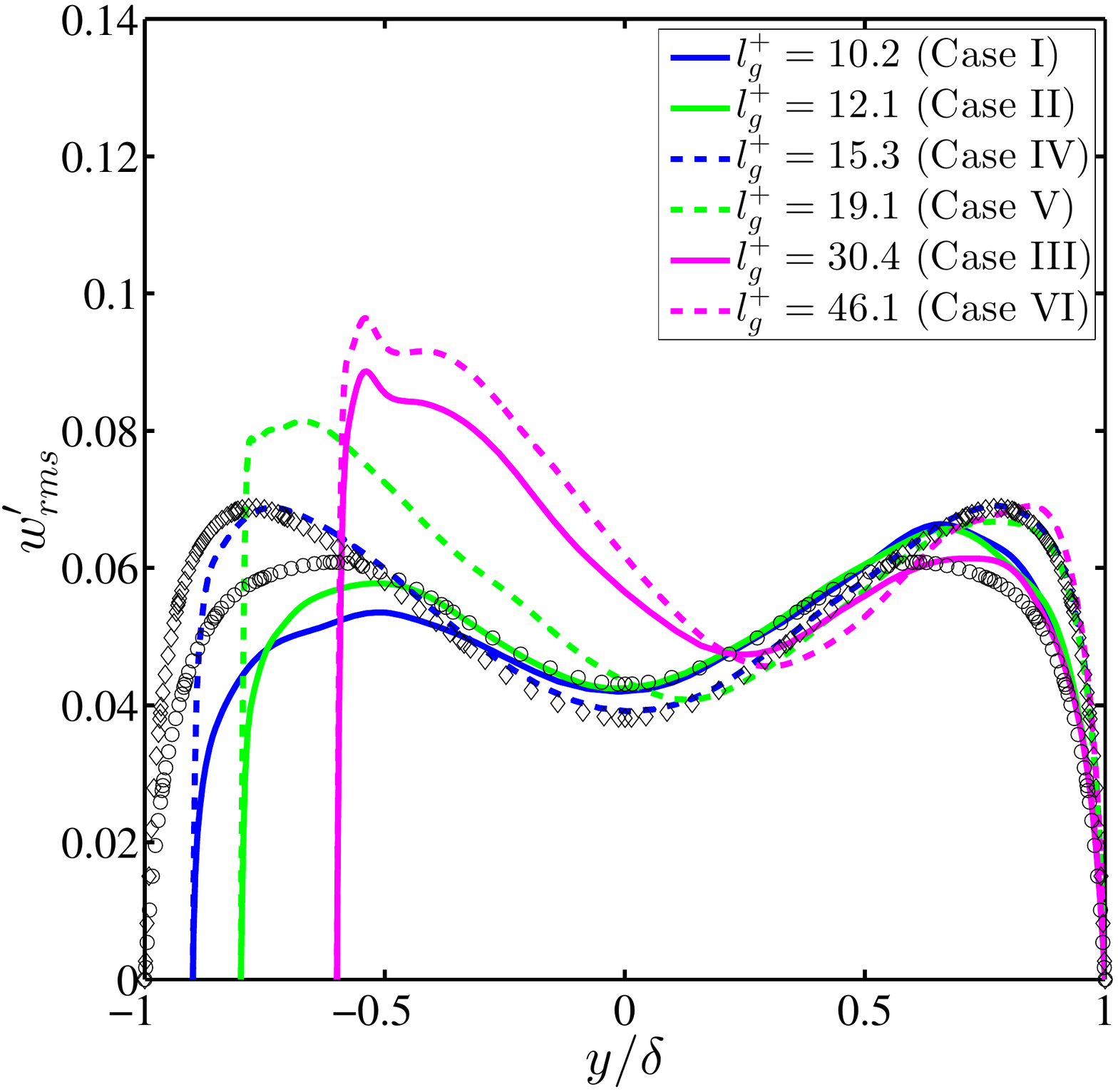}
			\put(21.2,70){\parbox{0.2\linewidth}{\linespread{1}\centering\scriptsize hump\\ $\downarrow$}}
			\end{overpic}
			\label{WrmslgplusTip}
		}
	\end{minipage}
	\hfill
	\begin{minipage}[c]{0.495\textwidth}
		\centering
		\subfigure[][]
		{
			% trim option's parameter order: left bottom right top
			\includegraphics[trim = 0.5mm 50mm 10mm 51mm, clip, width = 0.95\textwidth]{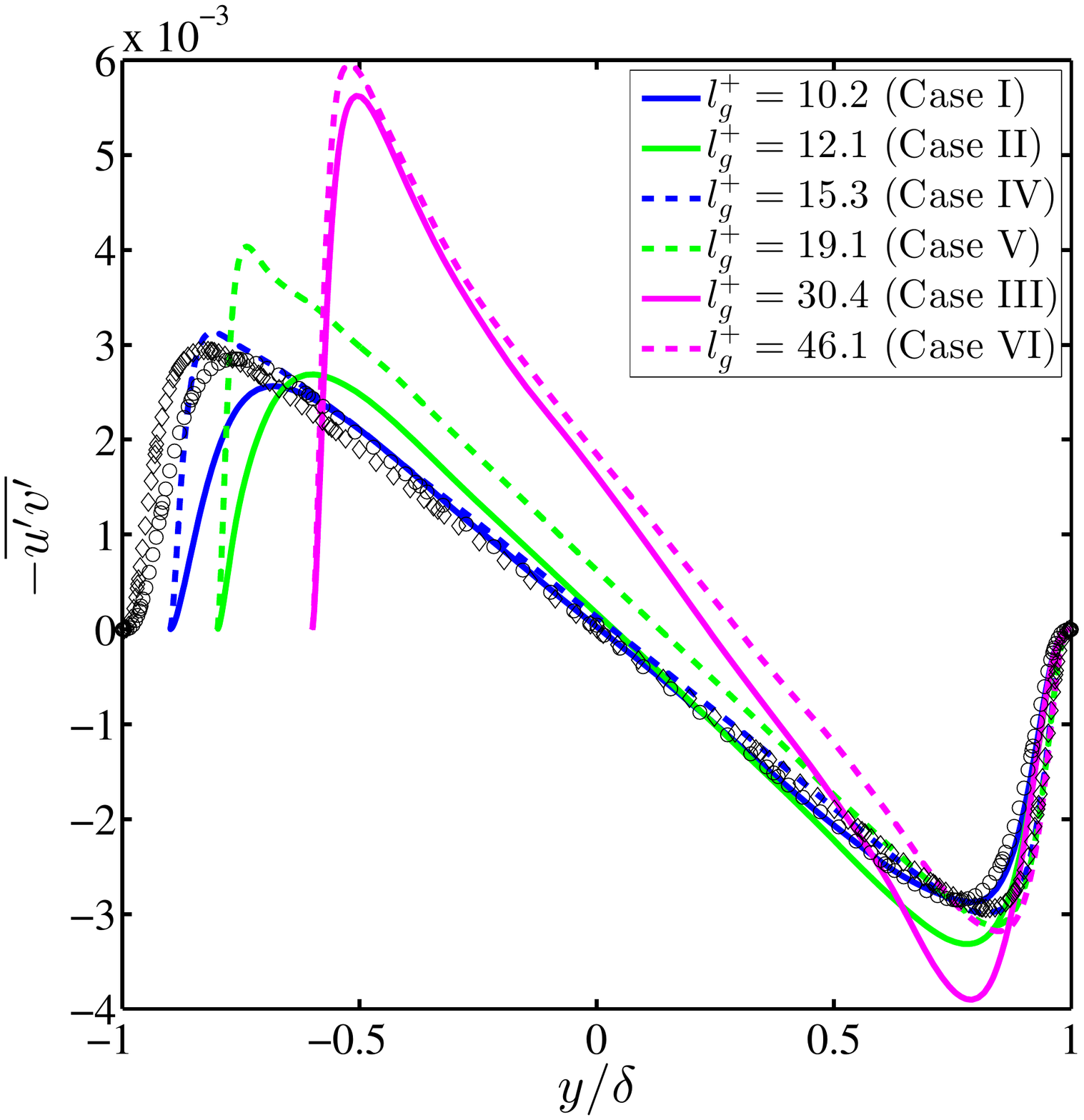}
			\label{UVmslgplusTip}
		}
	\end{minipage}
	
	\caption
		{
			Profiles of turbulent fluctuations of all six V-groove riblet configurations across the channel above the riblet tip:
			\subref{UrmslgplusTip} $u'_{rms}$, 
			\subref{VrmslgplusTip} $v'_{rms}$, 
			\subref{WrmslgplusTip} $w'_{rms}$, and
			\subref{UVmslgplusTip} $-\overline{u'v'}$.
			Profiles drawn in solid and dashed lines represent configurations at $\Rem = 1842$ and 2800, respectively. Likewise, the markers ($\circ$) and ($\smalldiamond$) represent the baseline profiles acquired from the configuration with two smooth walls at $\Rem = 1842$ and 2800, respectively. Note that all profiles are scaled in outer units using the bulk velocity $\Um$.
			\label{ReyStresslgplusTip}
		}
\end{figure*}
%
%=====================================================================================================================================================================================================%
%

	Similar arguments can be derived when analyzing the corresponding profiles of turbulence 
fluctuations above the riblet tip provided in Fig.~\ref{ReyStresslgplusTip}, except for a 
few differences in Cases III and VI that possess the two largest values of $\lgp$. 
Firstly, it is observed that the peak magnitude of $u'$ near the riblet wall is comparable 
to the baseline configuration. They are found to augment the peak magnitude of $u'$ when 
their size becomes too large, i.e. at $\lgp \approx 30$ and $46$. Although the attenuation 
of the peak magnitude of $u'$ still scales with $\lgp$, the V-groove riblets are less 
effective in mitigating the peak magnitude of $u'$ directly above the riblet tip. 
Secondly, humps are formed on the profile of $w'_{rms}$ near the riblet tip at $y/\delta 
\approx -0.6$. Both of these findings suggest that $u'$ and $w'$ are intensified at the 
tip when the size of riblets is large. On the other hand, there are no humps formed on the 
profiles of $v'$ and $-\overline{u'v'}$ as opposed to those plotted above the riblet mid-
point and valley. However, the peak magnitude of $-\overline{u'v'}$ is noticeably higher, 
especially in those configurations with $\lgp > 15$.

\subsection{Variation of flow features with the size of riblets}
\label{sec:ScalingFlowFieldSize}

	Previously, \citet{Choi93}, \citet{Lee01}, and~\citet{El-Samni07} have concluded that 
the skin friction drag is augmented when near-wall streamwise vortices penetrate into the 
groove if $\Sp > 30$, and vice-versa. Based on the present set of riblet configurations, 
both Cases III and VI possess $\Sp > 30$, thereby exceeds the condition specified 
by~\citet{Choi93}. Figures~\ref{u-Streamlines} and~\ref{Omegayz} compare, respectively, 
the instantaneous streamlines pattern and streamwise vorticity $\omega_x$ contours of the 
six configurations on a cross-sectional plane at $x/\delta = 2.5$. These figures show that 
the behavior of near-wall vortical structures somewhat scales with $\lgp$. Note that the 
visualizations for the riblet walls have been rescaled to visually the same spacing in 
both sets of figures. It is worthwhile to mention that the vortical regions identified by 
the streamlines and vorticity contours may not match entirely because the near-wall 
streamwise vortices are continuously shifting their positions with time~\citep{Perry82A}.
%
%=====================================================================================================================================================================================================%
%
%\newlength{\yzwidth}
\def\yzwidth{0.9\textwidth}
\begin{figure*}[t]
	\begin{minipage}[c]{0.495\textwidth}
		\raggedleft
		\subfigure[][]
		{
			% trim option's parameter order: left bottom right top
			\includegraphics[trim = 34mm 60mm 34mm 63mm, clip, width = \yzwidth]{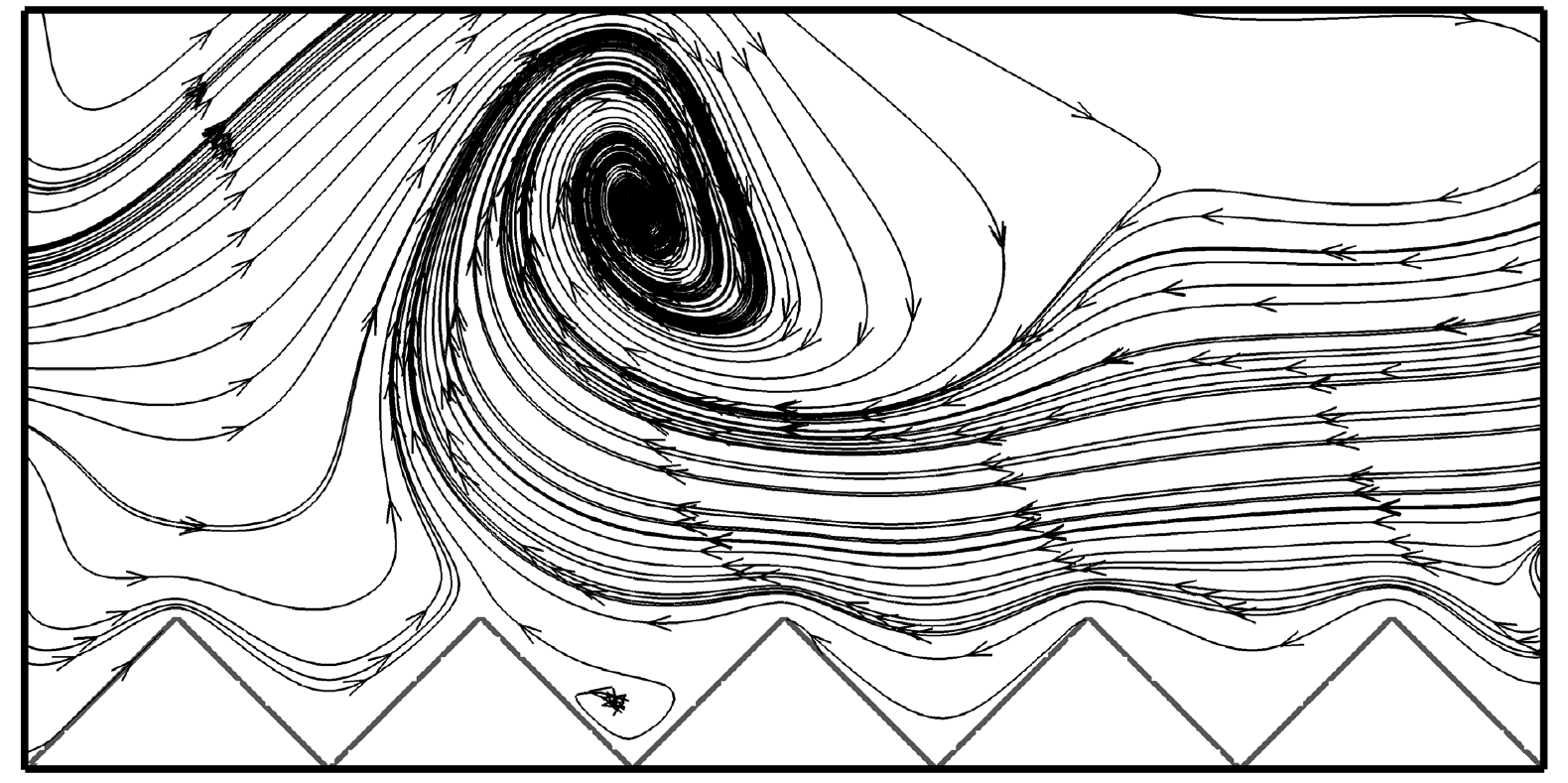}
			\label{Uyz1842-s0p2-h0p1}
		}
	\end{minipage}
	\hfill
	\begin{minipage}[c]{0.495\textwidth}
		\raggedright
		\subfigure[][]
		{
			% trim option's parameter order: left bottom right top
			\includegraphics[trim = 34mm 60mm 34mm 63mm, clip, width = \yzwidth]{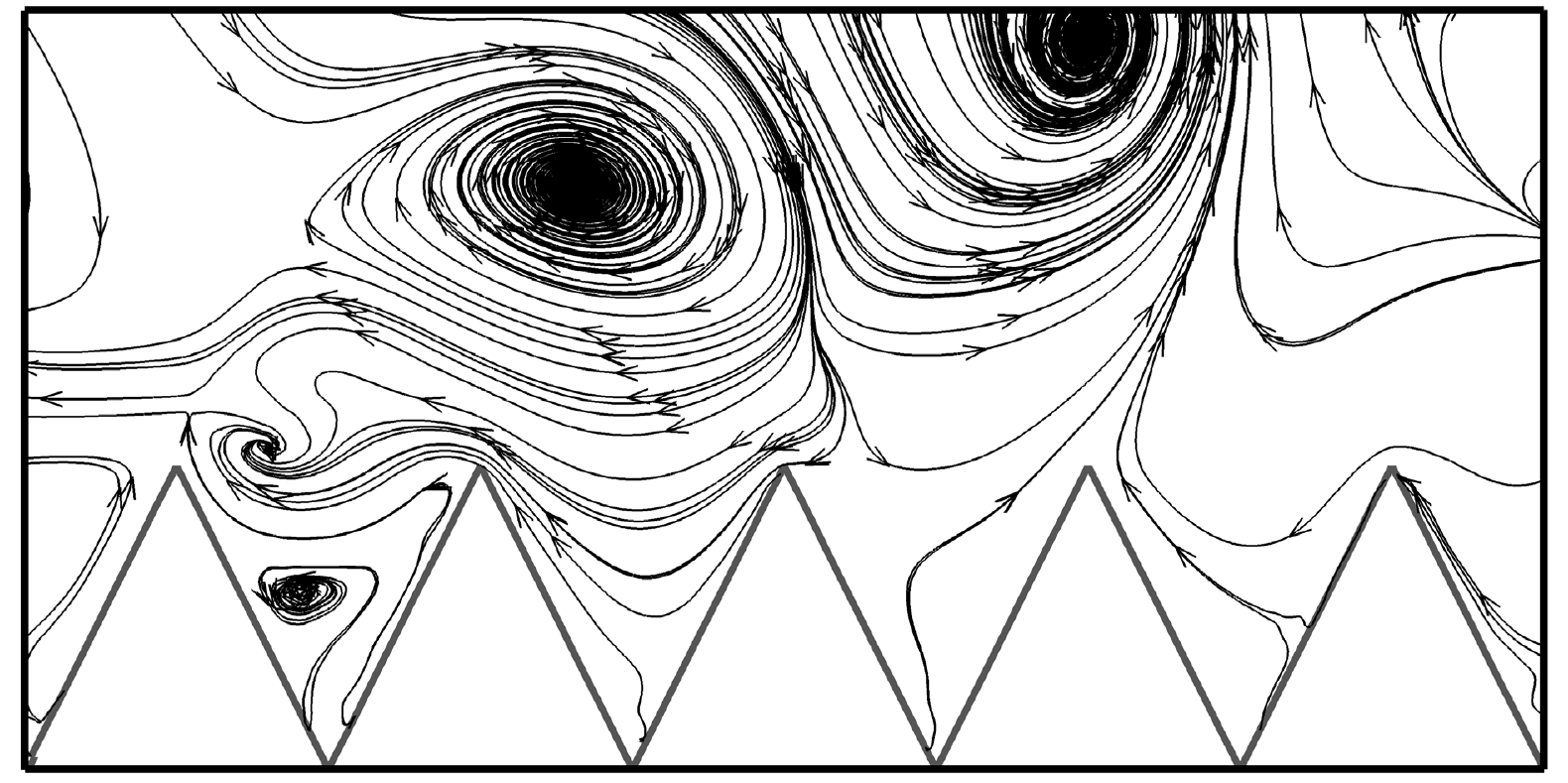}
			\label{Uyz1842-s0p2-h0p2}
		}
	\end{minipage}
	
	\begin{minipage}[c]{0.495\textwidth}
		\raggedleft
		\subfigure[][]
		{
			% trim option's parameter order: left bottom right top
			\includegraphics[trim = 34mm 60mm 34mm 63mm, clip, width = \yzwidth]{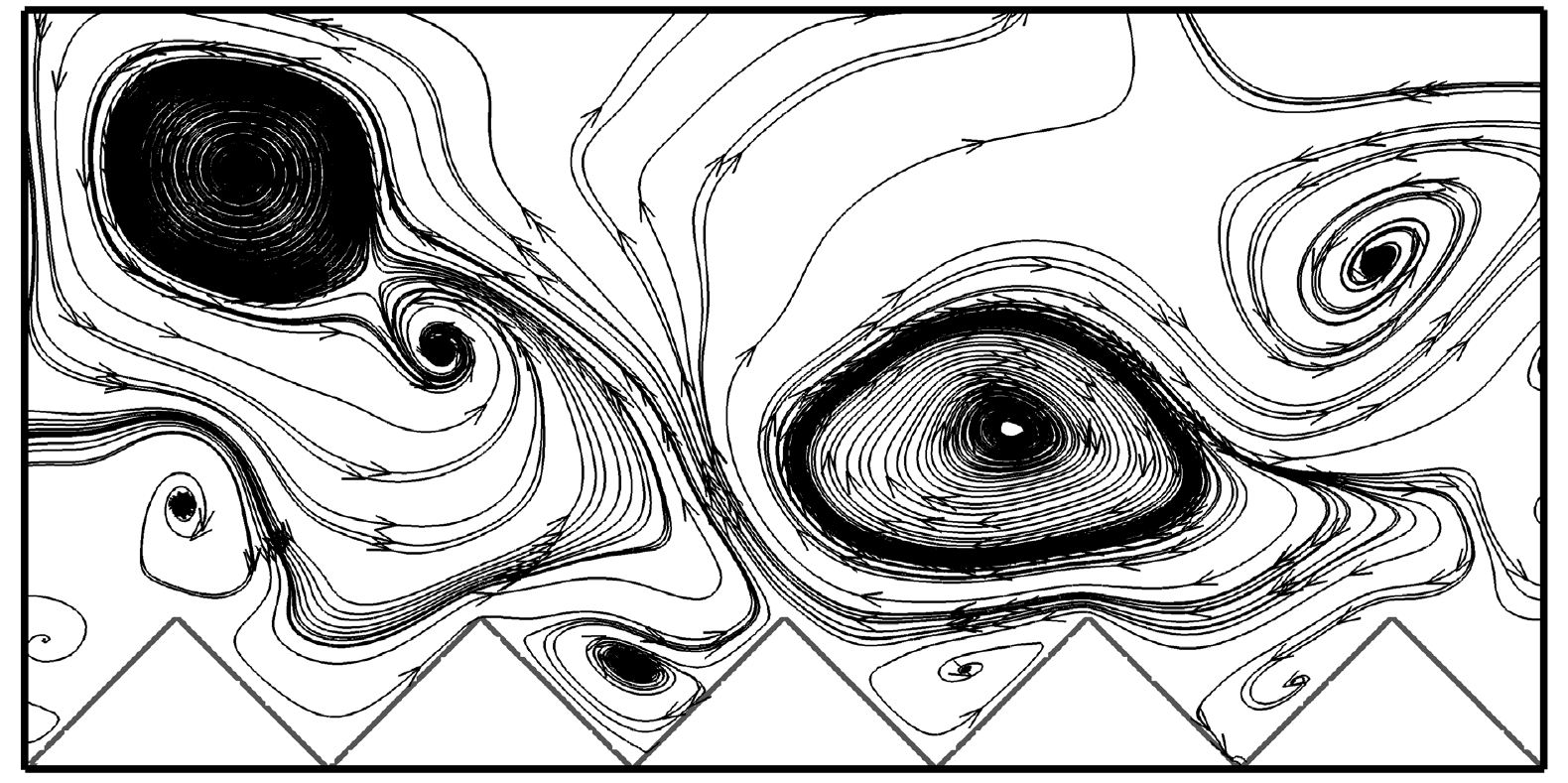}
			\label{Uyz2800-s0p2-h0p1}
		}
	\end{minipage}
	\hfill
	\begin{minipage}[c]{0.495\textwidth}
		\raggedright
		\subfigure[][]
		{
			% trim option's parameter order: left bottom right top
			\includegraphics[trim = 34mm 60mm 34mm 63mm, clip, width = \yzwidth]{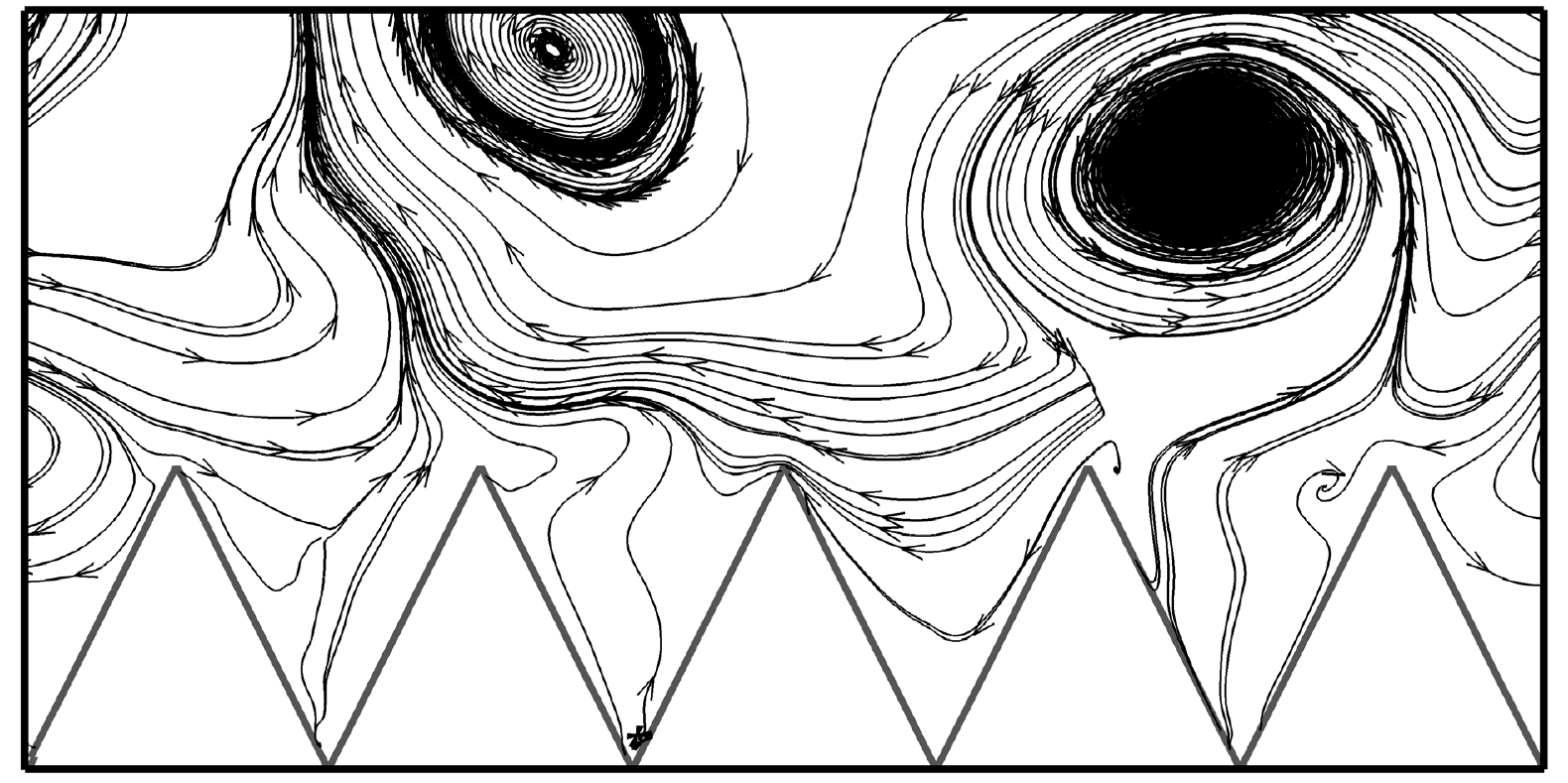}
			\label{Uyz2800-s0p2-h0p2}
		}
	\end{minipage}
	
	\begin{minipage}[c]{0.495\textwidth}
		\raggedleft
		\subfigure[][]
		{
			% trim option's parameter order: left bottom right top
			\includegraphics[trim = 34mm 60mm 34mm 63mm, clip, width = \yzwidth]{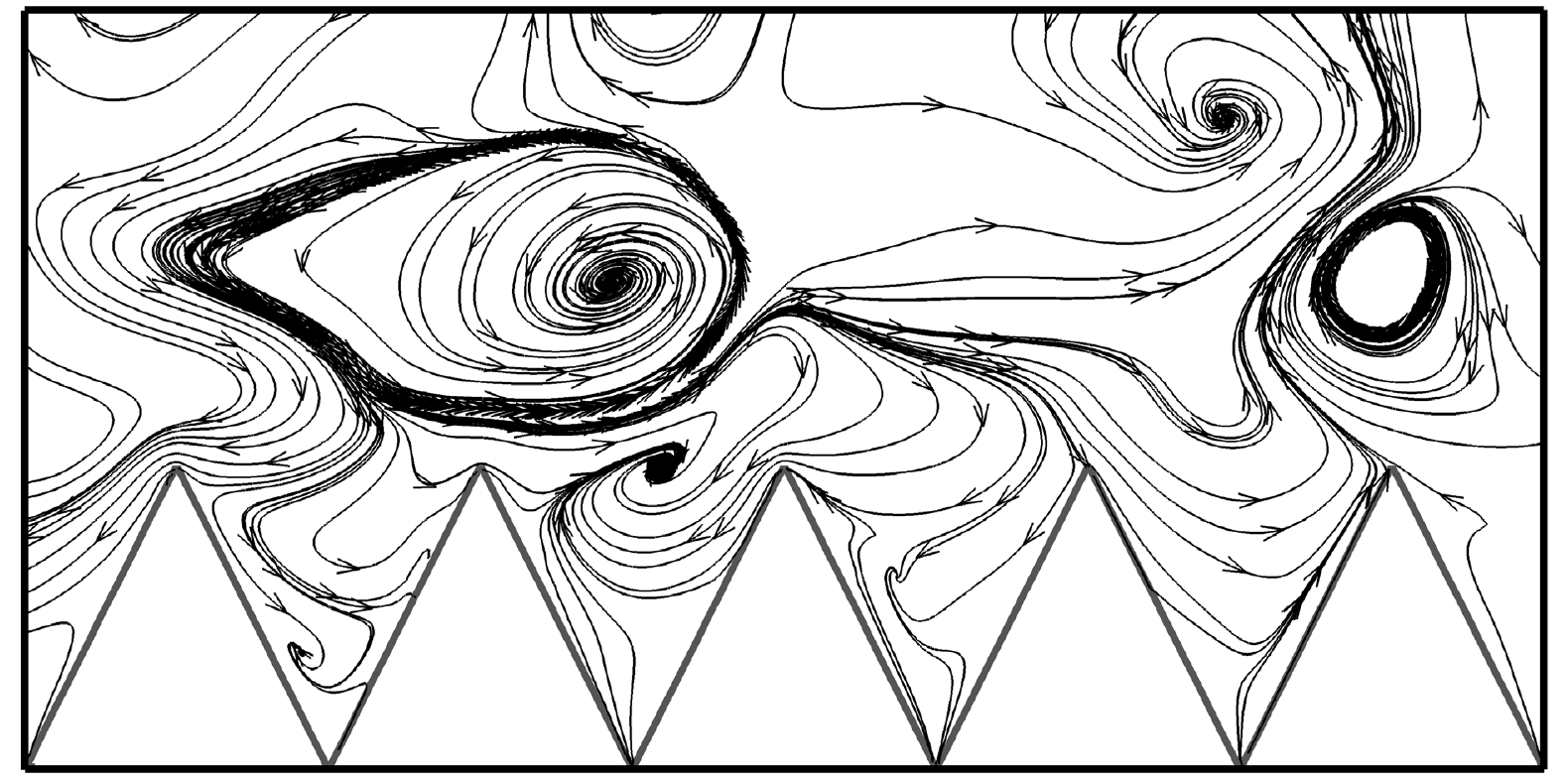}
			\label{Uyz1842-s0p4-h0p4}
		}
	\end{minipage}
	\hfill
	\begin{minipage}[c]{0.495\textwidth}
		\raggedright
		\subfigure[][]
		{
			% trim option's parameter order: left bottom right top
			\includegraphics[trim = 34mm 60mm 34mm 63mm, clip, width = \yzwidth]{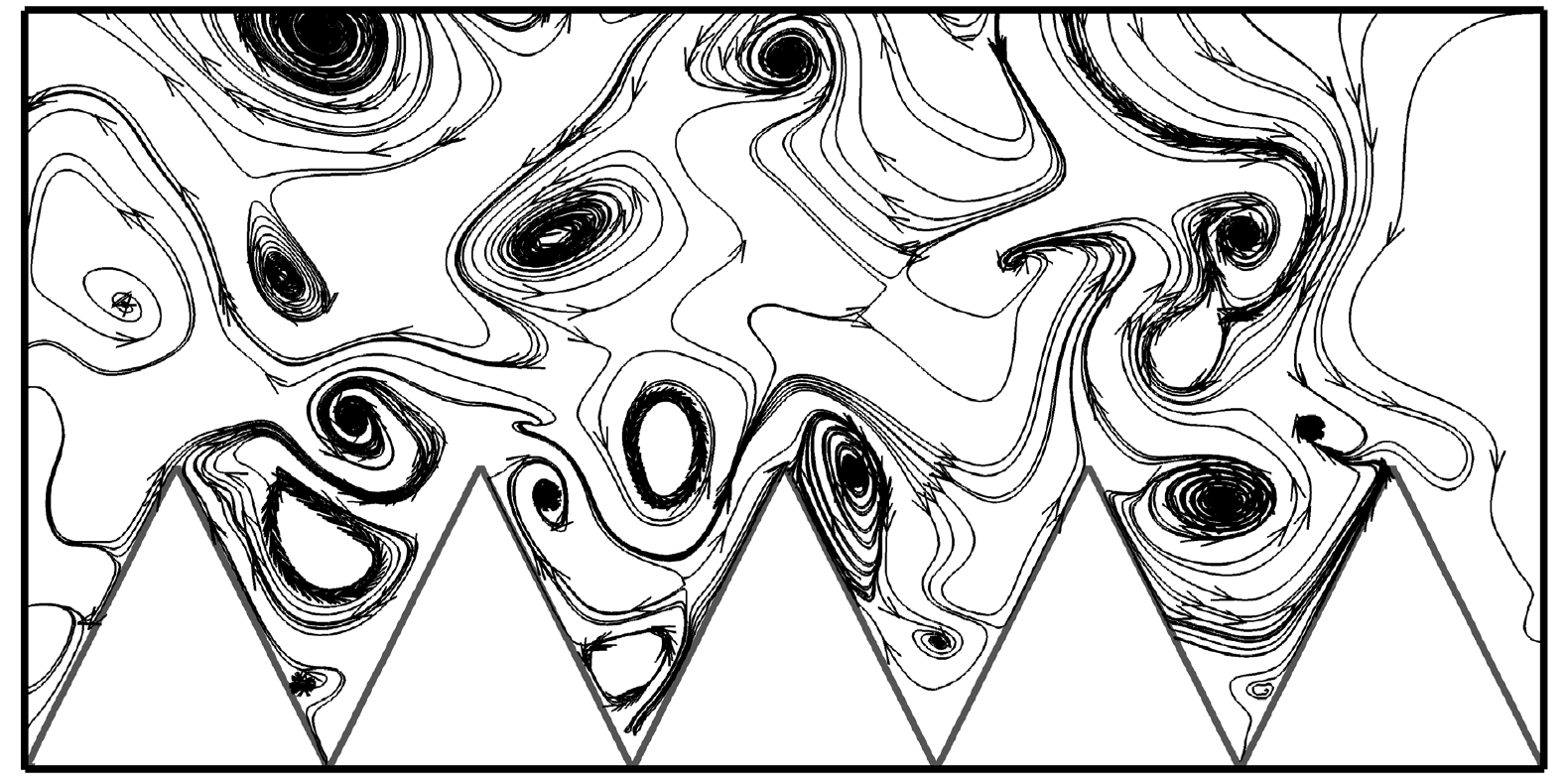}
			\label{Uyz2800-s0p4-h0p4}
		}
	\end{minipage}
	
	\caption
		{
			Comparison of instantaneous near-wall streamlines pattern on a $yz$-plane at $x/\delta = 2.5$ of the six riblet configurations near the riblet wall:
			\subref{Uyz1842-s0p2-h0p1} $\lgp \approx 10$ (Case I),
			\subref{Uyz1842-s0p2-h0p2} $\lgp \approx 12$ (Case II),
			\subref{Uyz2800-s0p2-h0p1} $\lgp \approx 15$ (Case IV),
			\subref{Uyz2800-s0p2-h0p2} $\lgp \approx 19$ (Case V),
			\subref{Uyz1842-s0p4-h0p4} $\lgp \approx 30$ (Case III), and 
			\subref{Uyz2800-s0p4-h0p4} $\lgp \approx 46$ (Case VI).
			Note that the figures are rescaled to visually the same riblet spacing $s/\delta$ for comparison purpose.
			\label{u-Streamlines}
		}
\end{figure*}	
%
%=====================================================================================================================================================================================================%
%
\begin{figure*}[t]
	\begin{minipage}[c]{0.495\textwidth}
		\raggedleft
		\subfigure[][]
		{
			% trim option's parameter order: left bottom right top
			\includegraphics[trim = 34mm 60mm 34mm 63mm, clip, width = \yzwidth]{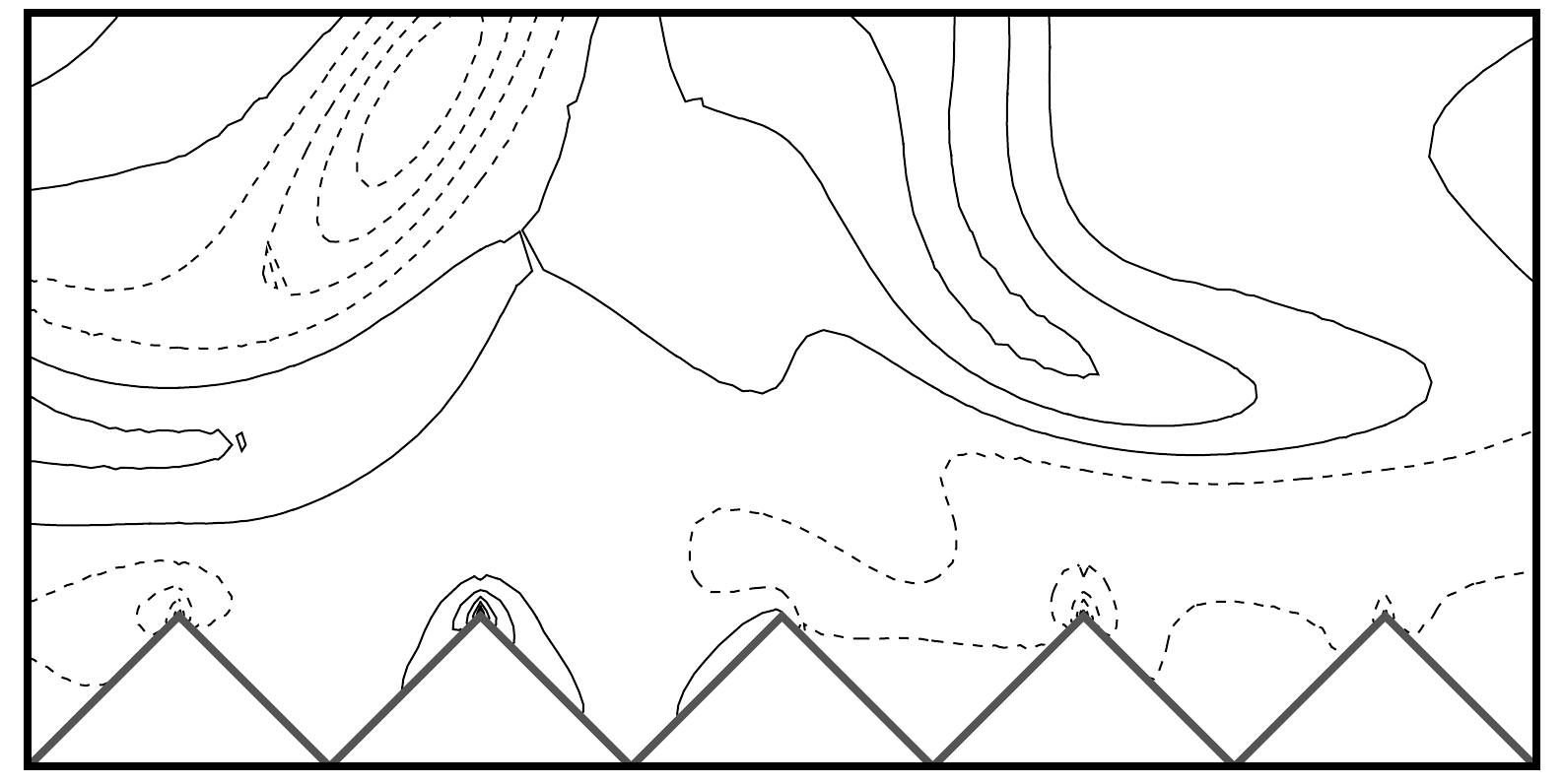}
			\label{Omegayz1842-s0p2-h0p1}
		}
	\end{minipage}
	\hfill
	\begin{minipage}[c]{0.495\textwidth}
		\raggedright
		\subfigure[][]
		{
			% trim option's parameter order: left bottom right top
			\includegraphics[trim = 34mm 60mm 34mm 63mm, clip, width = \yzwidth]{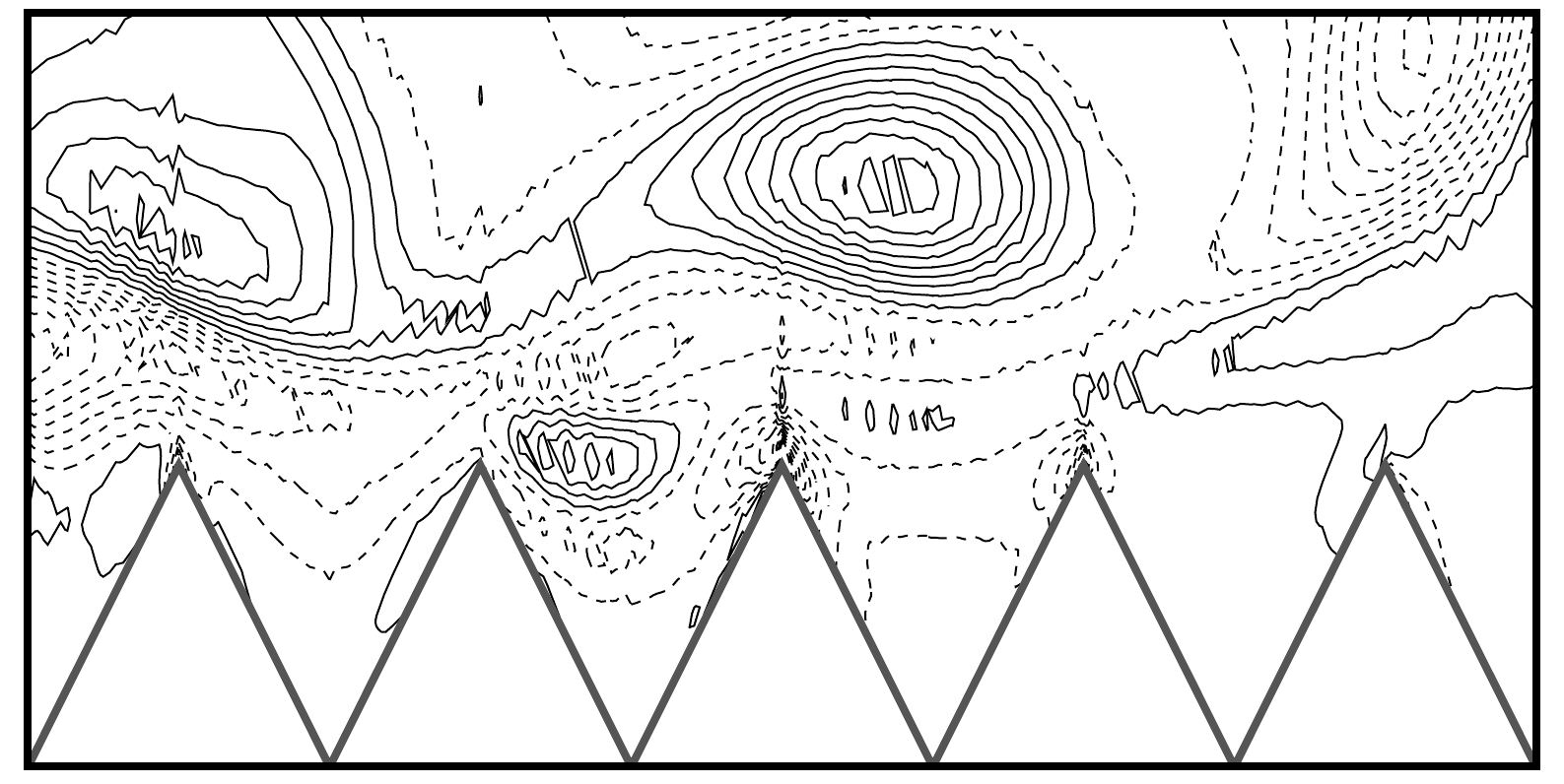}
			\label{Omegayz1842-s0p2-h0p2}
		}
	\end{minipage}
	
	\begin{minipage}[c]{0.495\textwidth}
		\raggedleft
		\subfigure[][]
		{
			% trim option's parameter order: left bottom right top
			\includegraphics[trim = 34mm 60mm 34mm 63mm, clip, width = \yzwidth]{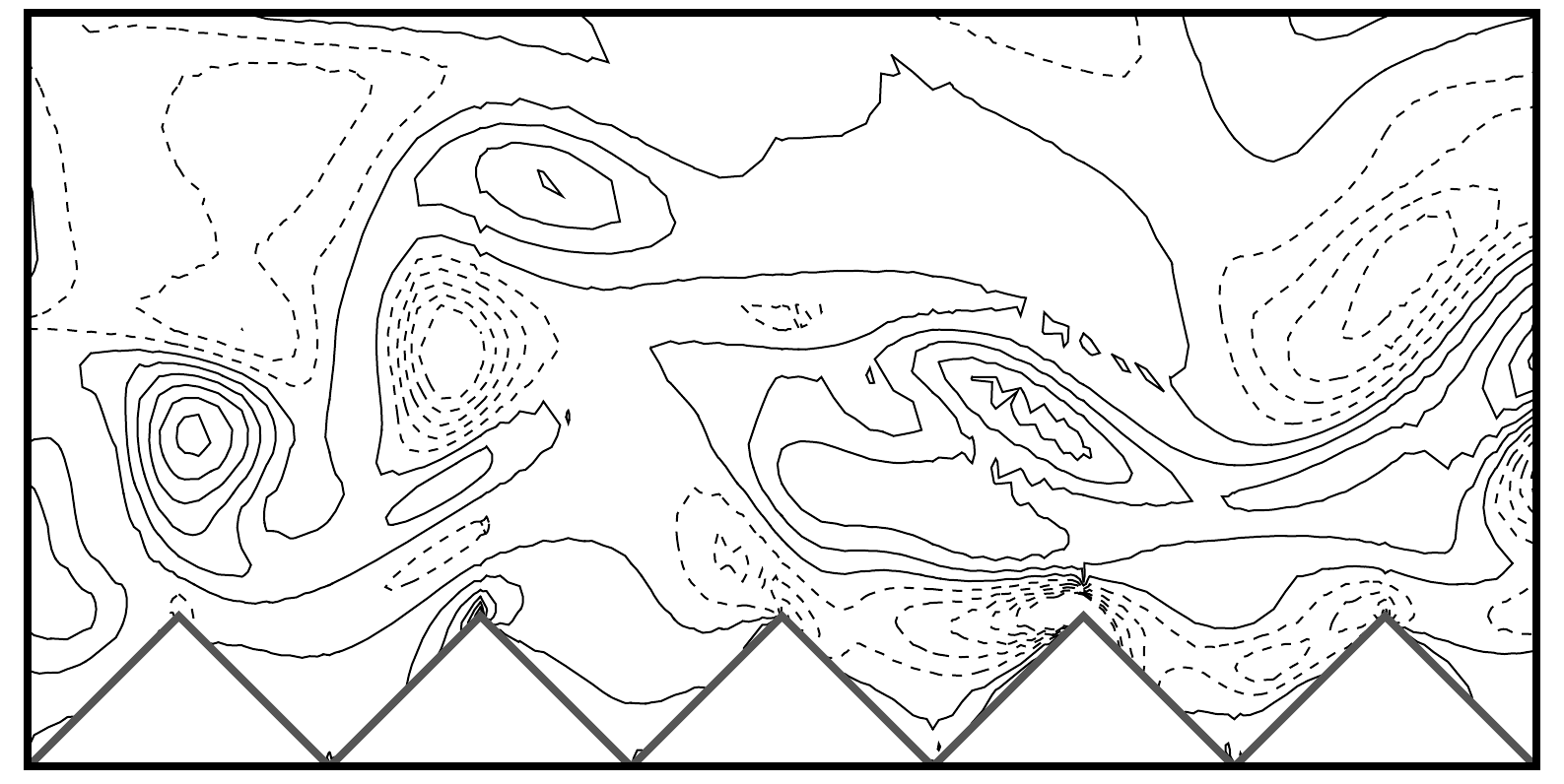}
			\label{Omegayz2800-s0p2-h0p1}
		}
	\end{minipage}
	\hfill
	\begin{minipage}[c]{0.495\textwidth}
		\raggedright
		\subfigure[][]
		{
			% trim option's parameter order: left bottom right top
			\includegraphics[trim = 34mm 60mm 34mm 63mm, clip, width = \yzwidth]{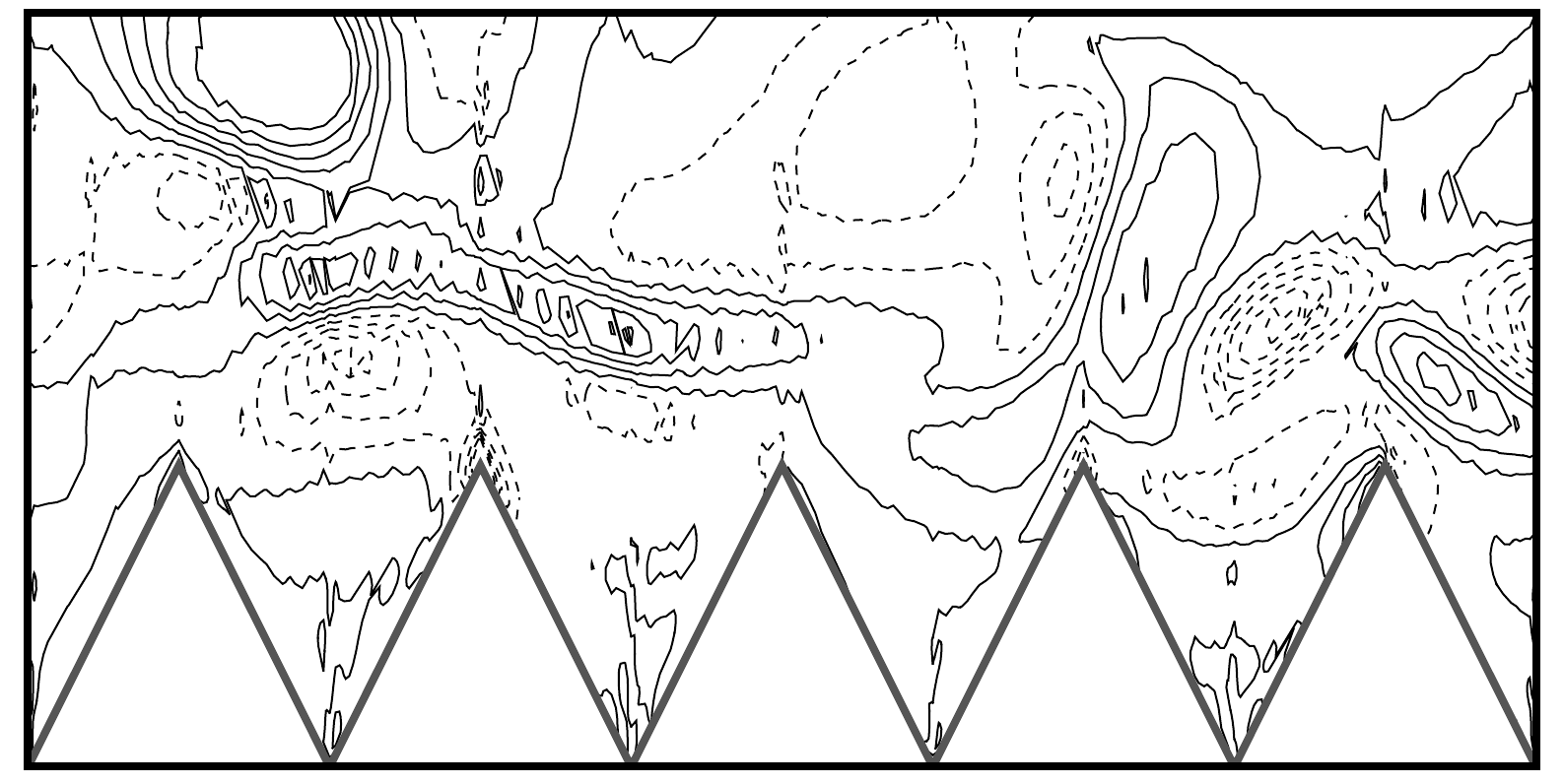}
			\label{Omegayz2800-s0p2-h0p2}
		}
	\end{minipage}
	
	\begin{minipage}[c]{0.495\textwidth}
		\raggedleft
		\subfigure[][]
		{
			% trim option's parameter order: left bottom right top
			\includegraphics[trim = 34mm 60mm 34mm 63mm, clip, width = \yzwidth]{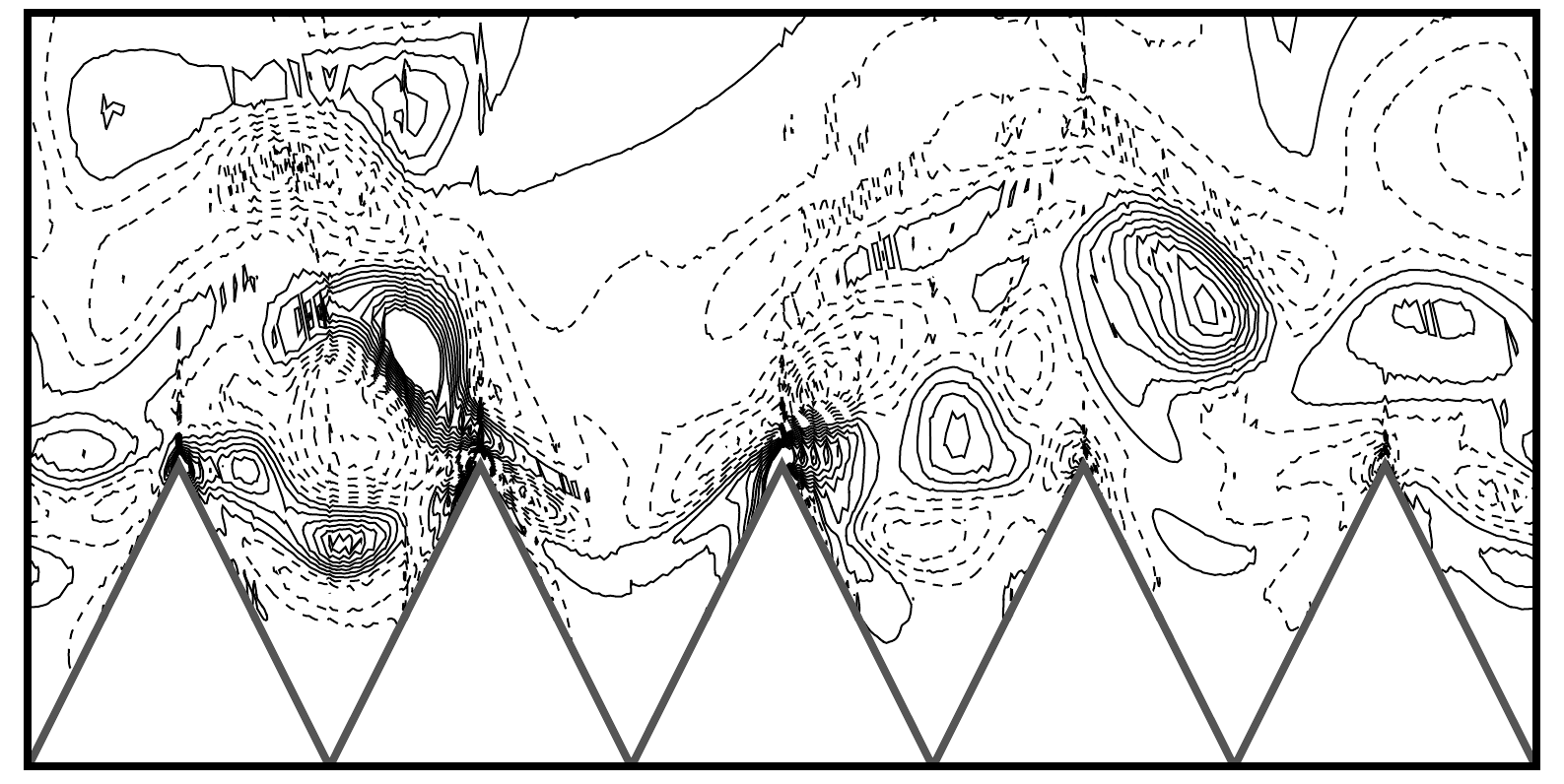}
			\label{Omegayz1842-s0p4-h0p4}
		}
	\end{minipage}
	\hfill
	\begin{minipage}[c]{0.495\textwidth}
		\raggedright
		\subfigure[][]
		{
			% trim option's parameter order: left bottom right top
			\includegraphics[trim = 34mm 60mm 34mm 63mm, clip, width = \yzwidth]{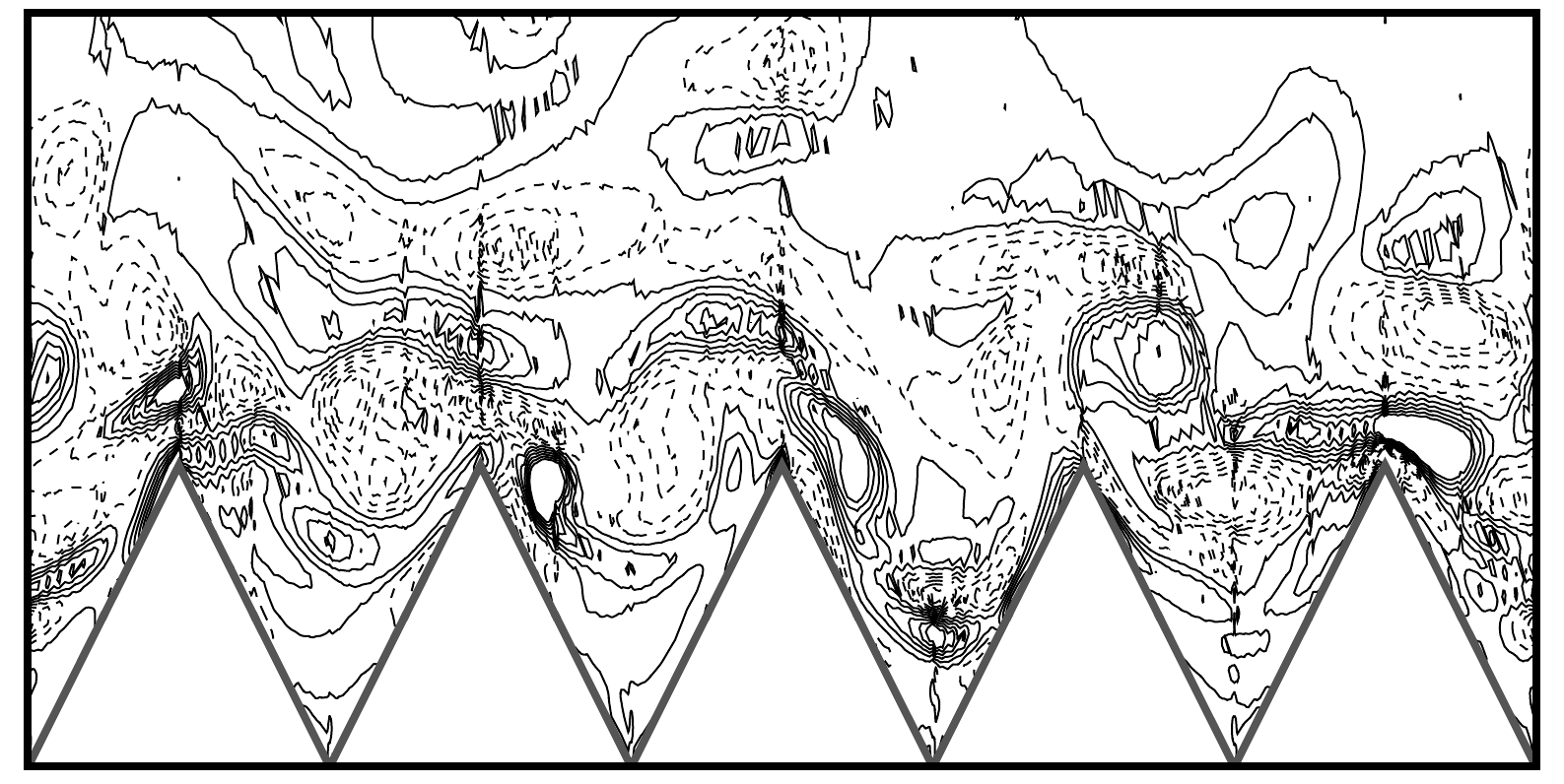}
			\label{Omegayz2800-s0p4-h0p4}
		}
	\end{minipage}
	
	\caption
		{
			Comparison of instantaneous near-wall streamwise vorticity $\omega_x$ pattern on a $yz$-plane at $x/\delta = 2.5$, between the top (smooth) wall and the bottom (riblet) wall:
			\subref{Omegayz1842-s0p2-h0p1} $\lgp \approx 10$ (Case I),
			\subref{Omegayz1842-s0p2-h0p2} $\lgp \approx 12$ (Case II),
			\subref{Omegayz2800-s0p2-h0p1} $\lgp \approx 15$ (Case IV),
			\subref{Omegayz2800-s0p2-h0p2} $\lgp \approx 19$ (Case V),
			\subref{Omegayz1842-s0p4-h0p4} $\lgp \approx 30$ (Case III), and 
			\subref{Omegayz2800-s0p4-h0p4} $\lgp \approx 46$ (Case VI).
			Note that the flow is into the plane ($+x$ direction), and dashed lines indicate negative vorticity (counterclockwise rotation). All the figures have the same number of contour levels evenly distributed within the same range, and they are rescaled to visually the same riblet spacing $s/\delta$ for comparison purpose.
			\label{Omegayz}
		}
\end{figure*}	
%
%=====================================================================================================================================================================================================%
%

	In Cases I and II, the average size vortices have a tendency to stay above the riblets. 
Figure~\ref{Omegayz} reveals that the bulk of the groove is free from vortical fluid 
motions. Thus, a small reduction in the peak magnitudes of cross-flow velocity 
fluctuations $v'$ and $w'$ is achieved in these two cases. The analysis by~\citet{Garcia11} 
also arrived at the same conclusion that the lodging of vortices inside the grooves does 
not happen in the neighborhood of performance optimum, i.e. around $\lgp \approx11$. As 
$\lgp$ gets larger, there are increasingly more vortical structures that reside in the 
groove. As seen in Fig.~\ref{Omegayz}, there are more distinct vortical regions in the 
groove when $\lgp > 30$, and these regions penetrate deeper into the groove as $\lgp$ 
increases to 46. As a result, these lodged vortices interact more frequently, and with a 
larger extent of the groove surface. In this regard, a notable rise of turbulent 
fluctuations would be expected, and it is reflected on the flow statistics presented in 
the preceding section. In addition, it is likely that the mutual interactions among the 
vortices are invigorated because of being confined within the groove, and all along 
subjected to the strong non-linear inertial effects exerted by the overlying flow 
structures. Naturally, the more complex and energetic turbulent motions would augment the 
friction drag significantly.

	 One can also relate the intensified turbulent motions to the appearance of a second 
peak on the profile of $-\overline{u'v'}$ for Cases III and VI as shown in 
figures~\ref{UVmslgplusMid} and~\ref{UVmslgplusVal}. The same goes for the hump that 
formed on the profile of $u'_{rms}$ for Case VI, see figure~\ref{UrmslgplusVal}. 
On the other hand, the fact that discernible humps are only formed on the profiles of 
$w'_{rms}$ above the tip demonstrates that riblets are still rather effective in impeding 
spanwise turbulent motions in the groove. However, Fig.~\ref{WrmslgplusVal} reveals that 
there is a marginal enhancement of $w'_{rms}$ in the groove in Cases III and VI with 
$\lgp > 30$, which manifest as a slight hump on the statistical profiles. Although there 
are strong indications that the lodging of vortices lead to intensified, and dynamically 
complex near-wall turbulent motions, its implications on the viscous drag cannot be 
established definitively. One would have to isolate and study the impact of their actions 
around riblets under a controlled setting.
%
%=====================================================================================================================================================================================================%
%
\begin{figure*}[t]
	\begin{minipage}[l]{0.325\textwidth}
		\raggedleft
		\subfigure[][]
		{
			% trim option's parameter order: left bottom right top
			\includegraphics[trim = 40mm 10mm 40mm 7mm, clip, width = 0.92\textwidth]{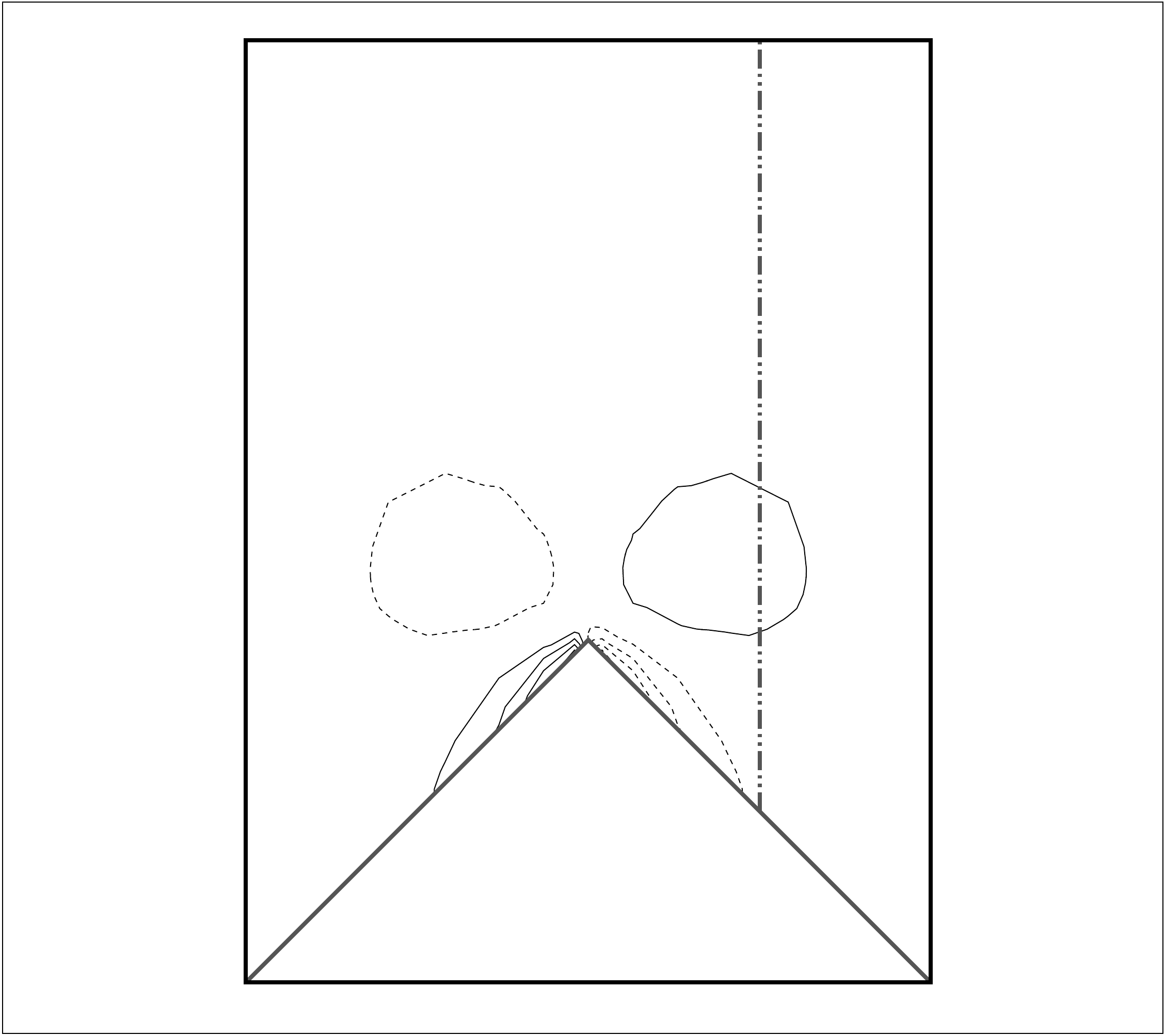}
			\label{Omegayz1842-s0p2-h0p1-avg}
		}
	\end{minipage}
	\hfill
	\begin{minipage}[c]{0.325\textwidth}
		\centering
		\subfigure[][]
		{
			% trim option's parameter order: left bottom right top
			\includegraphics[trim = 40mm 10mm 40mm 7mm, clip, width = 0.92\textwidth]{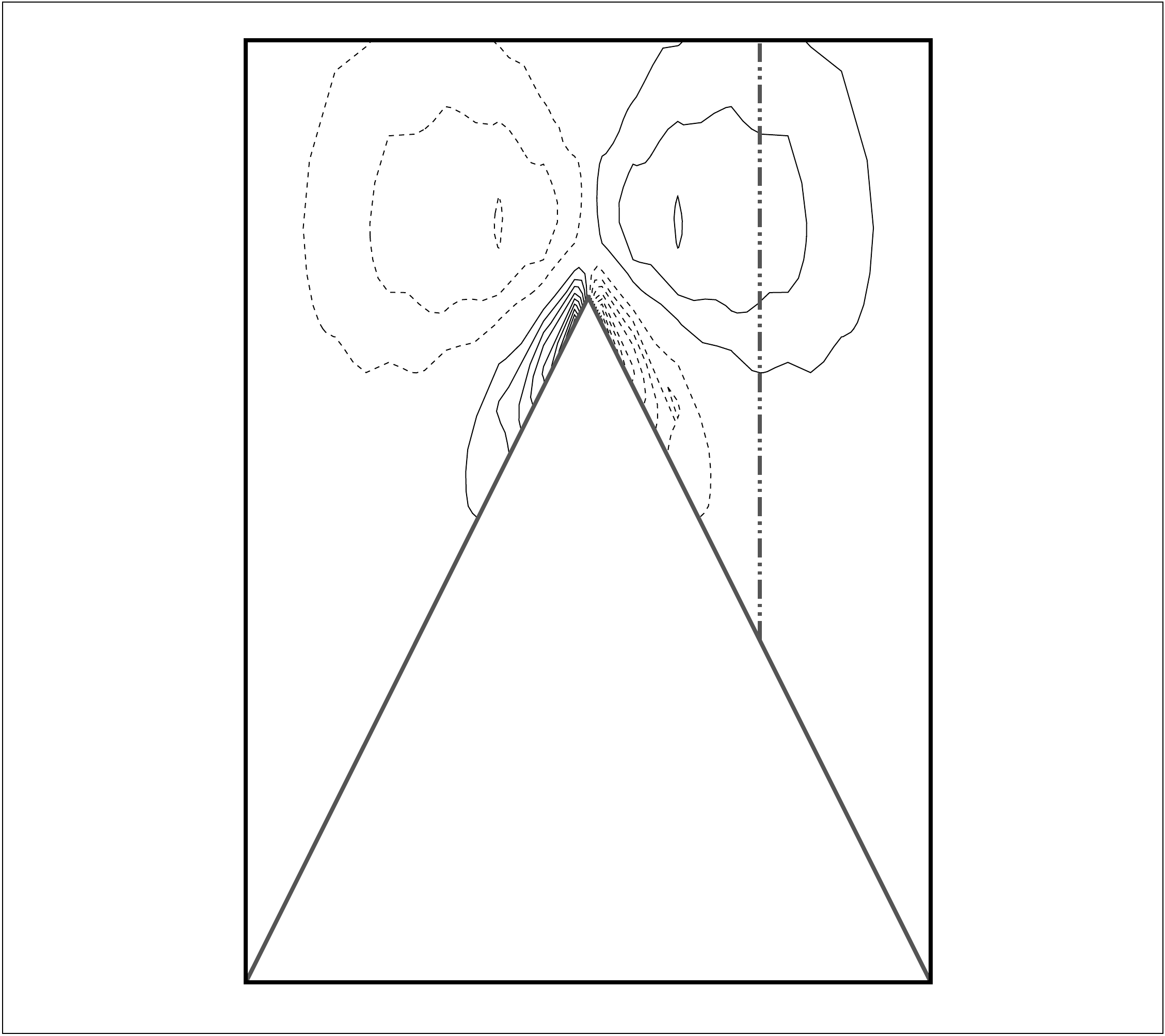}
			\label{Omegayz1842-s0p2-h0p2-avg}
		}
	\end{minipage}
	\hfill
	\begin{minipage}[r]{0.325\textwidth}
		\raggedright
		\subfigure[][]
		{
			% trim option's parameter order: left bottom right top
			\includegraphics[trim = 40mm 10mm 40mm 7mm, clip, width = 0.92\textwidth]{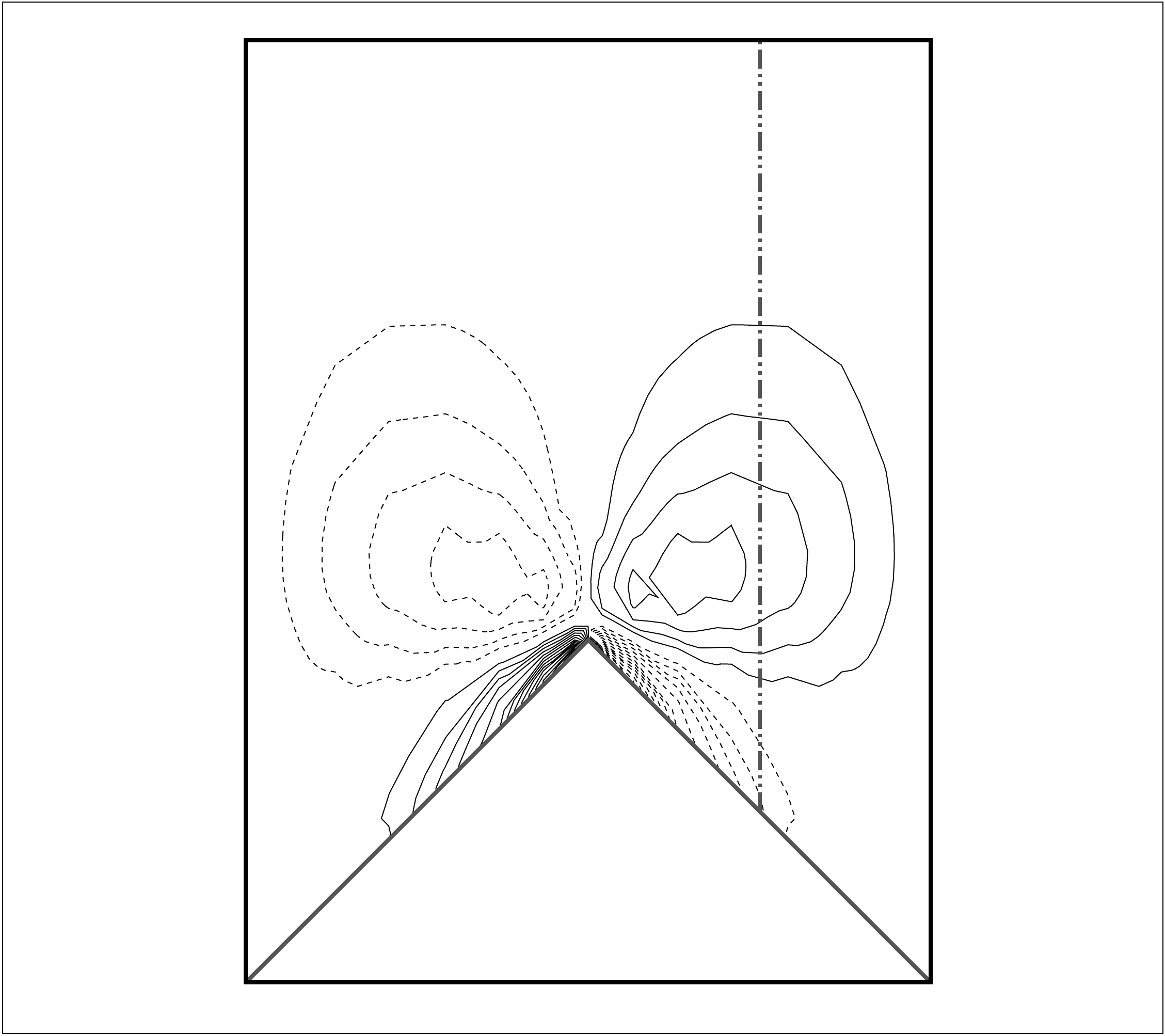}
			\label{Omegayz2800-s0p2-h0p1-avg}
		}
	\end{minipage}
	
	\begin{minipage}[l]{0.325\textwidth}
		\raggedleft
		\subfigure[][]
		{
			% trim option's parameter order: left bottom right top
			\includegraphics[trim = 40mm 10mm 40mm 7mm, clip, width = 0.92\textwidth]{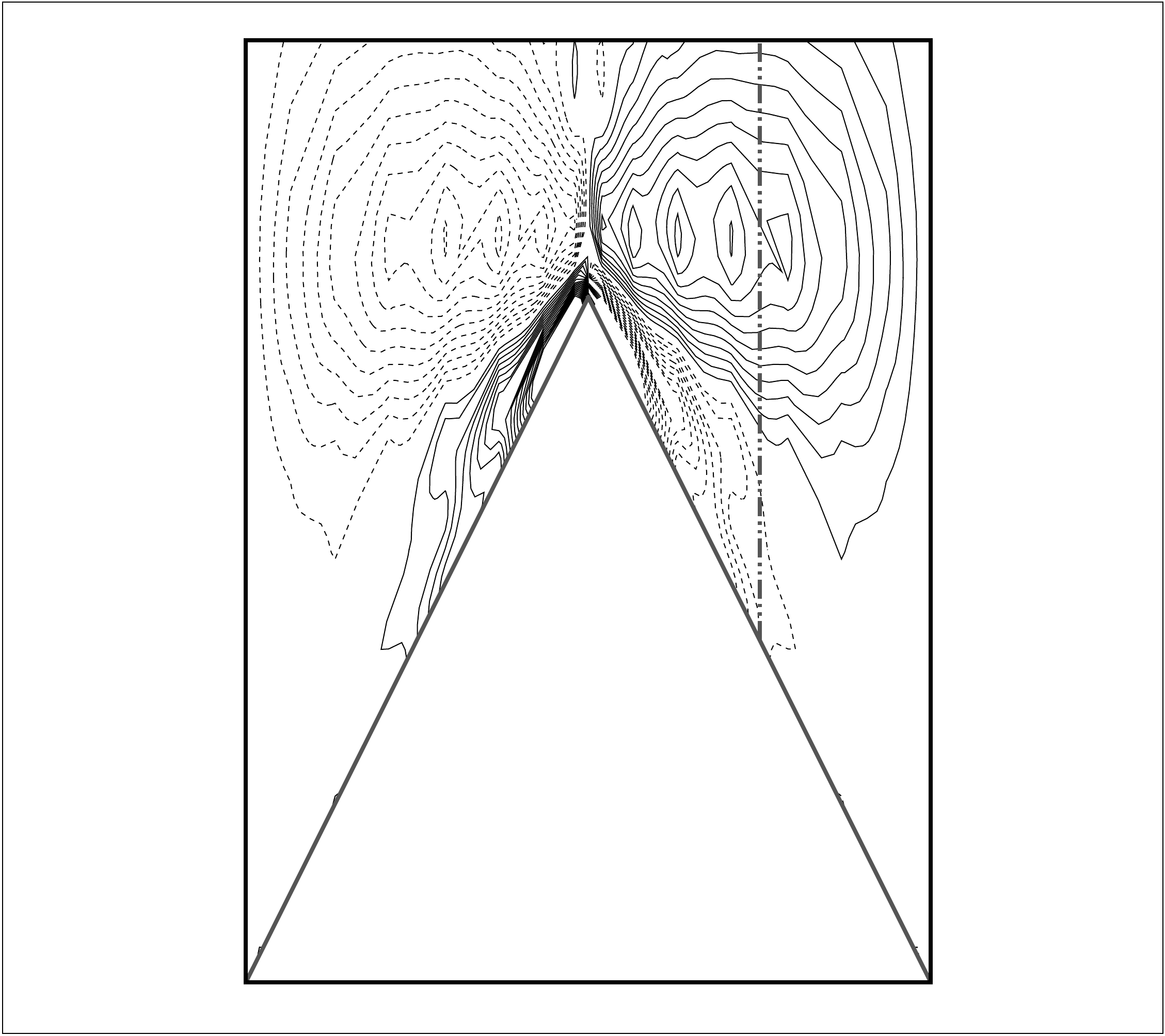}
			\label{Omegayz2800-s0p2-h0p2-avg}
		}
	\end{minipage}
	\hfill
	\begin{minipage}[c]{0.325\textwidth}
		\centering
		\subfigure[][]
		{
			% trim option's parameter order: left bottom right top
			\includegraphics[trim = 40mm 10mm 40mm 7mm, clip, width = 0.92\textwidth]{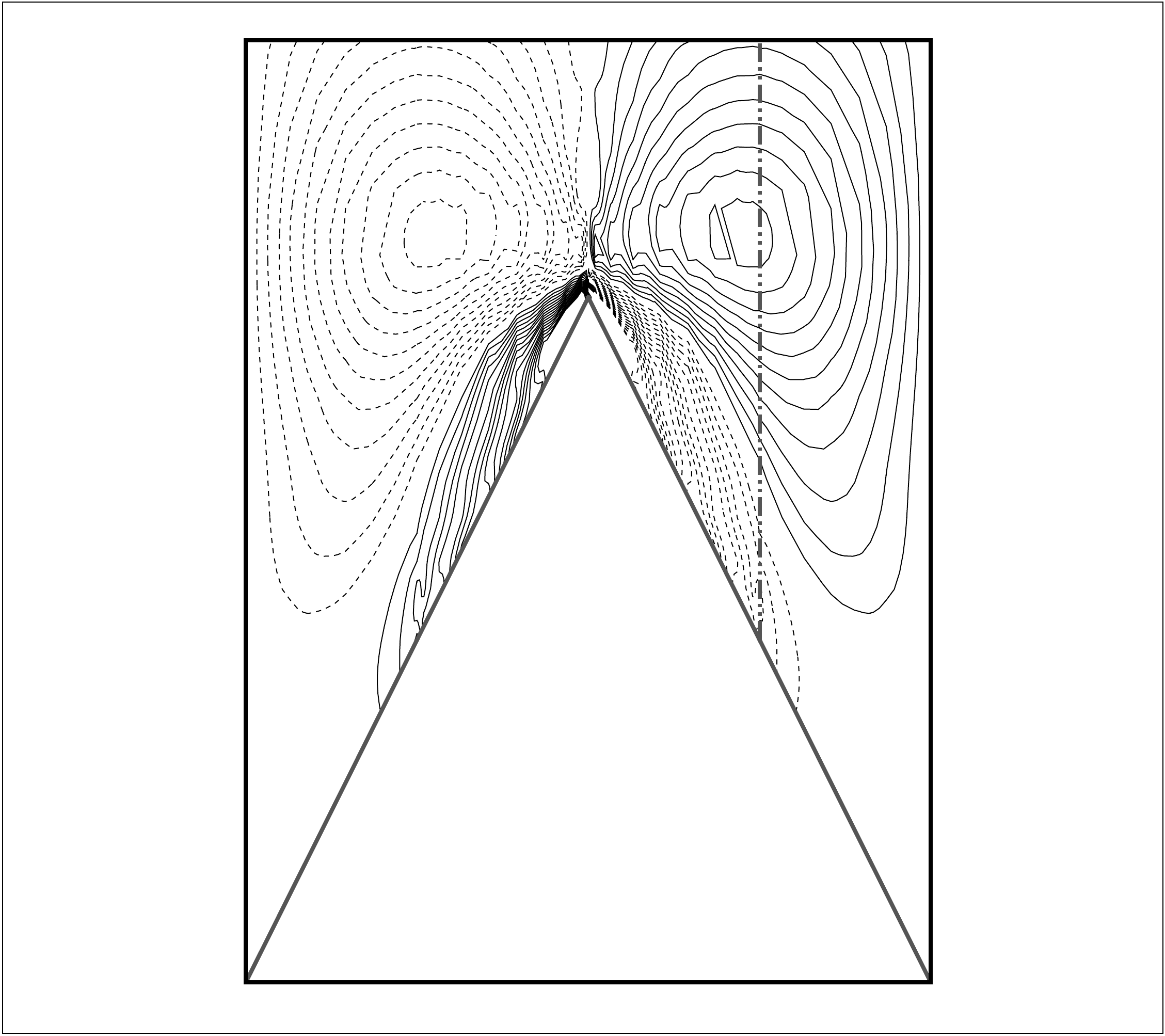}
			\label{Omegayz1842-s0p4-h0p4-avg}
		}
	\end{minipage}
	\hfill
	\begin{minipage}[r]{0.325\textwidth}
		\raggedright
		\subfigure[][]
		{
			% trim option's parameter order: left bottom right top
			\includegraphics[trim = 40mm 10mm 40mm 7mm, clip, width = 0.92\textwidth]{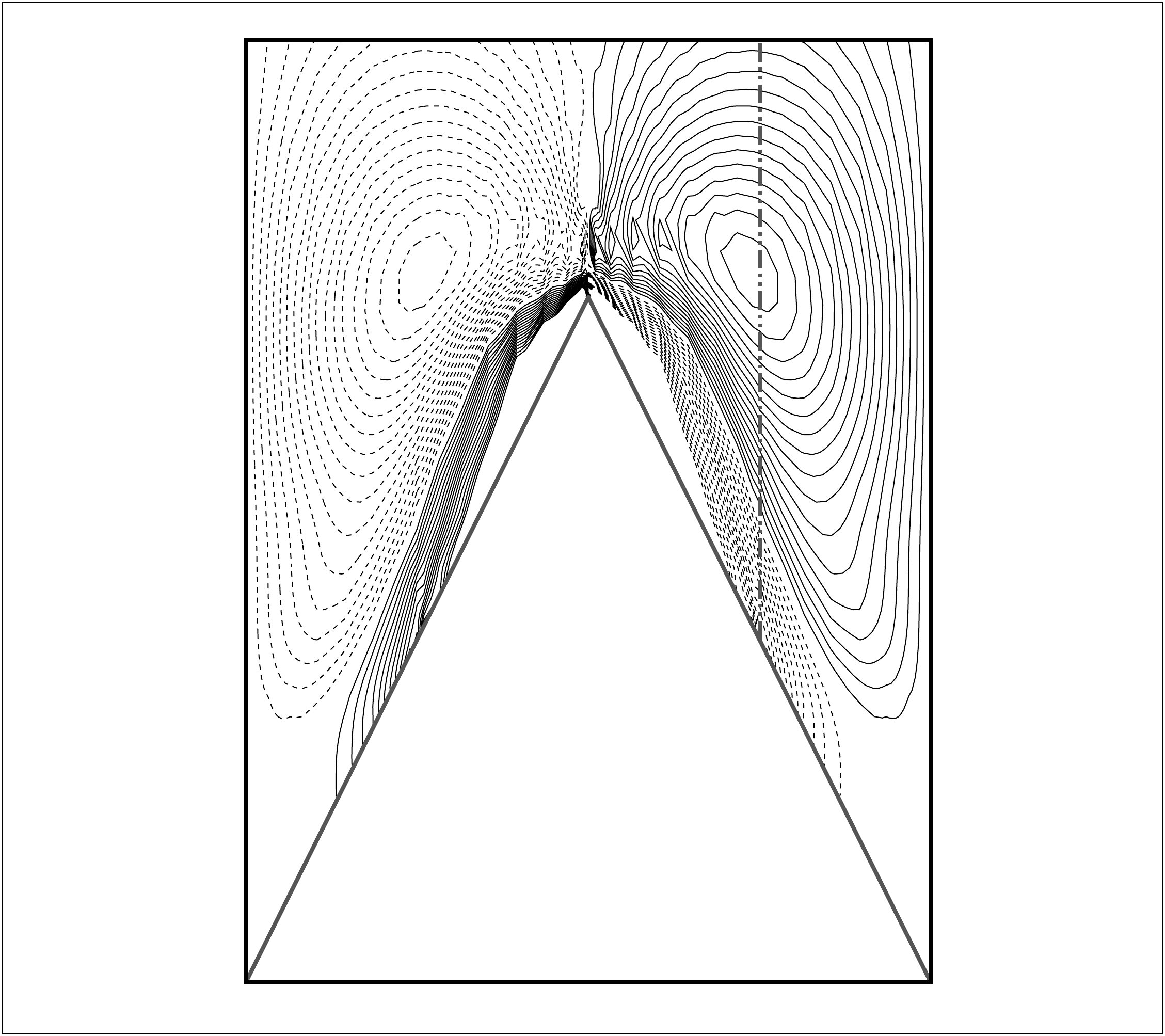}
			\label{Omegayz2800-s0p4-h0p4-avg}
		}
	\end{minipage}
			
	\caption
		{
			Comparison of near-wall time-averaged streamwise vorticity contour $\overline{\omega}_x$ near the bottom (riblet) wall: 
			\subref{Omegayz1842-s0p2-h0p1-avg} $\lgp \approx 10$ (Case I),
			\subref{Omegayz1842-s0p2-h0p2-avg} $\lgp \approx 12$ (Case II),
			\subref{Omegayz2800-s0p2-h0p1-avg} $\lgp \approx 15$ (Case IV),
			\subref{Omegayz2800-s0p2-h0p2-avg} $\lgp \approx 19$ (Case V),
			\subref{Omegayz1842-s0p4-h0p4-avg} $\lgp \approx 30$ (Case III), and 
			\subref{Omegayz2800-s0p4-h0p4-avg} $\lgp \approx 46$ (Case VI).
			Note that dashed lines indicate negative vorticity (counterclockwise rotation). Figures are rescaled to visually the same riblet spacing $s/\delta$ for comparison purpose, and the contours have the same number of levels and range. The vertical lines (\protect\parbox{28pt}{\textcolor{tecplotgray}{\hdashrule[0ex]{20pt}{1.5pt}{8pt 3pt 2pt 2pt 2pt 3pt}\hdashrule[0ex]{8pt}{1.5pt}{}}}) indicate the riblet mid-point.
			\label{Omegayz-avg}
		}
\end{figure*}
%
%=====================================================================================================================================================================================================%
% 

	Figure~\ref{Omegayz-avg} presents the contours of streamwise component of time-averaged 
vorticity $\overline{\omega}_x$ of the six configurations to look at the impact of 
increasing $\lgp$ on the mean secondary flow. The first observation is that the strength 
of the secondary vorticity indicated by the density of the contours increases with $\lgp$. 
In particular, it strengthens drastically when $\lgp$ transitions from 15 to 19, which 
coincides with the divergent of the profiles depicted in Fig.~\ref{RMSPercentBaseline}. 
Secondly, the secondary flow also tends to reach deeper into the groove as $\lgp$ 
increases. As the secondary flow appears to draw high momentum fluid from the core region 
towards the riblet valley, it could cause the instantaneous lodging of near-wall flow 
structures in the groove. In this regard, it can be postulated that the increasingly 
complex near-wall fluid motions that begins when $\lgp > 15$ are associated with the 
strengthening of the mean secondary flow. In turn, these phenomena seem to correlate with 
the higher viscous drag.
%
%=====================================================================================================================================================================================================%
%
\begin{figure*}[t]
	\begin{minipage}[l]{0.325\textwidth}
		\raggedleft
		\subfigure[][]
		{
			% trim option's parameter order: left bottom right top
			\includegraphics[trim = 40mm 9.5mm 40mm 7mm, clip, width = 0.92\textwidth]{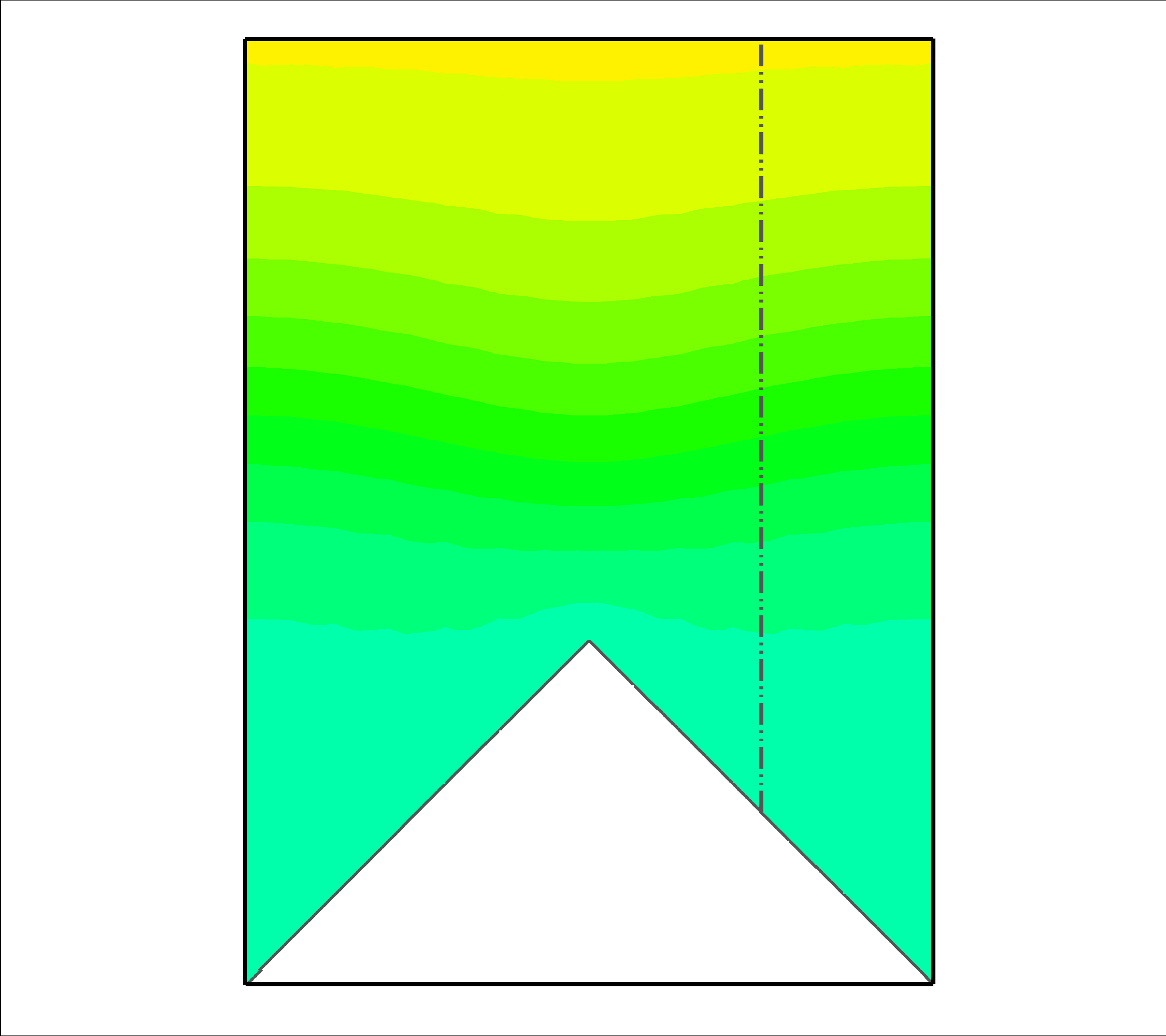}
			\label{ReyStressyz1842-s0p2-h0p1-avg}
		}
	\end{minipage}
	\hfill
	\begin{minipage}[c]{0.325\textwidth}
		\centering
		\subfigure[][]
		{
			% trim option's parameter order: left bottom right top
			\includegraphics[trim = 40mm 9.5mm 40mm 7mm, clip, width = 0.92\textwidth]{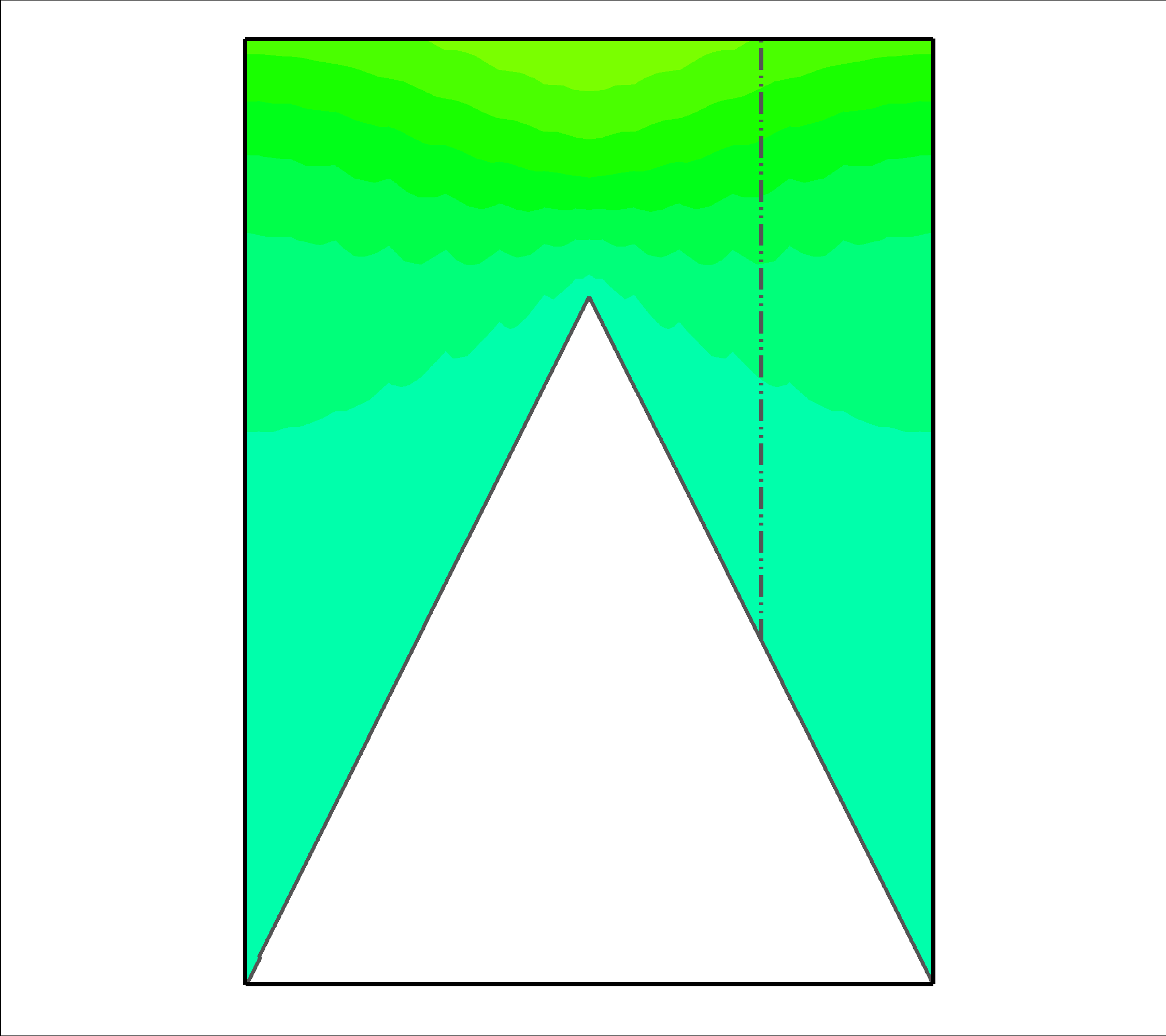}
			\label{ReyStressyz1842-s0p2-h0p2-avg}
		}
	\end{minipage}
	\hfill
	\begin{minipage}[r]{0.325\textwidth}
		\raggedright
		\subfigure[][]
		{
			% trim option's parameter order: left bottom right top
			\includegraphics[trim = 40mm 9.5mm 40mm 7mm, clip, width = 0.92\textwidth]{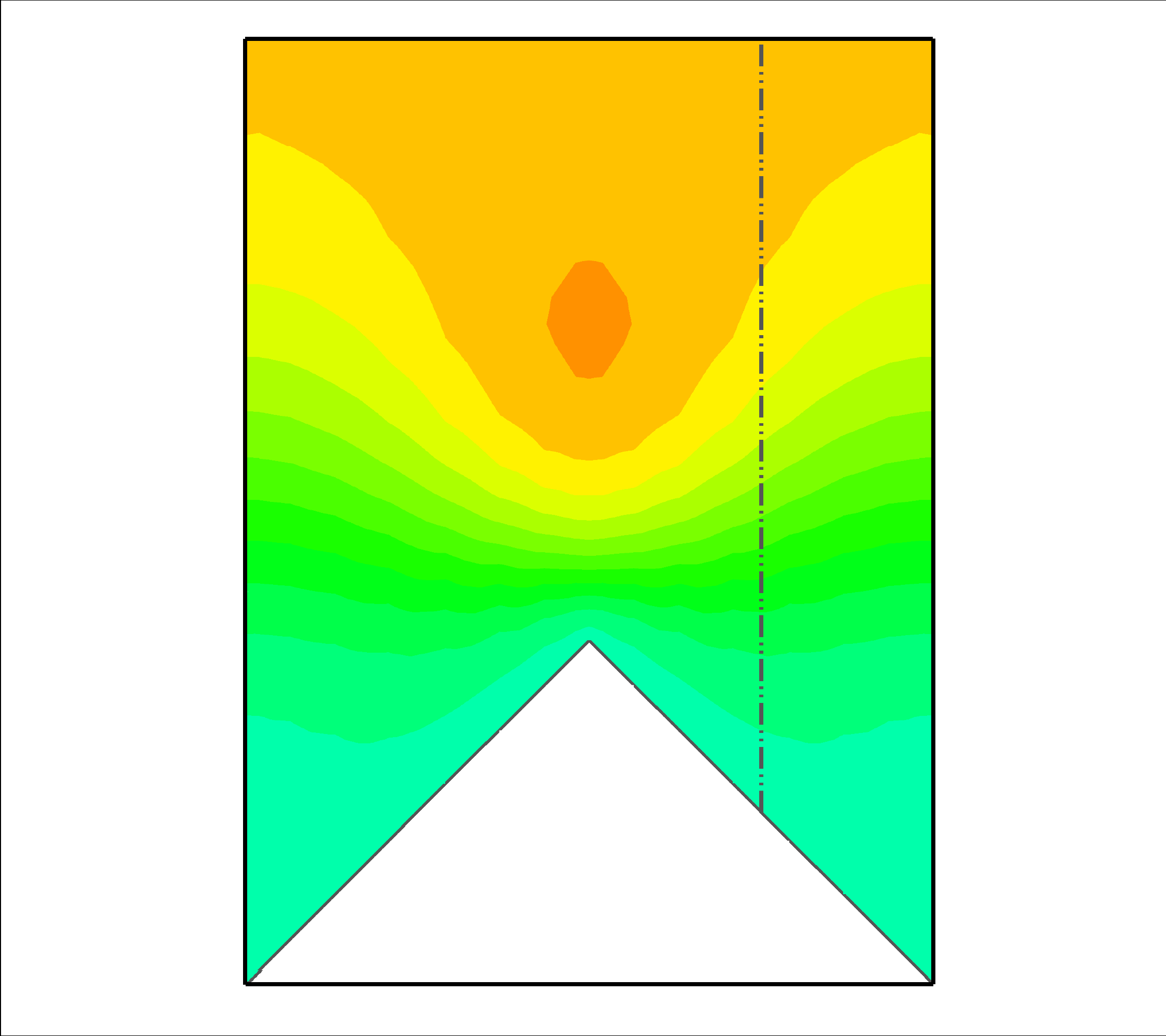}
			\label{ReyStressyz2800-s0p2-h0p1-avg}
		}
	\end{minipage}
	
	\begin{minipage}[l]{0.325\textwidth}
		\raggedleft
		\subfigure[][]
		{
			% trim option's parameter order: left bottom right top
			\includegraphics[trim = 40mm 9.5mm 40mm 7mm, clip, width = 0.92\textwidth]{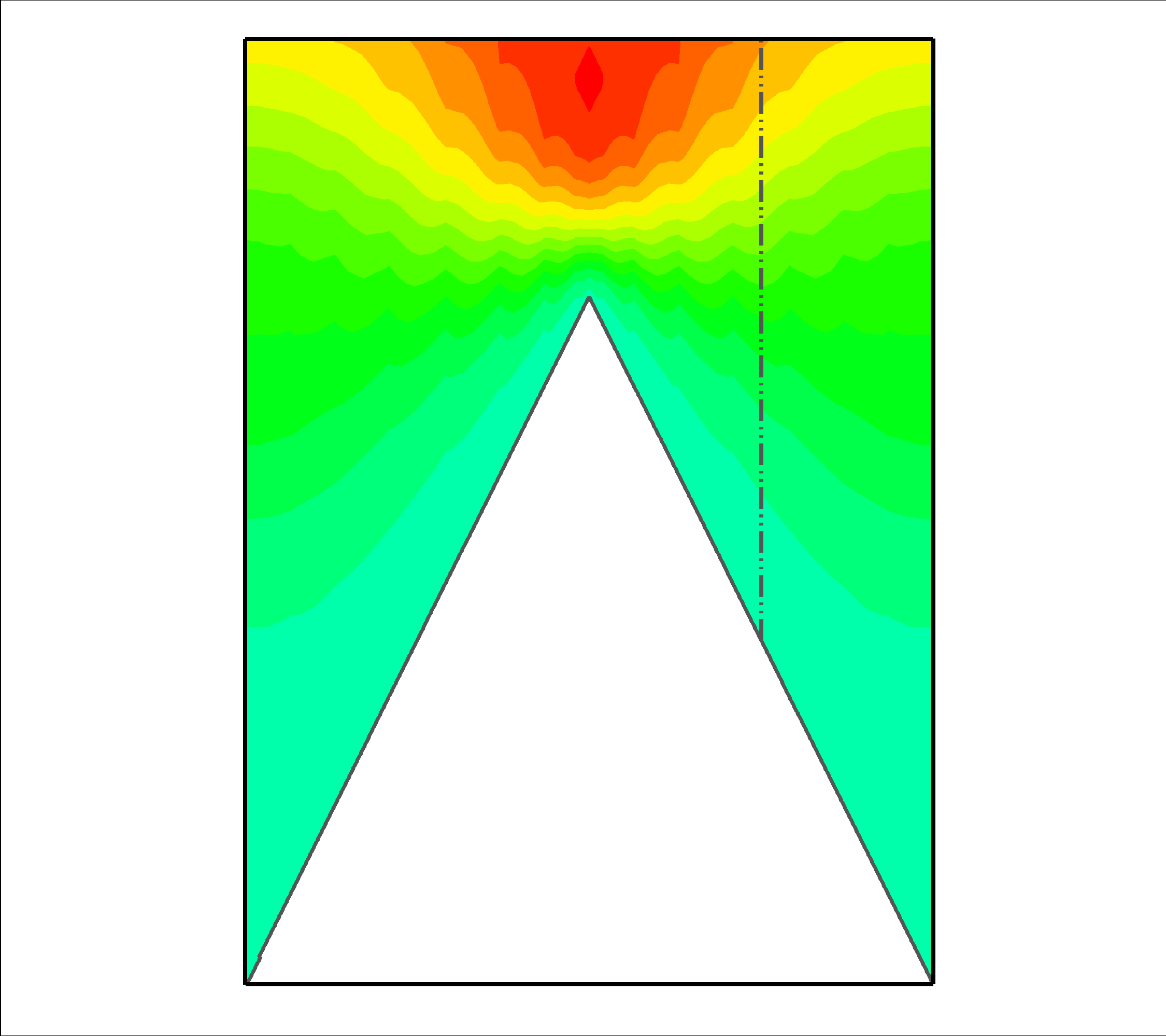}
			\label{ReyStressyz2800-s0p2-h0p2-avg}
		}
	\end{minipage}
	\hfill
	\begin{minipage}[c]{0.325\textwidth}
		\centering
		\subfigure[][]
		{
			% Insert 'grid, tics=10' in the square bracket to show the grid in 10% intervals.
			% trim option's parameter order: left bottom right top
			\begin{overpic}[trim = 40mm 9.5mm 40mm 7mm, clip, width = 0.92\textwidth]{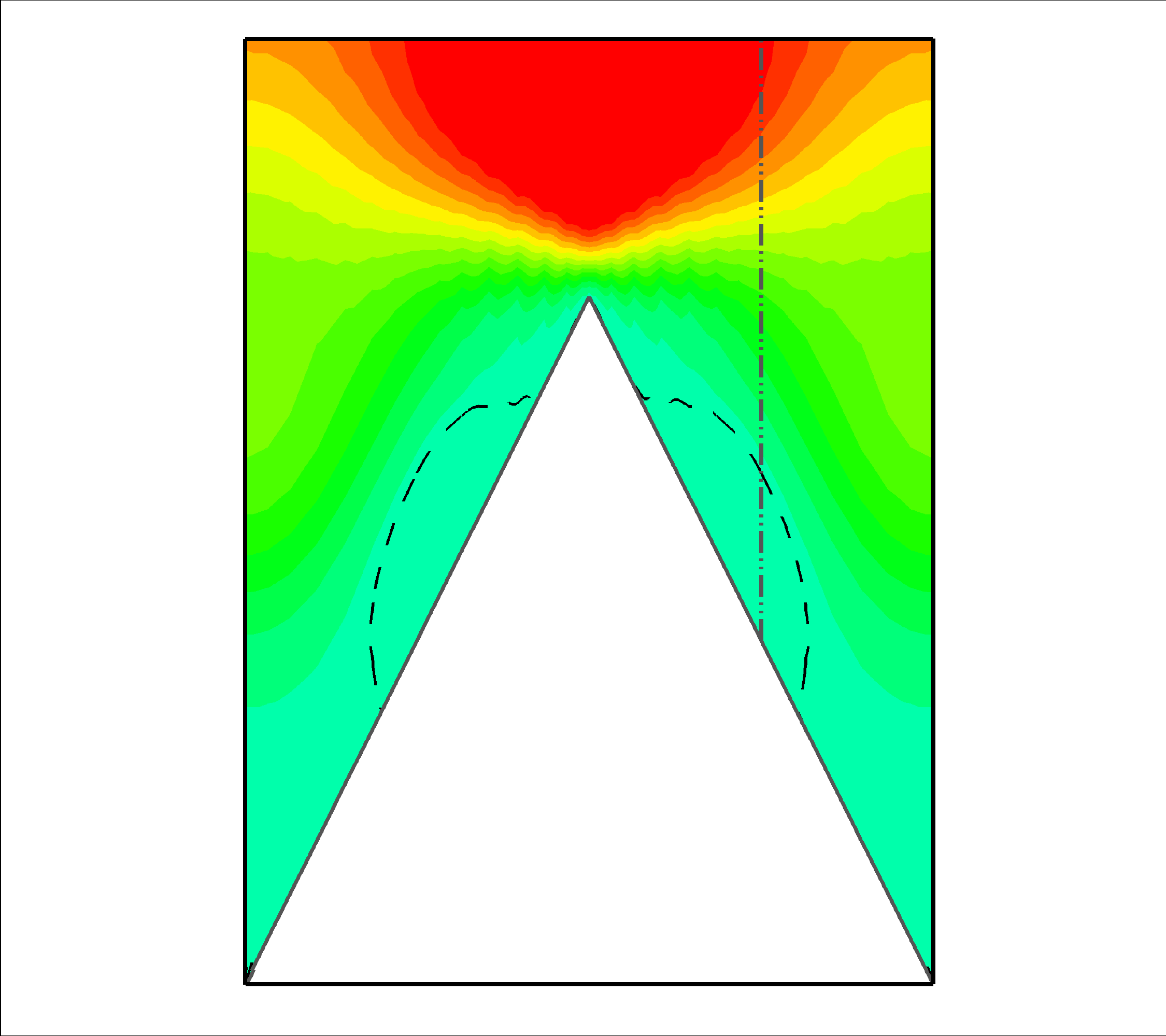}
				\put(22,27.5){\parbox{0.4275\linewidth}{\linespread{1}\centering\large $\Bigg\uparrow$\hfill$\Bigg\uparrow$\\ \scriptsize Regions with inverted sign of $-\overline{u'v'}$}}
			\end{overpic}
			\label{ReyStressyz1842-s0p4-h0p4-avg}
		}
	\end{minipage}
	\hfill
	\begin{minipage}[r]{0.325\textwidth}
		\raggedright
		\subfigure[][]
		{
			% Insert 'grid, tics=10' in the square bracket to show the grid in 10% intervals.
			% trim option's parameter order: left bottom right top
			\begin{overpic}[trim = 40mm 9.5mm 40mm 7mm, clip, width = 0.92\textwidth]{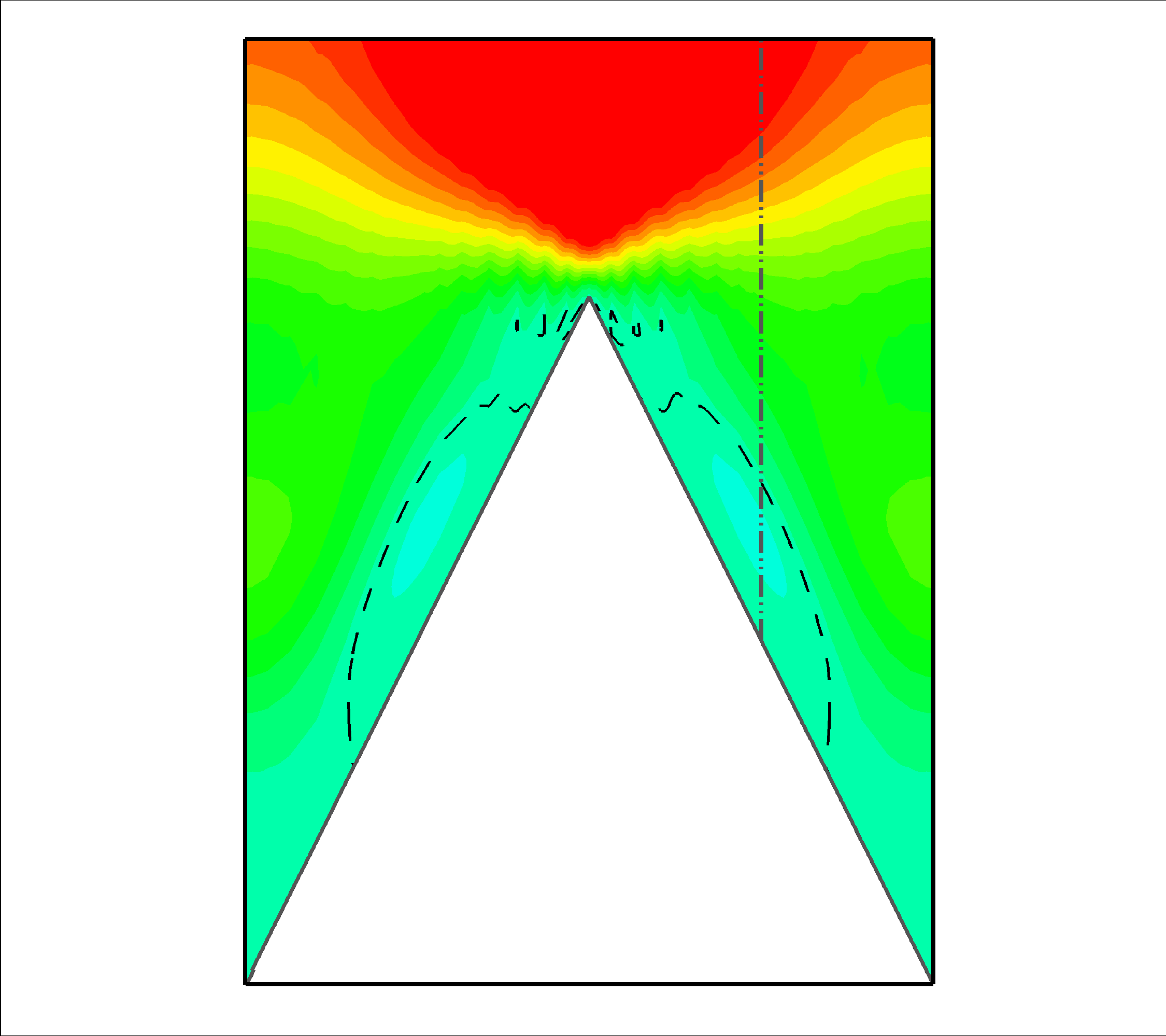}
				\put(22,27.5){\parbox{0.4275\linewidth}{\linespread{1}\centering\large $\Bigg\uparrow$\hfill$\Bigg\uparrow$\\ \scriptsize Regions with inverted sign of $-\overline{u'v'}$}}
			\end{overpic}
			\label{ReyStressyz2800-s0p4-h0p4-avg}
		}
	\end{minipage}
			
	\caption
		{
			Comparison of time-averaged Reynolds shear stress $-\overline{u'v'}$ near the bottom (riblet) wall: 
			\subref{ReyStressyz1842-s0p2-h0p1-avg} $\lgp \approx 10$ (Case I),
			\subref{ReyStressyz1842-s0p2-h0p2-avg} $\lgp \approx 12$ (Case II),
			\subref{ReyStressyz2800-s0p2-h0p1-avg} $\lgp \approx 15$ (Case IV),
			\subref{ReyStressyz2800-s0p2-h0p2-avg} $\lgp \approx 19$ (Case V),
			\subref{ReyStressyz1842-s0p4-h0p4-avg} $\lgp \approx 30$ (Case III), and 
			\subref{ReyStressyz2800-s0p4-h0p4-avg} $\lgp \approx 46$ (Case VI).
			Figures are rescaled to visually the same riblet spacing $s/\delta$ for comparison purpose, and the contours have the same number of levels and range. The vertical lines (\protect\parbox{28pt}{\textcolor{tecplotgray}{\hdashrule[0ex]{20pt}{1.5pt}{8pt 3pt 2pt 2pt 2pt 3pt}\hdashrule[0ex]{8pt}{1.5pt}{}}}) indicate the riblet mid-point, whereas the contour lines (\protect\parbox{30pt}{\textcolor{black}{\hdashrule[0ex]{22pt}{1.5pt}{8pt 3pt}\hdashrule[0ex]{8pt}{1.5pt}{}}}) demarcate the level of $-\overline{u'v'} = 0$.
			\label{ReyStress-avg}
		}
\end{figure*}
%
%=====================================================================================================================================================================================================%
%

	Figure~\ref{ReyStress-avg} depicts the distribution of $-\overline{u'v'}$ around the 
six configurations of riblets. One can see that a sign change of $-\overline{u'v'}$, 
demarcated by the dashed contour lines, occurs in both Cases III ($\lgp \approx 30$) and 
VI ($\lgp \approx 46$). Hence, the humps on the corresponding statistical profiles 
depicted in Fig.~\ref{UVmslgplusMid} are the manifestation of such region since it lies 
in the proximity of the riblet mid-point where the profiles are extracted. A comparison 
between figures~\ref{Omegayz-avg} and~\ref{ReyStress-avg} shows that such region is found 
along the boundary where the mean secondary flow meets the opposing vortex induced on the 
riblet surface. Furthermore, the region penetrates deeper into the groove and its 
magnitude becomes larger as $\lgp$ increases from 30 to 46. Both features suggest that the 
sign inversion is associated with the induced mean secondary flow. In an earlier work, 
\citet{Crawford96} showed that the spanwise variation of the Reynolds stress component 
$-\overline{u'w'}$ can modify the gradient of $-\overline{u'v'}$, and the sign of 
$\partial \overline{u'w'}/\partial z$ has a direct connection with the wall-normal 
transport of fluid by the secondary flow. 

	\citet{Goldstein98} conjectured that the generation of secondary flow involves spanwise 
sloshing of the flow near the riblet elements due to the presence of spanwise velocity 
fluctuations $w'$ having a range of frequencies and amplitudes. Indeed, a recap of 
Fig.~\ref{RMSPercentBaseline} reveals that the peak magnitude of $w'_{rms}$ increases at 
a greater extent than the one in the wall-normal direction when $\lgp > 15$. One plausible 
explanation could be that larger riblets become less effective in impeding the lateral 
motions of near-wall flow structures. Moreover, since the groove spacings for Cases IV, 
III and VI (see table~\ref{tab:RibletPerformance}) all of which with $\lgp \gtrapprox 15$ 
are larger than 30 wall units, the lateral motions near-wall flow structures could be 
promoted rather than mitigated while they are evolving and mutually interacting over 
riblets. A related conjecture is that larger riblets give rise to stronger spanwise 
variation of the near-wall flow field, and it in turn leads to force imbalance that can 
induce cross-flow motions.
	
\subsection{Additional correlations with the size of riblets}
\label{sec:AddScalingSize}

	Apart from serving as a means to characterize the drag reduction performance, the present 
work intends to explore the potential of $\lgp$ in correlating the changes of other 
interesting flow properties with the size of riblets. Of particular interest is whether 
there exists a connection between the riblet size defined in terms of $\lgp$ and the 
overall turbulent flow field in the channel. In earlier works, it was postulated 
that mitigating Reynolds stresses could result in a significant skin friction drag 
reduction~\citep{Fukagata02}. Likewise, it has been demonstrated that the skin friction 
drag is predominantly contributed by the Reynolds stresses in flows at high Reynolds 
number~\citep{Iwamoto05}. As such, the present study would like to examine the implications 
of riblets on the fluctuating velocity field in the light of $\lgp$.

	Based on an energy consideration of the fluctuating velocity field, \citet{Marusic07} 
showed that the criterion for achieving sustained sub-laminar drag in a channel flow with 
a fixed volume flux is equivalent to exceeding the volume flux of laminar flow in a 
channel flow driven by a fixed pressure gradient. The criterion is given as:
%
%=====================================================================================================================================================================================================%
%
\begin{align}
	\Gamma > \left<\left|\nabla\boldsymbol{u} \right|^2 \right> + \Rem^2\left<\left[\hspace{0.15em} \overline{u'v'} - \left<\overline{u'v'} \right>\hspace{0.15em}\right]^2 \right>
	\label{eq:GammaCritEq}
\end{align}
%
%=====================================================================================================================================================================================================%
%
where $\Gamma$ is a parameter related to the choice of control scheme. The overbar denotes 
averaging in time and along the homogeneous directions. $<>$ denotes bulk averaging over 
the channel cross-section. $\boldsymbol{u}$ is the velocity fluctuations vector. Note that 
all flow quantities in expression~\eqref{eq:GammaCritEq} have been normalized accordingly 
by the length scale $\delta$ and velocity scale $\Um$. 

	Let us denote the right hand side of expression~\eqref{eq:GammaCritEq} as $\Gamma^*$. 
In its original derivation, $\Gamma^*$ denotes the threshold for blowing or suction 
techniques to achieve sustainable sub-laminar drag in a channel flow. The larger is the 
value of $\Gamma^*$, the more difficult it is to damp out the turbulent 
stresses~\citep{Marusic07}. In the present context, this parameter can thus be employed to 
quantify the state of turbulence in the presence of riblets, which is predominantly 
influenced by the quantity $-\overline{u'v'}$. Such effect is captured by the second term 
on the right hand side of equation~\eqref{eq:GammaCritEq}. Although there is an additional 
non-zero component of Reynolds stresses, i.e. $-\overline{u'w'}$, due to the spanwise 
inhomogeneity near the riblet wall, its net effect is nullify after taking a bulk-averaged 
of the flow field because of the symmetry of V-groove riblets.

\begin{figure*}[t]
	\centering
	\begin{minipage}[c]{0.495\textwidth}
		\subfigure[][]
		{
			% trim option's parameter order: left bottom right top
			\includegraphics[trim = 41mm 4mm 57.5mm 15mm, clip, width = 0.965\textwidth]{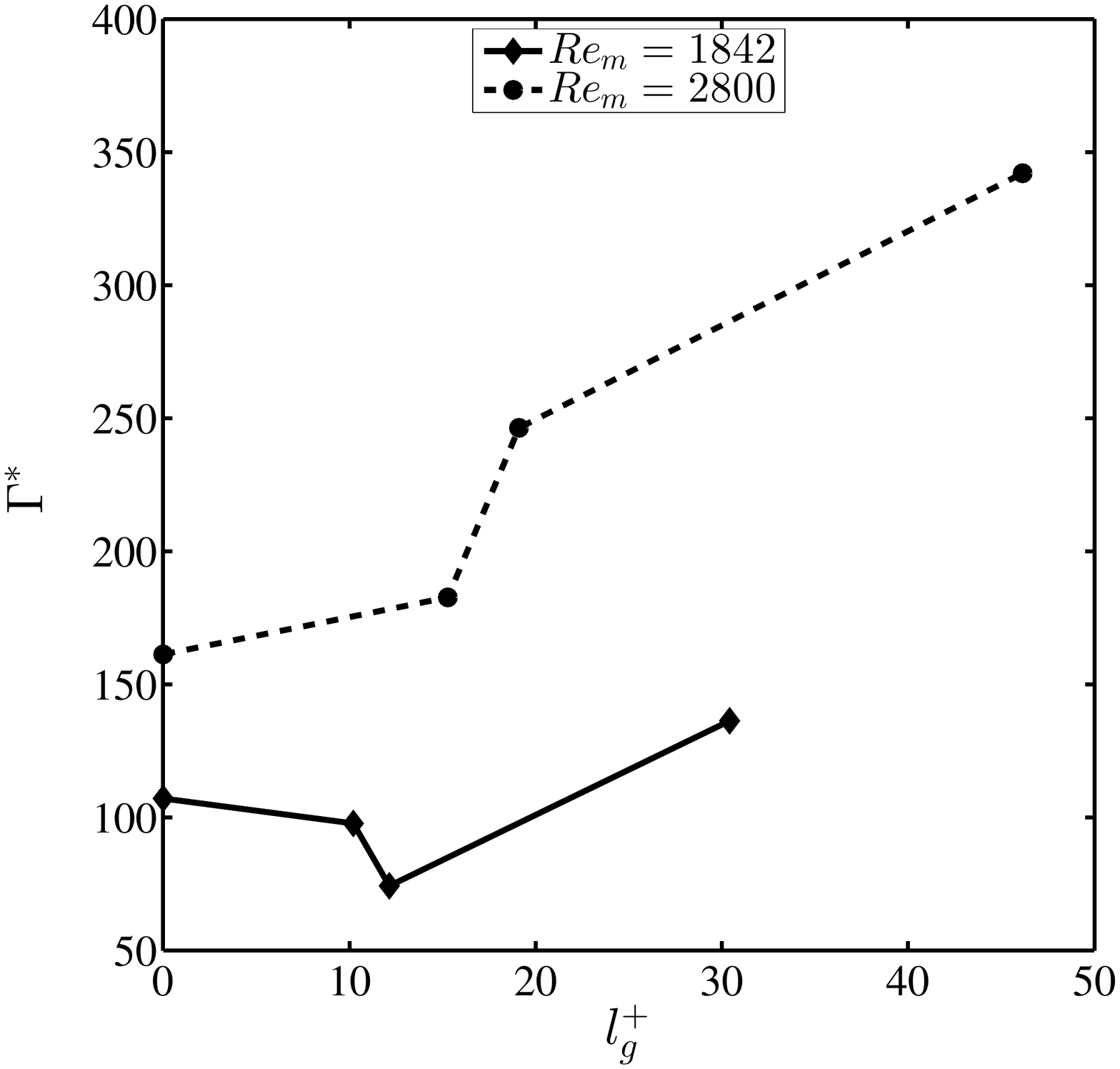}	
			\label{GammaCrit}
		}
	\end{minipage}
	\centering
	\begin{minipage}[c]{0.495\textwidth}
		\subfigure[][]
		{
			% trim option's parameter order: left bottom right top
			\includegraphics[trim = 41mm 4mm 57.5mm 15mm, clip, width = 0.965\textwidth]{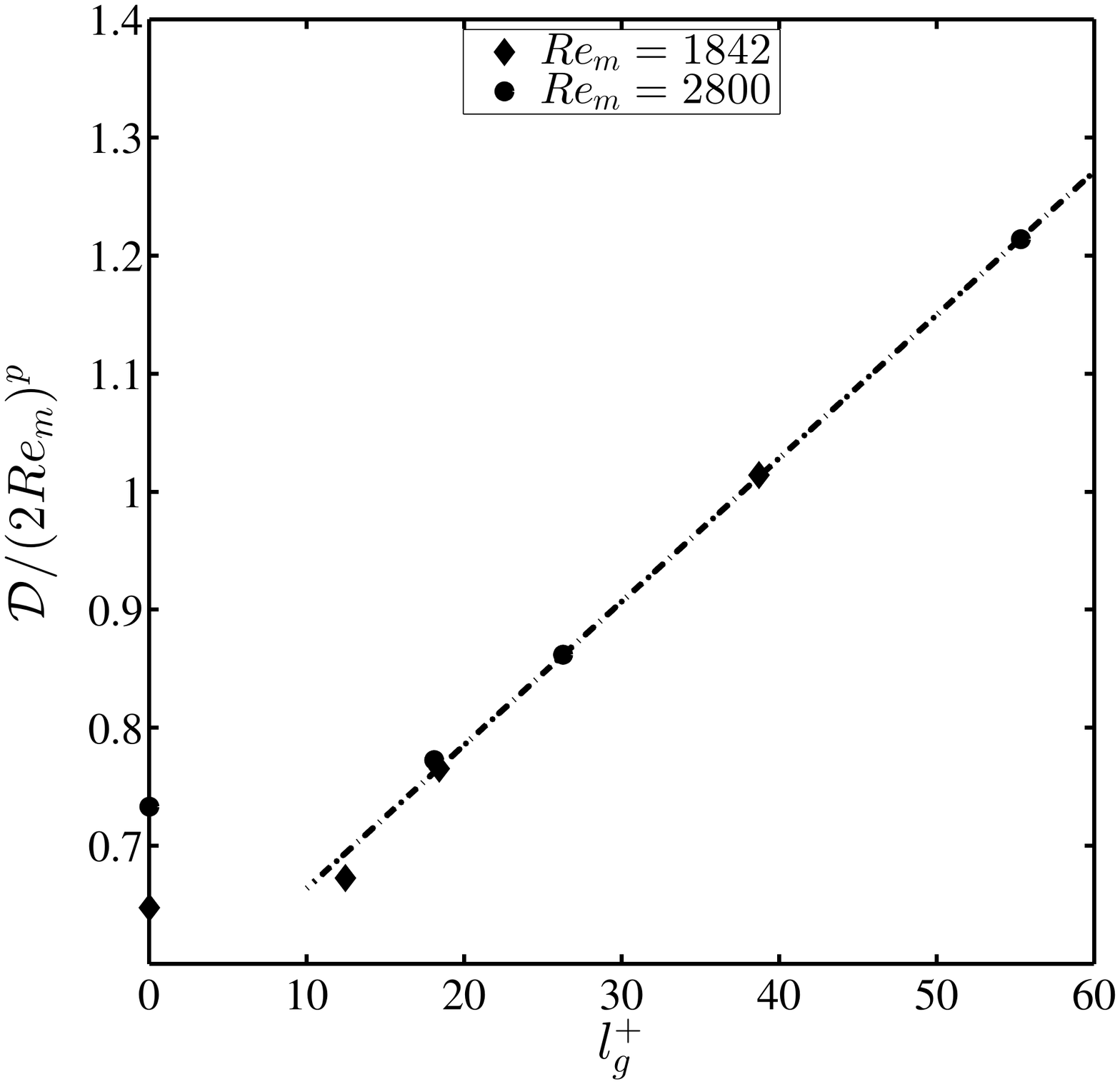}	
			\label{MeanVelMag-Re}
		}
	\end{minipage}
	
	\caption[Profiles of $\Gamma^*$ and $\mathcal{D}/(\Rem)^p$ $(p = 0.26)$ against $\lgp$.]
		{
			\subref{GammaCrit}	   Profiles of $\Gamma^*$ against $\lgp$, and
			\subref{MeanVelMag-Re} Profiles of $\mathcal{D}/(\Rem)^p$, where $p = 0.26$, against $\lgp$.
			Note that the values of $\lgp$ in Fig.~\ref{MeanVelMag-Re} are computed based on the surface-averaged friction velocity on the smooth (top) wall.
			\label{MeanVelMagAll}
		}
\end{figure*}

	Figure~\ref{GammaCrit} shows the profiles of $\Gamma^*$ against $\lgp$. Firstly, the 
baseline value of $\Gamma^*$ at $\Rem = 2800$ is greater than the one at $\Rem = 1842$. 
This is consistent with a more chaotic flow field at higher Reynolds numbers.  Note that 
the baseline configurations mentioned here are simulated with a domain size of $5\delta 
\times 2\delta \times 2\delta$ to be consistent with the riblet configurations. It is 
found that the variations of both percentage drag reduction and $\Gamma^*$ of all the 
six simulated riblet configurations follow the same trend. For the cases where $\lgp 
\approx 10$ and 12, both representing the drag-reducing configurations, the values of 
$\Gamma^*$ are lower than the baseline value. On the contrary, the rest of the 
drag-increasing configurations have values of $\Gamma^*$ that are greater than the 
corresponding baseline values. The similar trend confirms that any attempt to reduce 
the viscous drag would have to involve the manipulation of $\overline{u'v'}$. 
%
%=====================================================================================================================================================================================================%
%
%
%=====================================================================================================================================================================================================%
%

	However, it is noted that the computed percentage drag reduction at $\lgp \approx 10$ 
is greater than the one at $\lgp \approx 12$. The lower $\Gamma^*$ value in Case II is 
not surprising given that the mitigation of $u'$ is more effective when $h/s = 1$, and 
Case II should reduce slightly greater drag according to experimental 
measurements~\citep{Bechert97}. In addition, the percentage drag reduction is subjected 
to an uncertainty of $\pm 1\%$. Disregarding this slight anomaly, $\Gamma^*$ could serve 
as an alternative indicator of the effectiveness of different riblet configurations in 
manipulating the state of turbulence in a channel flow.

	Next, a parameter $\mathcal{D}$ involving the mean velocity gradient or deformation 
tensor is defined:
%
%=====================================================================================================================================================================================================%
%
\begin{align}
	\mathcal{D} = \left<\left|\nabla \overline{\boldsymbol{V}} \right|^2 \right>
\end{align}	
%
%=====================================================================================================================================================================================================%
%
This quantity provides an indication of the degree of deformation in the mean flow. It can 
be perceived as a quantity related to the energy dissipation of the mean flow by viscosity. 
Figure~\ref{MeanVelMag-Re} shows a graphical view of $\mathcal{D}$ normalized by a power 
law of the Reynolds number, i.e. $(\Rem)^p$, where $p = 0.26$. In this case, values of 
$\lgp$ are computed using the surface-averaged friction velocity on the smooth (top) wall. 
The curves at two distinct Reynolds numbers appear to collapse onto a single curve under 
such normalization. Figure~\ref{MeanVelMag-Re} also illustrates that $\mathcal{D}$ is 
linearly proportional to $\lgp$. In other words, the deformation in the mean flow 
intensified with a greater degree of surface manipulation on the bottom wall. This is 
reasonable since larger V-groove riblets should impose stronger gradients on the mean 
flow.

	The choice of power law index $p = 0. 26$ in Fig.~\ref{MeanVelMag-Re} seems to 
coincide with the Blasius scaling used in the empirical relation concerning the skin 
friction coefficient at low Reynolds number~\citep{Dean78}: $C_f = 0.073 \Rem^{-0.25}$. 
More recently, a theoretical analysis~\citep{Yakhot10} shows that $C_f \propto 
Re^{-\theta}$, where $\theta = 0.26 \sim 0.27$. One possible reason why the power law 
index $p = 0.26$ is found to fit the available data well is probably because $\mathcal{D}$ 
and $C_f$ are both derived from the same time-averaged velocity gradient tensor. Hence, 
the Blasius scaling should collapse the profiles at two distinct Reynolds numbers, despite 
the presence of riblets in the channel. Having said that, it should be noted that the 
number of riblet configurations reported in this work may not be a representative set of 
data. Nevertheless, the present work has explored and demonstrated the potential and 
appeal of using the alternative length scale $\lgp$ in describing the turbulent channel 
flow subjected to the manipulation of V-groove riblets.
	
\section{Conclusions}

	A parametric DNS study comprises of six V-groove riblet configurations of different 
sizes has been carried out in a low Reynolds number turbulent channel flow setting. The 
simulations intend to investigate the evolution of the flow dynamics when subjected to 
a systematic change in the riblet size measured by $\lgp$. In all, the present set of 
simulations at either $\Rem = 1842$ or $2800$ has confirmed that $\lgp$ can also provide 
a better characterization of the flow field over V-groove riblets, apart from the 
improved scaling of drag reduction curves. Even when the ratio of $h/s$ is not constant, 
the flow dynamics is found to evolve systematically with $\lgp$, which could not be 
achieved when using either $\Sp$ or $\Hp$ alone. In this respect, $\lgp$ is more 
suitable as the effective Reynolds number that governs the local flow physics near 
V-groove riblets. By analyzing the six riblet configurations of various $\lgp$, it is 
found that the flow dynamics has a dependency on the nature of flow features arising 
from fluid-riblet interaction.

	At the lower end of the range of $\lgp$ considered, the profiles of turbulence statistics 
mostly resemble the baseline plane channel flow, except for a systematic shift in 
accordance with the groove size. Although some lodging of near-wall flow structures may 
happen from time to time, the fluid motions in the groove are less affected by such 
phenomenon. At the same time, the mean secondary flow is weaker, while its region of 
influence is narrower, and further away from the groove. The likelihood of flow structures 
to lodge in the riblet groove appears to be proportional to $\lgp$. Likewise, the strength 
and the degree of penetration of the induced mean secondary flow scale with $\lgp$. In 
particular, the secondary flow becomes increasingly prominent when $\lgp > 15 \sim 19$, 
and this coincides with the more notable enhancement of spanwise velocity fluctuations 
$w'_{rms}$. 

	The lodging of flow structures in the groove and the induced mean secondary flow seem 
to be inter-related and act hand-in-hand to cause complex and intense near-wall turbulent 
motions, which manifest as humps on the statistical profiles. Additionally, their 
increasing dominance appears to correlate with the rise in viscous drag. In those 
configurations with $\lgp > 30$, the pronounced impact of secondary flow produces a sign 
inversion of the Reynolds stress in some part of the groove, while the induced downwash 
may in part cause the entrainment of flow structures inside the groove. Two new 
correlations in terms of $\lgp$ are examined to illustrate the potential and appeal of 
$\lgp$ in correlating the riblet size and the regulated flow field properties.

	Since the present simulations have only considered six configurations of V-groove 
riblets, more extensive assessments that consider riblets of various sizes and shapes, and 
at a higher range of Reynolds numbers would be desirable. Such study can potentially help 
to pinpoint the important flow features or mechanisms that dictate the flow attributes, 
and to provide useful knowledge for utilizing structured surface pattern in a wide range 
of fluid flow applications. One interesting issue is to determine the geometrical 
parameters, apart from the groove size, that affect the lodging of near-wall flow 
structures and the generation of mean secondary flow, as well as their extent of 
influence on the fluid motions in the groove. More importantly, their impact on the 
viscous drag should be assessed via suitable means.

\vspace{\baselineskip}

	The authors wish to acknowledge supports from the National University of Singapore 
and the Ministry of Education, Singapore.

%\bibliography{Reference}

%

\end{document}